%% file: main.tex
\documentclass[10pt,twocolumn,letterpaper]{article}

\usepackage[pagenumbers]{cvpr} %

\input{preamble}

\definecolor{cvprblue}{rgb}{0.21,0.49,0.74}
\usepackage[pagebackref,breaklinks,colorlinks,allcolors=cvprblue]{hyperref}
\usepackage[capitalize]{cleveref}
\crefname{question}{Question}{Questions}

\def\paperID{*****} %
\def\confName{CVPR}
\def\confYear{2026}

\title{What Is the Optimal Ranking Score Between Precision and Recall?\\ We Can Always Find It and It Is Rarely $\scoreFOne$}

\author{S\'ebastien Pi\'erard, Adrien Deli\`ege, and Marc Van Droogenbroeck\\
Montefiore Institute, University of Li\`ege, Li\`ege, Belgium\\
{\tt\small \{S.Pierard,Adrien.Deliege,M.VanDroogenbroeck\}@uliege.be}
}

\begin{document}

\input{macros}

\input{new_macros}
\global\long\def\linkIWDD{\href{https://mivia.unisa.it/iwddcontest2026}{IWDD}\xspace}
\global\long\def\linkCOVIDCT{\href{https://covid-ct.grand-challenge.org/Evaluation/}{COVID-CT}\xspace}
\global\long\def\linkPAIP{\href{https://paip2020.grand-challenge.org/rank/}{PAIP}\xspace}
\global\long\def\linkECDP{\href{https://ecdp2020.grand-challenge.org/Rules/}{ECDP}\xspace}
\global\long\def\linkUNICORN{\href{https://unicorn.grand-challenge.org/tasks/}{UNICORN}\xspace}
\global\long\def\linkTNSCUI{\href{https://tn-scui2020.grand-challenge.org/evaluation_metric/}{TN-SCUI}\xspace}
\global\long\def\linkTDTEETHSEG{\href{https://3dteethseg.grand-challenge.org/Evaluation/}{3D-TEETH-SEG}\xspace}
\global\long\def\linkDRAC{\href{https://drac22.grand-challenge.org/Evaluation/}{DRAC}\xspace}
\global\long\def\linkMIDOG{\href{https://midog2025.grand-challenge.org/evaluation/track-1-mitosis-detection-final-submission-phase/leaderboard/}{MIDOG}\xspace}
\global\long\def\linkCADARRE{\href{https://cada-rre.grand-challenge.org/Assessment/}{CADA-RRE}\xspace}
\global\long\def\linkVAND{\href{https://sites.google.com/view/vand30cvpr2025/challenge?authuser=0}{VAND}\xspace}
\global\long\def\linkABAW{\href{https://affective-behavior-analysis-in-the-wild.github.io/8th}{ABAW}\xspace}

\maketitle
\input{sections/0_abstract}
\input{sections/1_introduction}
\input{sections/2_related_work}
\input{sections/3_0_theory}

\input{sections/4_0_case_studies}

\input{sections/5_conclusion}

\clearpage
{
    \small

\input{main.bbl}
}

\newpage
\onecolumn
\appendix
\input{sections/A_0_supplementary}

\end{document}

%% file: preamble.tex
\newenvironment{myImportantResult}{
	\color{cvprblue}
	\em
}
{}

\usepackage{xargs}

\newcommand{\mysection}[1]{\vspace{2pt}\noindent\textbf{#1}}

\newcommand{\aka}{\emph{a.k.a.}\xspace}

\newcommand{\first}{1\textsuperscript{st}\xspace}
\newcommand{\second}{2\textsuperscript{nd}\xspace}
\newcommand{\third}{3\textsuperscript{rd}\xspace}

\newcommand{\roc}{ROC\xspace}
\newcommand{\pca}{PCA\xspace}
\newcommand{\bgs}{BGS\xspace}

\newcommand{\trivialOrdering}{\protect\usym{2720}\xspace}
\newcommand{\correctOrdering}{\protect\usym{2713}\xspace}
\newcommand{\wrongOrdering}{\protect\usym{2717}\xspace}

\newcommand{\CDnetPlatform}{\href{http://jacarini.dinf.usherbrooke.ca/results2014}{changedetection.net}\xspace}

\newcommand{\CDnetMMXIV}{\emph{CDnet 2014}\xspace}

\usepackage{rotating}
\usepackage{nicefrac}
\usepackage{stix}
\usepackage{pifont}
\usepackage{pifont}
\usepackage{utfsym}
\usepackage{listings} %
\usepackage{romanbar}

\usepackage{amsthm}

\newtheoremstyle{exampstyle}
  {0.5\topsep} %
  {0.5\topsep} %
  {} %
  {} %
  {\bfseries} %
  {.} %
  {.5em} %
  {} %

\theoremstyle{exampstyle}

\graphicspath{{drawings/}}

\usepackage{soul}

\ExplSyntaxOn
\clist_const:Nn \l_soul_protect_clist { cite , citep , ref , eqref }
\NewDocumentCommand{\robusthl}{m}{
  \tl_set:Nn \l_tmpa_tl { #1 }
  \clist_map_inline:Nn \l_soul_protect_clist {
    \regex_replace_all:nnN { (\c{##1}[^{}]*\{[^{}]*\}) } %
                           { \c{mbox}\{\1\} } \l_tmpa_tl
  }
  \hl\l_tmpa_tl
}
\ExplSyntaxOff

\usepackage{titletoc}

%% file: macros.tex
\newcommand{\paperA}{paper~A~\cite{Pierard2024Foundations-arxiv}\xspace}
\newcommand{\paperB}{paper~B~\cite{Pierard2024TheTile-arxiv}\xspace}
\newcommand{\paperC}{paper~C~\cite{Halin2024AHitchhikers-arxiv}\xspace}
\newcommand{\PaperA}{Paper~A~\cite{Pierard2024Foundations-arxiv}\xspace}
\newcommand{\PaperB}{Paper~B~\cite{Pierard2024TheTile-arxiv}\xspace}
\newcommand{\PaperC}{Paper~C~\cite{Halin2024AHitchhikers-arxiv}\xspace}

\global\long\def\sampleSpace{\Omega}%
\global\long\def\aSample{\omega}%
\global\long\def\eventSpace{\Sigma}%
\global\long\def\anEvent{E}%
\global\long\def\measurableSpace{(\sampleSpace,\eventSpace)}%
\global\long\def\expectedValueSymbol{\mathbf{E}}%

\global\long\def\aPerformance{P}%
\global\long\def\allPerformances{\mathbb{\aPerformance}_{\measurableSpace}}%
\global\long\def\aSetOfPerformances{\Pi}%
\global\long\def\randVarSatisfaction{S}%
\global\long\def\aScore{X}%
\global\long\def\allScores{\mathbb{\aScore}_{\measurableSpace}}%
\newcommandx\domainOfScore[1][usedefault, addprefix=\global, 1=\aScore]{\mathrm{dom}(#1)}%
\global\long\def\evaluation{\mathrm{eval}}%
\global\long\def\opFilter{\mathrm{filter}_\randVarImportance}%
\global\long\def\opNoSkill{\mathrm{no\text{\textendash{}}skill}}%
\global\long\def\opPriorShift{\mathrm{shift}_{\pi\rightarrow\pi'}}%
\global\long\def\opChangePredictedClass{\mathrm{change}_{\randVarPredictedClass}}%
\global\long\def\opChangeGroundtruthClass{\mathrm{change}_{\randVarGroundtruthClass}}%
\global\long\def\opSwapGroundtruthAndPredictedClasses{\mathrm{swap}_{\randVarGroundtruthClass\leftrightarrow\randVarPredictedClass}}%
\global\long\def\opSwapClasses{\mathrm{swap}_{\classNeg\leftrightarrow\classPos}}%
\global\long\def\allWorstPerformances{\frownie}
\global\long\def\allBestPerformances{\smiley}

\global\long\def\randVarGroundtruthClass{Y}%
\global\long\def\randVarPredictedClass{\hat{Y}}%
\global\long\def\allClasses{\mathbb{C}}%
\global\long\def\aClass{c}%
\global\long\def\classNeg{c_-}%
\global\long\def\classPos{c_+}%
\global\long\def\sampleTN{tn}%
\global\long\def\sampleFP{fp}%
\global\long\def\sampleFN{fn}%
\global\long\def\sampleTP{tp}%
\global\long\def\eventTN{\{\sampleTN\}}%
\global\long\def\eventFP{\{\sampleFP\}}%
\global\long\def\eventFN{\{\sampleFN\}}%
\global\long\def\eventTP{\{\sampleTP\}}%
\global\long\def\scorePTN{PTN}%
\global\long\def\scorePFP{PFP}%
\global\long\def\scorePFN{PFN}%
\global\long\def\scorePTP{PTP}%
\global\long\def\scoreAccuracy{A}%
\global\long\def\scoreExpectedSatisfaction{\aScore_{\randVarSatisfaction}}%
\global\long\def\scoreTNR{TNR}%
\global\long\def\scoreFPR{FPR}%
\global\long\def\scoreTPR{TPR}%
\global\long\def\scoreFNR{FNR}%
\global\long\def\scoreNPV{NPV}%
\global\long\def\scoreFOR{FOR}%
\global\long\def\scorePPV{PPV}%
\global\long\def\scorePrecision{\scorePPV}%
\global\long\def\scoreFDR{FDR}%
\global\long\def\scoreJaccardNeg{J_-}%
\global\long\def\scoreJaccardPos{J_+}%
\global\long\def\scoreCohenKappa{\kappa}%
\global\long\def\scoreScottPi{\pi}%
\global\long\def\scoreFleissKappa{\kappa}%
\global\long\def\scoreBalancedAccuracy{BA}%
\global\long\def\scoreWeightedAccuracy{WA}%
\global\long\def\scoreYoudenJ{J_Y}
\global\long\def\scorePLR{PLR}%
\global\long\def\scoreNLR{NLR}%
\global\long\def\scoreOR{OR}%
\global\long\def\scoreSNPV{SNPV}%
\global\long\def\scoreSPPV{SPPV}%
\global\long\def\scoreACP{ACP}%
\global\long\def\scoreFOne{F_{1}}%
\newcommandx\scoreFBeta[1][usedefault, addprefix=\global, 1=\beta]{F_{#1}}%
\global\long\def\scoreFOne{\scoreFBeta[1]}%
\global\long\def\priorpos{\pi_+}%
\global\long\def\priorneg{\pi_-}%
\global\long\def\scoreBiasIndex{BI}%
\global\long\def\ratepos{\tau_+}%
\global\long\def\rateneg{\tau_-}%
\global\long\def\scoreACP{ACP}%
\global\long\def\scorePFour{P_4}%
\global\long\def\normalizedConfusionMatrix{C}%
\global\long\def\scoreConfusionMatrixDeterminant{|\normalizedConfusionMatrix|}

\global\long\def\allEntities{\mathbb{E}}%
\global\long\def\entitiesToRank{\mathbb{E}}%
\global\long\def\anEntity{\epsilon}%
\global\long\def\randVarImportance{I}%
\global\long\def\randVarCanonicalImportance{\randVarImportance_{a,b}}
\global\long\def\canonicalRankingScore{\rankingScore[\randVarCanonicalImportance]}
\newcommandx\rankingScore[1][usedefault, addprefix=\global, 1=\randVarImportance]{R_{#1}}%
\global\long\def\canonicalRankingScore{\rankingScore[\randVarImportance_{a,b}]}
\global\long\def\scoreVUT{VUT}%
\global\long\def\tileCurvePriors{\gamma_\pi}
\global\long\def\tileCurveRates{\gamma_\tau}
\global\long\def\relWorseOrEquivalent{\lesssim}%
\global\long\def\relBetterOrEquivalent{\gtrsim}%
\global\long\def\relEquivalent{\sim}%
\global\long\def\relBetter{>}%
\global\long\def\relWorse{<}%
\global\long\def\relIncomparable{\not\lesseqqgtr}%
\global\long\def\rank{\mathrm{rank}_\entitiesToRank}%
\global\long\def\ordering{\relWorseOrEquivalent}%
\global\long\def\invertedOrdering{\relBetterOrEquivalent}%

\global\long\def\LScityscapes{\ding{171}}%
\global\long\def\LSade{\ding{170}}%
\global\long\def\LSvoc{\ding{169}}%
\global\long\def\LScoco{\ding{168}}%

\global\long\def\indicatorSymbol{\mathbf{1}}%
\global\long\def\realNumbers{\mathbb{R}}%
\global\long\def\aRelation{\mathcal{R}}%
\global\long\def\achievableByCombinations{\Phi}%
\global\long\def\allConvexCombinations{\mathrm{conv}}%
\newcommand{\indep}{\perp \!\!\! \perp}

\global\long\def\cityscapes{\LScityscapes{}~Cityscapes}
\global\long\def\ade{\LSade{}~ADE20K}
\global\long\def\voc{\LSvoc{}~Pascal VOC 2012}
\global\long\def\coco{\LScoco{}~COCO-Stuff 164k}

\newcommand{\MethodDesigner}{Bernadette}
\newcommand{\Benchmarker}{Leonard}
\newcommand{\AppDeveloper}{Howard}
\newcommand{\TheoreticalAnalyst}{Sheldon}

\newcommand{\tile}{Tile\xspace}
\newcommand{\tiles}{Tiles\xspace}
\newcommand{\valueTile}{Value Tile\xspace}
\newcommand{\valueTiles}{Value Tiles\xspace}
\newcommand{\baselineTile}{Baseline Value Tile\xspace}
\newcommand{\baselineTiles}{Baseline Value Tiles\xspace}
\newcommand{\SOTATile}{State-of-the-Art Value Tile\xspace}
\newcommand{\SOTATiles}{State-of-the-Art Value Tiles\xspace}
\newcommand{\noSkillTile}{No-Skill Tile\xspace}
\newcommand{\noSkillTiles}{No-Skill Tiles\xspace}
\newcommand{\skillTile}{Relative-Skill Tile\xspace}
\newcommand{\skillTiles}{Relative-Skill Tiles\xspace}
\newcommand{\correlationTile}{Correlation Tile\xspace}
\newcommand{\correlationTiles}{Correlation Tiles\xspace}
\newcommand{\rankingTile}{Ranking Tile\xspace}
\newcommand{\rankingTiles}{Ranking Tiles\xspace}
\newcommand{\entityTile}{Entity Tile\xspace}
\newcommand{\entityTiles}{Entity Tiles\xspace}

\global\long\def\aNonSkilledPerformance{\aPerformance_{\indep}}%
\global\long\def\allNonSkilledPerformances{\mathbb{\aPerformance}^{\randVarGroundtruthClass\indep\randVarPredictedClass}_{\measurableSpace}}%

\global\long\def\allPriorFixedPerformances{\mathbb{\aPerformance}^{\priorpos}_{\measurableSpace}}%

\newcommand{\comma}{\,,}
\newcommand{\point}{\,.}

\newcommandx\unconditionalProbabilisticScore[1]{\aScore_{#1}^{U}}%
\global\long\def\formulaPTN{\unconditionalProbabilisticScore{\eventTN}}%
\global\long\def\formulaPFP{\unconditionalProbabilisticScore{\eventFP}}%
\global\long\def\formulaPFN{\unconditionalProbabilisticScore{\eventFN}}%
\global\long\def\formulaPTP{\unconditionalProbabilisticScore{\eventTP}}%
\global\long\def\formulapriorneg{\unconditionalProbabilisticScore{\{\sampleTN,\sampleFP\}}}%
\global\long\def\formulapriorpos{\unconditionalProbabilisticScore{\{\sampleFN,\sampleTP\}}}%
\global\long\def\formularateneg{\unconditionalProbabilisticScore{\{\sampleTN,\sampleFN\}}}%
\global\long\def\formularatepos{\unconditionalProbabilisticScore{\{\sampleFP,\sampleTP\}}}%
\global\long\def\formulaAccuracy{\unconditionalProbabilisticScore{\{\sampleTN,\sampleTP\}}}%

\newcommandx\conditionalProbabilisticScore[2]{\aScore_{#1 \vert #2}^{C}}%
\global\long\def\formulaTNR{\conditionalProbabilisticScore{\{\sampleTN\}}{\{\sampleTN,\sampleFP\}}}%
\global\long\def\formulaTPR{\conditionalProbabilisticScore{\{\sampleTP\}}{\{\sampleFN,\sampleTP\}}}%
\global\long\def\formulaNPV{\conditionalProbabilisticScore{\{\sampleTN\}}{\{\sampleTN,\sampleFN\}}}%
\global\long\def\formulaPPV{\conditionalProbabilisticScore{\{\sampleTP\}}{\{\sampleFP,\sampleTP\}}}%
\global\long\def\formulaJaccardNeg{\conditionalProbabilisticScore{\{\sampleTN\}}{\{\sampleTN,\sampleFP,\sampleFN\}}}%
\global\long\def\formulaJaccardPos{\conditionalProbabilisticScore{\{\sampleTP\}}{\{\sampleFP,\sampleFN,\sampleTP\}}}%

\global\long\def\scoreBennettS{S}

%% file: new_macros.tex
\global\long\def\scoreSkewInsensitiveVersionFOne{SIVF}
\global\long\def\scoreOptimalTradeoff{\scoreFBeta[*]}%
\global\long\def\scorePrecision{Pr}%
\global\long\def\scoreRecall{Re}%
\global\long\def\scoreIoU{IoU}%
\newcommand{\aClassifier}{c}
\global\long\def\spearman{\rho}%
\global\long\def\kendall{\tau}%
\newcommand{\aDistance}{d}
\newcommand{\distSpearman}{\aDistance_\spearman}
\newcommand{\distKendall}{\aDistance_\kendall}
\newcommand{\setI}{\aSetOfPerformances_1}
\newcommand{\setII}{\aSetOfPerformances_2(ptn)}
\newcommand{\setIII}{\aSetOfPerformances_3(\priorpos)}
\newcommand{\setIV}{\aSetOfPerformances_4(\priorpos)}
\newcommand{\setV}{\aSetOfPerformances_5(\priorpos)}
\newcommand{\aDistributionOfPerformances}{\mathcal{P}}
\newcommand{\distriI}{\aDistributionOfPerformances_1}
\newcommand{\distriII}{\aDistributionOfPerformances_2}
\newcommand{\distriIII}{\aDistributionOfPerformances_3}
\newcommand{\distriIV}{\aDistributionOfPerformances_4}
\newcommand{\distriV}{\aDistributionOfPerformances_5}
\newcommand{\optimality}{\mathcal{O}}

%% file: sections/0_abstract.tex
\begin{abstract}
Ranking methods or models based on their performance is of prime importance but is tricky because performance is fundamentally multidimensional. In the case of classification, precision and recall are scores with probabilistic interpretations that are both important to consider and complementary. The rankings induced by these two scores are often in partial contradiction. In practice, therefore, it is extremely useful to establish a compromise between the two views to obtain a single, global ranking. Over the last fifty years or so, it has been proposed to take a weighted harmonic mean, known as the F-score, F-measure, or $\scoreFBeta$. Generally speaking, by averaging basic scores, we obtain a score that is intermediate in terms of \emph{values}. However, there is no guarantee that these scores lead to meaningful \emph{rankings} %
and no guarantee that the rankings are good \emph{tradeoffs} between these base scores. Given the ubiquity of $\scoreFBeta$ scores in the literature, some clarification is in order. Concretely: (1)~We establish that $\scoreFBeta$-induced rankings are meaningful and define a shortest path between precision- and recall-induced rankings. (2)~We frame the problem of finding a tradeoff \emph{between} two scores as an \emph{optimization} problem expressed with Kendall rank correlations. We show that $\scoreFOne$ and its skew-insensitive version are far from being optimal in that regard. (3)~We provide theoretical tools and a closed-form expression to \emph{find} the optimal value for $\beta$ for any distribution or set of performances, and we illustrate their use on six case studies. Code is available at \url{https://github.com/pierard/cvpr-2026-optimal-tradeoff-precision-recall}.
\end{abstract}

%% file: sections/1_introduction.tex
\section{Introduction}

\input{figs/graphical_abstract}

The \emph{precision} $\scorePrecision$ (also called \emph{positive predictive value}) and \emph{recall} $\scoreRecall$ (also called \emph{true positive rate}) are of first importance in classification and other related problems. 
These scores have a straightforward interpretation, as they give the probabilities of taking the correct decision when the predicted or ground-truth class is positive, respectively~\cite{Goutte2005AProbabilistic}.

Unfortunately, working with several scores is often impractical. For this reason, it is common to average the values of $\scorePrecision$ and $\scoreRecall$. Most often, this is done with a harmonic mean, leading to the family of F-scores $\scoreFBeta$, with $\beta\ge0$:
\begin{equation}
    \scoreFBeta^{-1}
    = (1-b) \, \scorePrecision^{-1}
    + b \, \scoreRecall^{-1}
    \quad
    \textrm{ with }
    b = \nicefrac{\beta^2}{1+\beta^2}
    \point
    \label{eq:f-beta-as-harmonic-mean}
\end{equation}

These last decades, this family of scores has become very popular in the literature\footnote{
    With the requests ``F-measure + classification'', ``F-score + classification'', and ``F-beta + classification'', Google Scholar reported on July 23th 2025 that there was about $315{,}000$, $205{,}000$, and $9{,}080$ matching documents, respectively. Moreover, we estimated that about $10\%$ of the papers accepted at CVPR 2025 refer to or use $\scoreFOne$.
}. It forms a continuum between precision $\scorePrecision=\scoreFBeta[0]$ and recall $\scoreRecall=\scoreFBeta[\infty]$. The traditional (balanced) F-score, $\scoreFOne$, gives equal weights to $\scorePrecision$ and $\scoreRecall$. But is $\scoreFOne$ really the optimal tradeoff between precision and recall? The answer to this question depends on the point of view.

\textbf{The point of view of values.} The F-scores are obviously between $\scorePrecision$ and $\scoreRecall$ in the sense that, for any given performance $\aPerformance$ (\ie, a confusion matrix), the value $\scoreFBeta(\aPerformance)$ varies %
monotonically with $\beta$, between $\scorePrecision(\aPerformance)$ and $\scoreRecall(\aPerformance)$.

\textbf{The point of view of ranks.} This is the point of view that we develop. Although many authors use the $\scoreFOne$ or $\scoreFBeta$ scores to rank and compare methods, the claim that they are an appropriate compromise for ranking deserves to be studied. Several questions arise. How are the rankings induced by the $\scoreFBeta$ scores spatially organized? Are the $\scoreFBeta$ scores suitable to induce performance-based rankings? Do the $\scoreFBeta$ scores induce rankings that form a shortest path between those induced by $\scorePrecision$ and $\scoreRecall$?     For which $\beta$(s) do we obtain the optimal tradeoff between the rankings induced by $\scorePrecision$ and $\scoreRecall$?

The objective of this paper is to provide answers, teased in~\cref{fig:graphical_abstract}, to all these questions. For that purpose, we develop the theory, and we provide experimental results for various distributions and sets of two-class classification performances.
We summarize 
our contributions as follows:
\begin{enumerate}
    \item 
We establish that the $\scoreFBeta$-family is the right set of scores to consider when looking for an optimal tradeoff between $\scorePrecision$ and $\scoreRecall$ for ranking, as, on the one hand, all $\scoreFBeta$ lead to meaningful rankings according to the theory of performance-based ranking~\cite{Pierard2025Foundations} and, on the other hand, the rankings induced by $\scoreFBeta$ always form a shortest path between the rankings induced by $\scorePrecision$ and $\scoreRecall$.
\item
We show that neither $\scoreFOne$, traditionally considered as a balanced compromise between $\scorePrecision$ and $\scoreRecall$, nor its skew-insensitive version (denoted by $\scoreSkewInsensitiveVersionFOne$) introduced in~\cite{Flach2003TheGeometry} are acceptable tradeoffs for ranking, in general. These scores can lead to rankings far from optimal.
\item
We provide the theory and methods for defining and finding (with a closed-form expression) the optimal tradeoffs between the rankings induced by $\scorePrecision$ and $\scoreRecall$. These methods are then successfully applied to numerous sets and distributions of performances.
\end{enumerate}

%% file: figs/graphical_abstract.tex
\definecolor{colorCurve}{RGB}{179, 179, 179}
\newcommand{\markerCurve}{\textcolor{colorCurve}{$\smile$}\xspace}
\definecolor{colorPrecision}{RGB}{81, 158, 62}
\newcommand{\markerPrecision}{\textcolor{colorPrecision}{$\blacklozenge$}\xspace}
\definecolor{colorRecall}{RGB}{197, 58, 50}
\newcommand{\markerRecall}{\textcolor{colorRecall}{$\blacklozenge$}\xspace}
\definecolor{colorFOne}{RGB}{58, 117, 175}
\newcommand{\markerFOne}{\textcolor{colorFOne}{\ding{54}}\xspace}
\definecolor{colorSkewInsensitiveFOne}{RGB}{239, 134, 54}
\newcommand{\markerSkewInsensitiveFOne}{\textcolor{colorSkewInsensitiveFOne}{\ding{54}}\xspace}
\definecolor{colorTradeoff}{RGB}{141, 105, 184}
\newcommand{\markerTradeoff}{\textcolor{colorTradeoff}{$\blacklozenge$}\xspace}

\begin{figure}[t]
    \begin{centering}
        \includegraphics[width=\linewidth]{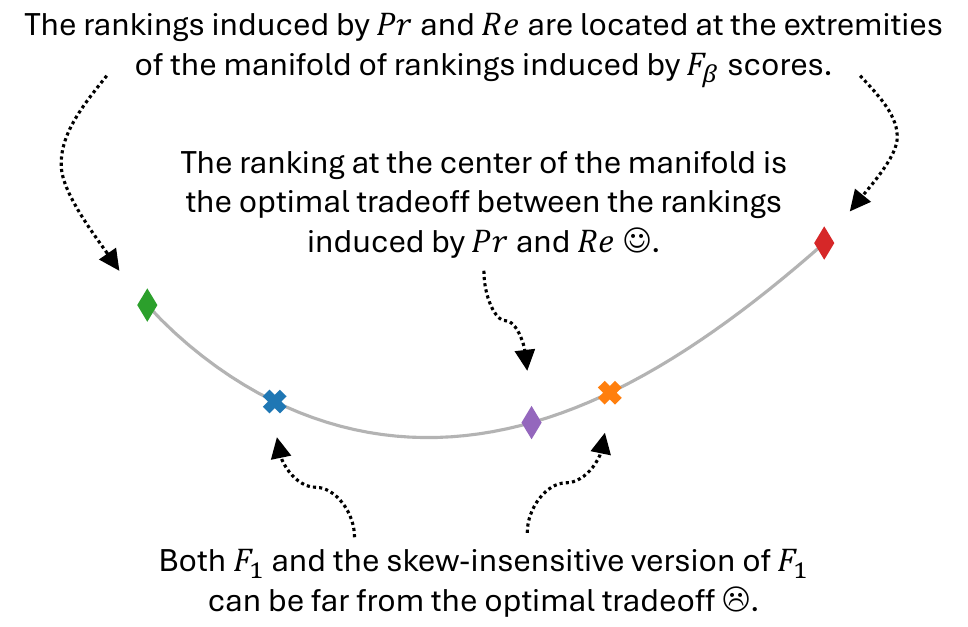}
    \par\end{centering}
    
    \caption{
        The manifold of rankings inducible with the $\scoreFBeta$ scores is a curve (drawn as \markerCurve) that depends on the set or distribution of performances that is considered. The distance along the curve is proportional to the number of swaps of neighboring classifiers needed to transform a ranking into another. The rankings induced by precision \markerPrecision and recall \markerRecall are at the extremities of the manifold. The rankings induced by the traditional (balanced) $\scoreFOne$ \markerFOne and by the skew-insensitive version of $\scoreFOne$ \markerSkewInsensitiveFOne \cite{Flach2003TheGeometry} can be anywhere, depending on the set or distribution of performances that are compared, and cannot be considered as the optimal tradeoff between precision and recall. We consider that the ranking located at equal distance, along the curve, from precision and recall is the optimal tradeoff \markerTradeoff.
        \label{fig:graphical_abstract}
    }
\end{figure}

%% file: sections/2_related_work.tex
\section{Related Work}

\subsection{Useful References}

The meaning of F-scores is discussed in countless papers; see, for example, the work by \citet{Christen2023AReview} for a recent review about the F-scores. A plethora of other scores have been defined for classification. They can be found in  several reviews~ \cite{Ballabio2018Multivariate,Canbek2017Binary,Choi2010Survey,Ferrer2022Analysis-arxiv,Ferri2009AnExperimental,Hossin2015AReview,Japkowicz2011Evaluating,Powers2011Evaluation,Sokolova2009ASystematic}, to cite only a few.

Regarding the ranking of methods or models based on their performances, theoretical foundations have been provided in~\cite{Pierard2025Foundations} and the theory has been particularized to the case of two-class crisp classification in~\cite{Pierard2024TheTile-arxiv,Halin2024AHitchhikers-arxiv}. When comparing performances with the same class priors, the isometrics depicted in the \roc (Receiver Operating Characteristic) space is a graphical representation of the performance ordering induced by a given score. The isometrics of $\scorePrecision$ and $\scoreFOne$ have been depicted in~\cite{Flach2003TheGeometry} for different class priors.

There are only a few studies of the correlations between the rankings induced by scores in classification. Both~\citet{Ferri2009AnExperimental} and~\citet{Liu2014AStrategy} used Pearson linear correlations and Spearman rank correlations. To the best of our knowledge, there is no study comparing classification scores based on Kendall rank correlations.

\subsection{Building Upon the Related Work}
\label{sec:building-upon-related-work}

\subsubsection{Sometimes, $\scoreFOne$ Mimics the Rankings Induced by $\scorePrecision$ or $\scoreRecall$ and Ignores the Other One}

\input{figs/roc_isometrics}

Let us consider the particular case in which the class priors are fixed and denote them by $\priorneg$ and $\priorpos$ for the negative and positive classes, respectively. Extending the work of~\cite{Flach2003TheGeometry}, we depict, in \cref{fig:roc_isometrics}, the isometrics of $\scorePrecision$, $\scoreFBeta$ for different values of $\beta$, $\scoreRecall$, and $\scoreSkewInsensitiveVersionFOne$ in the \roc space. For all these scores, we found that the isometrics form pencils of lines with vertices located at $(\scoreFPR,\scoreTPR)=(-\ell,0)$ with some $\ell \ge 0$. The performance orderings, and the resulting rankings, depend only on the location of the vertex, and thus on $\ell$. For $\scorePrecision$, $\ell=0$. For $\scoreRecall$, $\ell=\infty$. For $\scoreSkewInsensitiveVersionFOne$, $\ell=1$. And for $\scoreFBeta$,
\begin{equation}
    \ell = \beta^2 \frac{\priorpos}{\priorneg} = \beta^2 \frac{\priorpos}{1-\priorpos}
    \point
    \label{eq:ell}
\end{equation}
When $\priorpos \rightarrow 0$, we see that $\ell \rightarrow 0$ and that all $\scoreFBeta$ mimic the ranking of $\scorePrecision=\scoreFBeta[0]$ and ignore $\scoreRecall$. Similarly, when $\priorpos \rightarrow 1$, we see that $\ell \rightarrow \infty$ and that all $\scoreFBeta$ mimic the ranking of $\scoreRecall=\scoreFBeta[\infty]$, ignoring $\scorePrecision$. So, \begin{myImportantResult}a single value of $\beta$ cannot provide the optimal ranking for all priors\end{myImportantResult}.

\subsubsection{All $\scoreFBeta$ Scores Lead to Meaningful Rankings, Without Any Constraint on the Performances}

There are some conditions that a performance ordering must fulfill to be meaningful. They have been given in the form of axioms in %
\cite{Pierard2025Foundations} (reminded in suppl. mat.). We found that all these axioms are satisfied by the performance orderings induced by $\scoreFBeta$, $\forall \beta \ge 0$. In other words, \begin{myImportantResult}all $\scoreFBeta$  lead to meaningful performance orderings\end{myImportantResult}. 
It was not a foregone conclusion, because the scores that are, from the point of view of values, between two scores suitable for ranking are not guaranteed to be themselves suitable for ranking. For example, the performance orderings induced by the arithmetic and geometric means of $\scorePrecision$ and $\scoreRecall$ do not satisfy the axioms of performance-based ranking. As the ranking induced by $\scoreSkewInsensitiveVersionFOne$ is the same as the ranking induced by $\scoreFBeta$ when $\beta^2$ is equal to the skew ratio~\cite{Flach2003TheGeometry} $\nicefrac{\priorneg}{\priorpos}$, $\scoreSkewInsensitiveVersionFOne$ leads to a meaningful performance ordering when the priors are fixed. However, we found that it does not satisfy the axioms of \cite{Pierard2025Foundations} without this constraint.

%% file: figs/roc_isometrics.tex
\begin{figure*}
    \centering
    \begin{tabular}[b]{cccccccc|c}
        \begin{turn}{90}
            $\priorpos=0.1$
        \end{turn} &
        \includegraphics[scale=0.15]{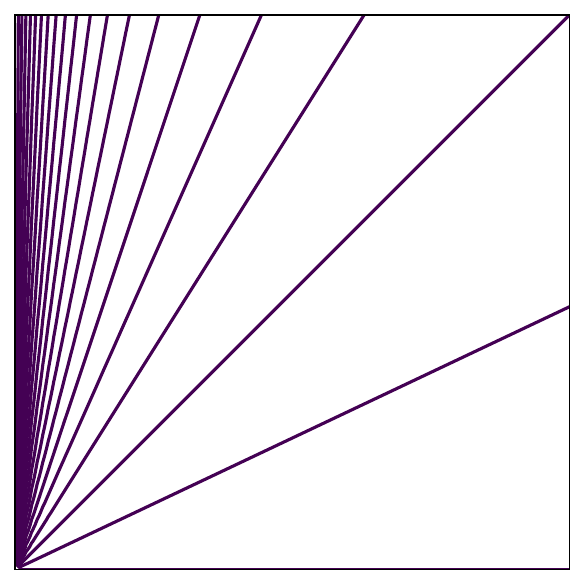} &
        \includegraphics[scale=0.15]{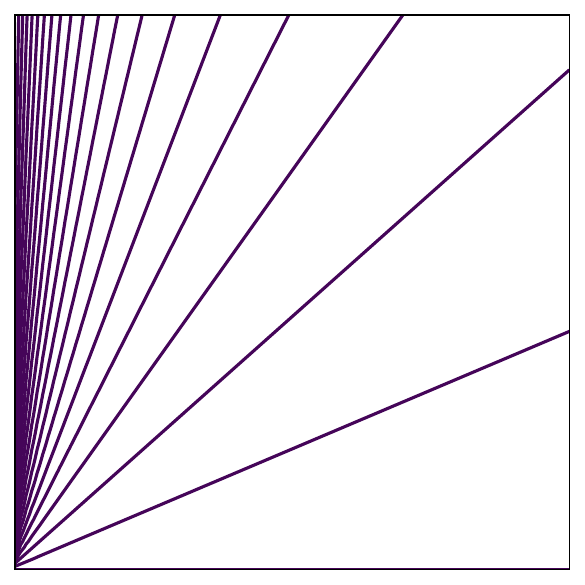} &
        \includegraphics[scale=0.15]{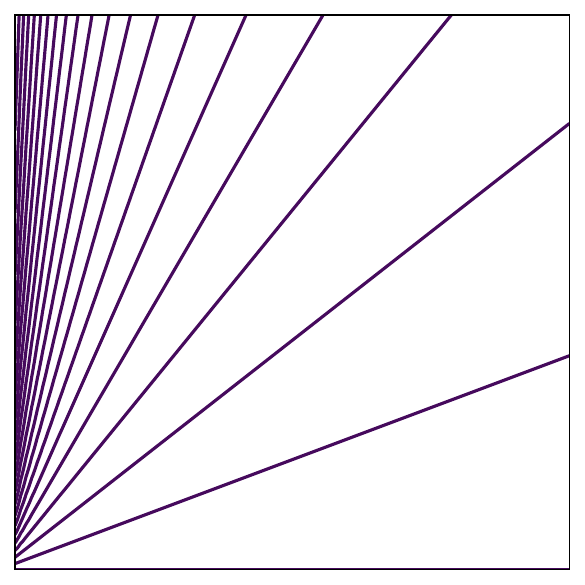} &
        \includegraphics[scale=0.15]{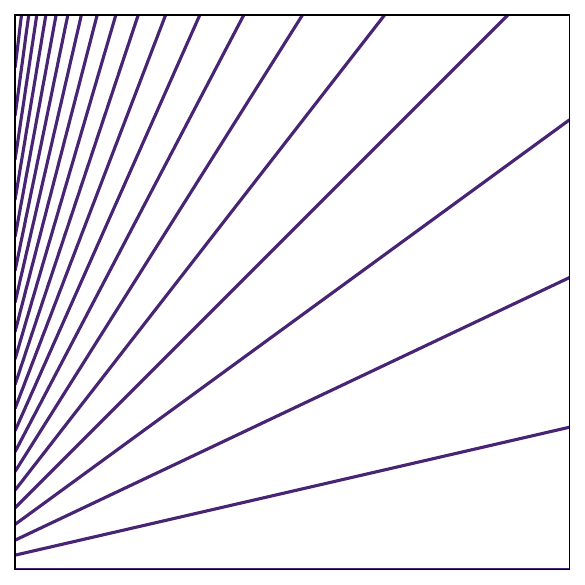} &
        \includegraphics[scale=0.15]{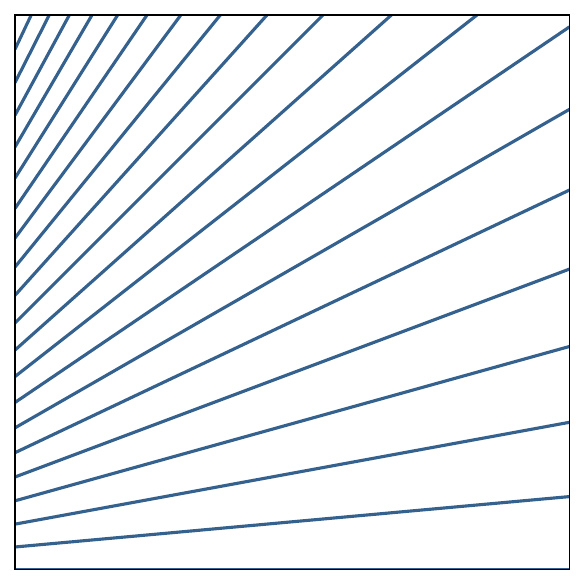} &
        \includegraphics[scale=0.15]{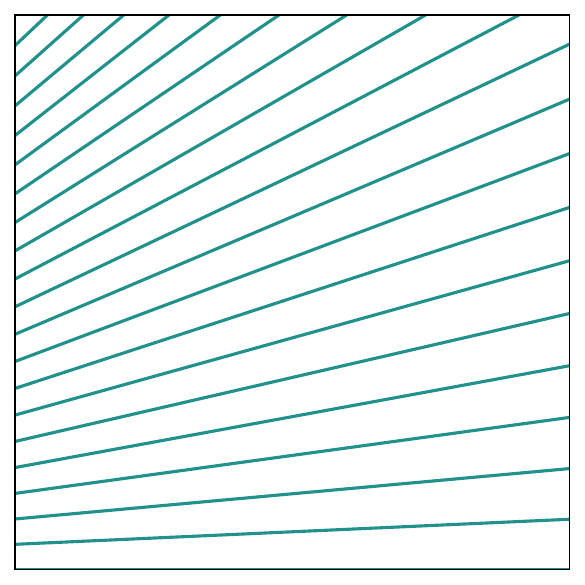} &
        \includegraphics[scale=0.15]{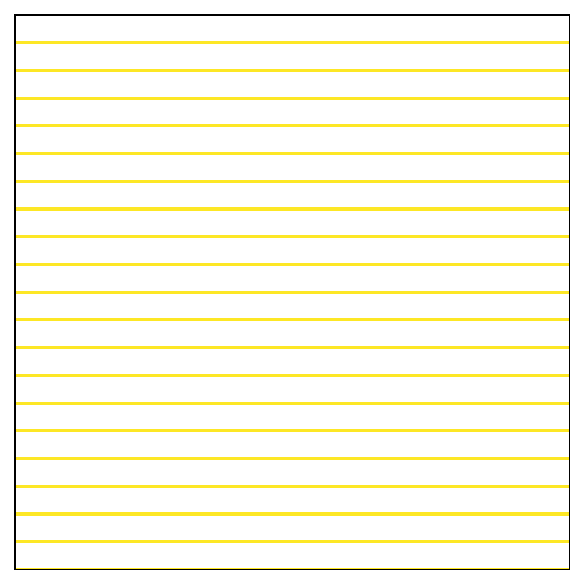} &
        \includegraphics[scale=0.15]{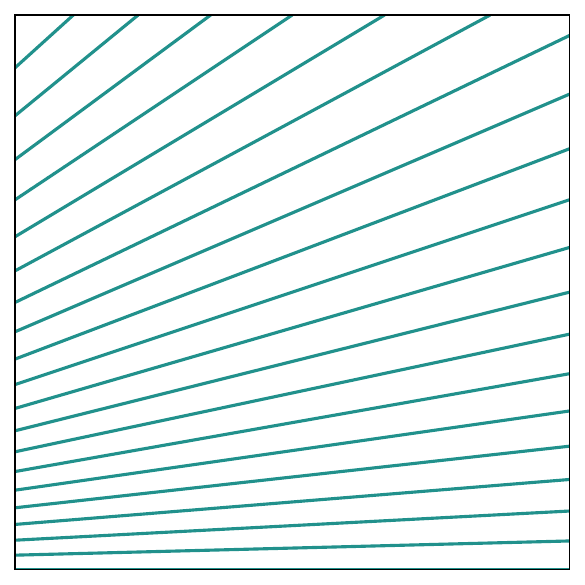} \\
        \begin{turn}{90}
            $\priorpos=0.5$
        \end{turn} &
        \includegraphics[scale=0.15]{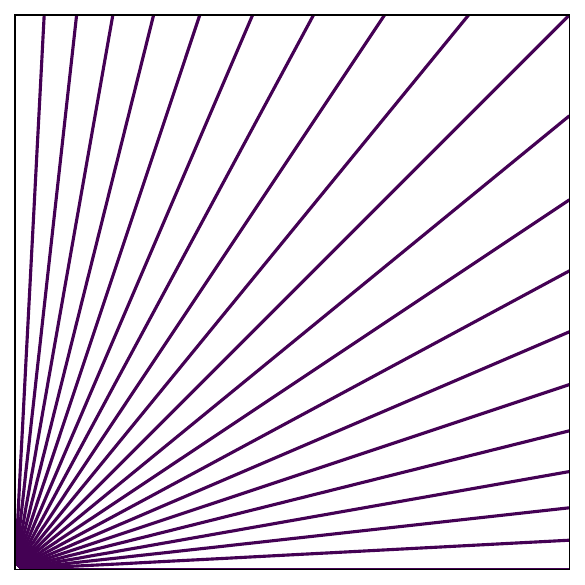} &
        \includegraphics[scale=0.15]{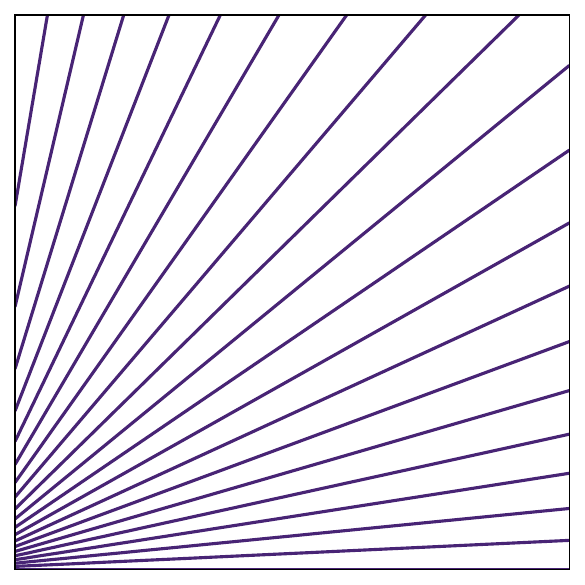} &
        \includegraphics[scale=0.15]{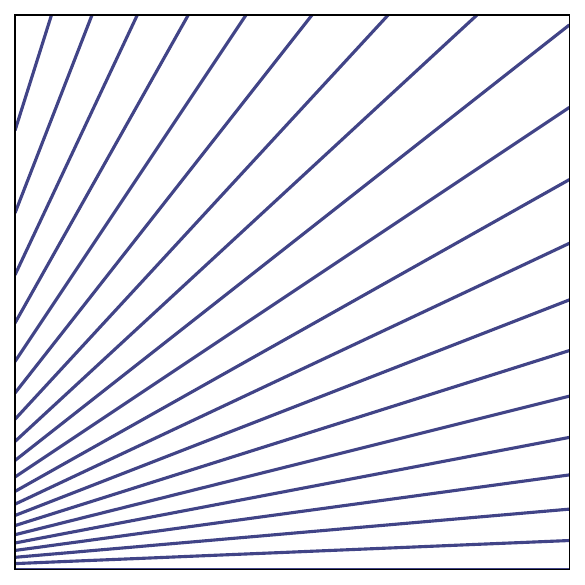} &
        \includegraphics[scale=0.15]{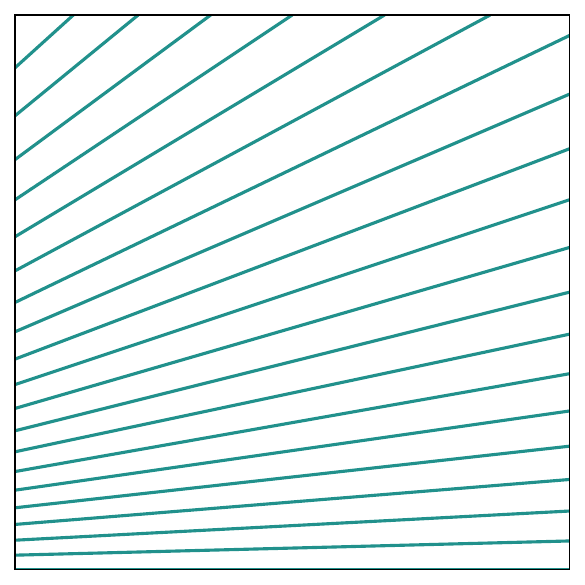} &
        \includegraphics[scale=0.15]{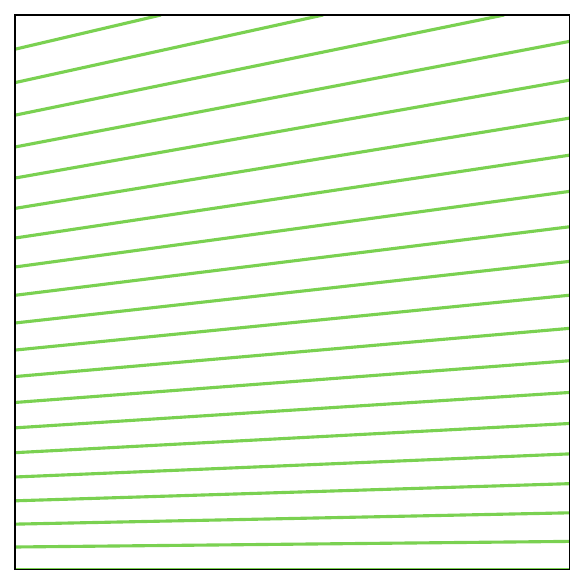} &
        \includegraphics[scale=0.15]{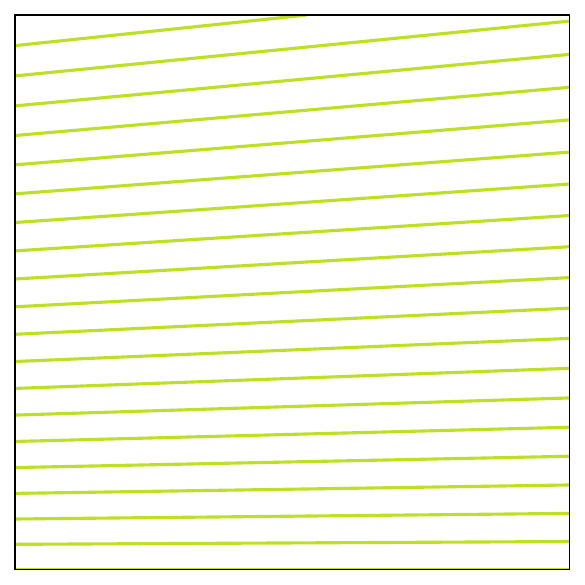} &
        \includegraphics[scale=0.15]{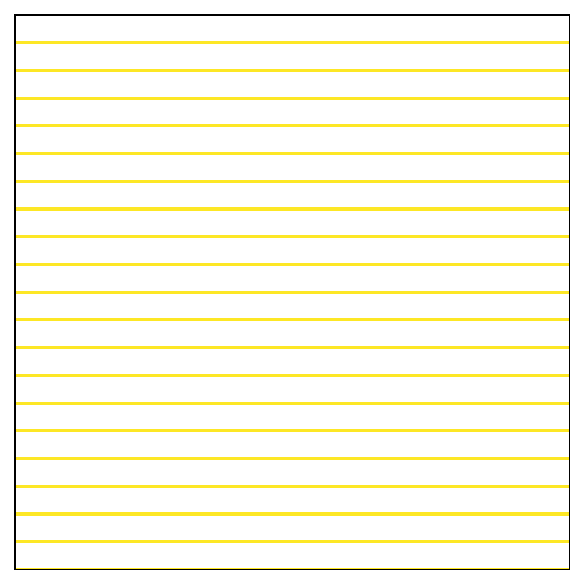} &
        \includegraphics[scale=0.15]{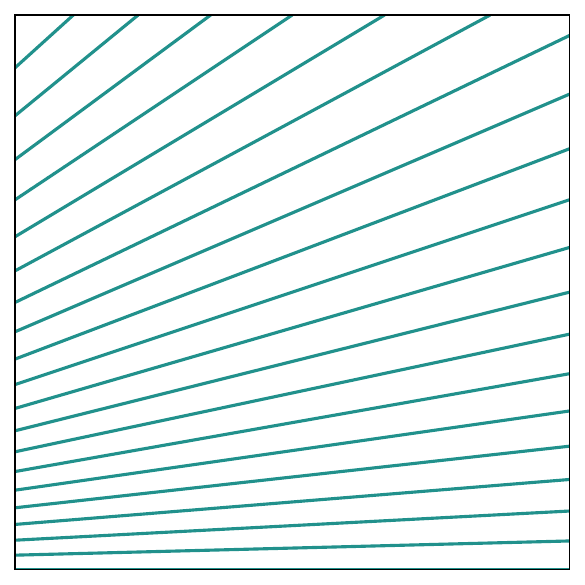} \\
        \begin{turn}{90}
            $\priorpos=0.9$
        \end{turn} &
        \includegraphics[scale=0.15]{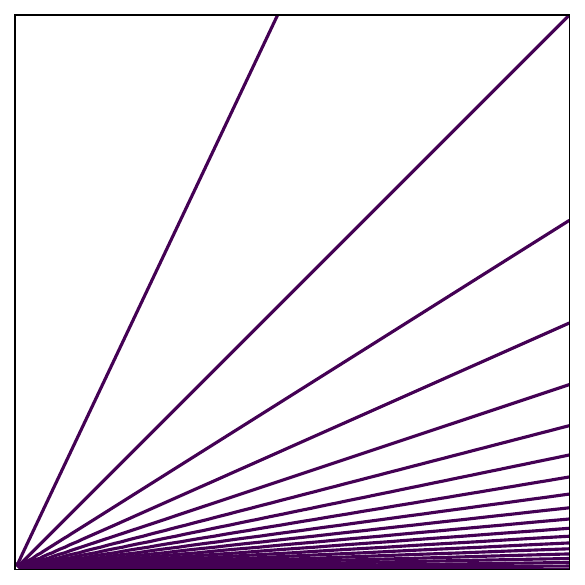} &
        \includegraphics[scale=0.15]{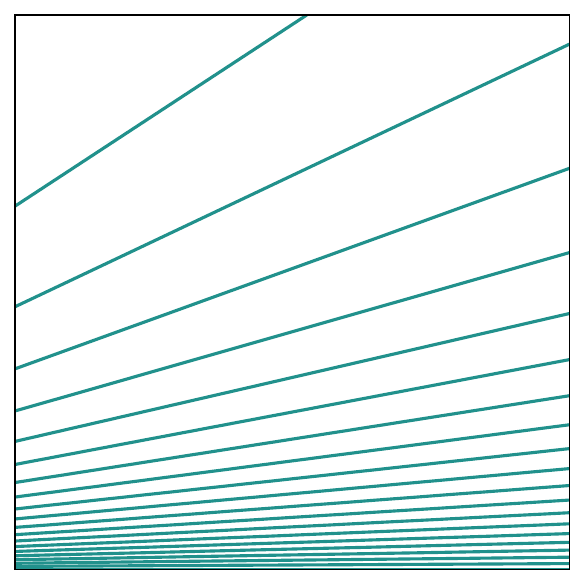} &
        \includegraphics[scale=0.15]{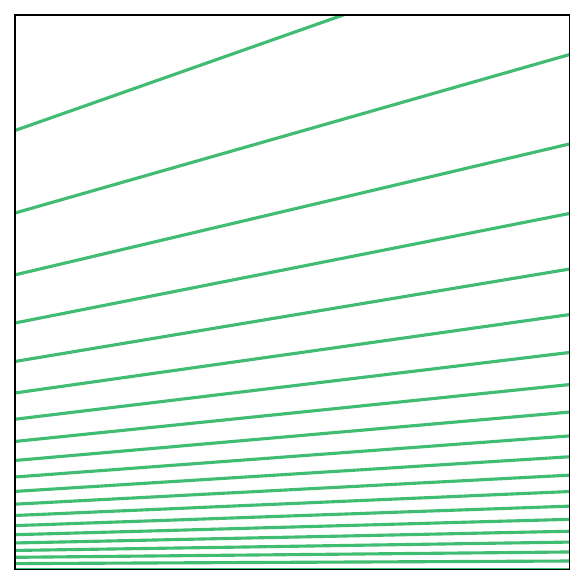} &
        \includegraphics[scale=0.15]{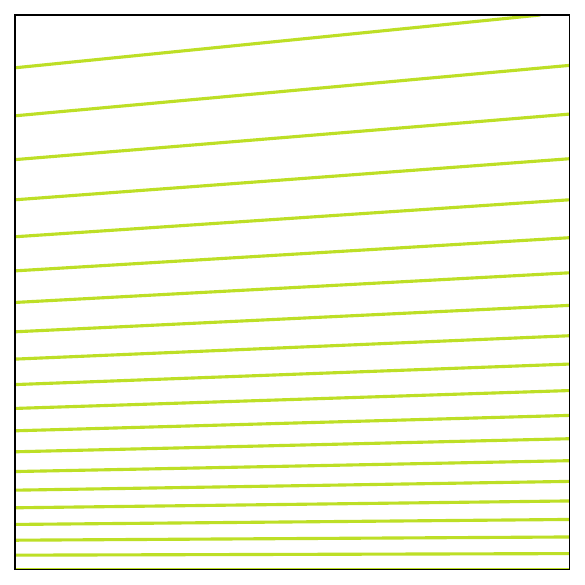} &
        \includegraphics[scale=0.15]{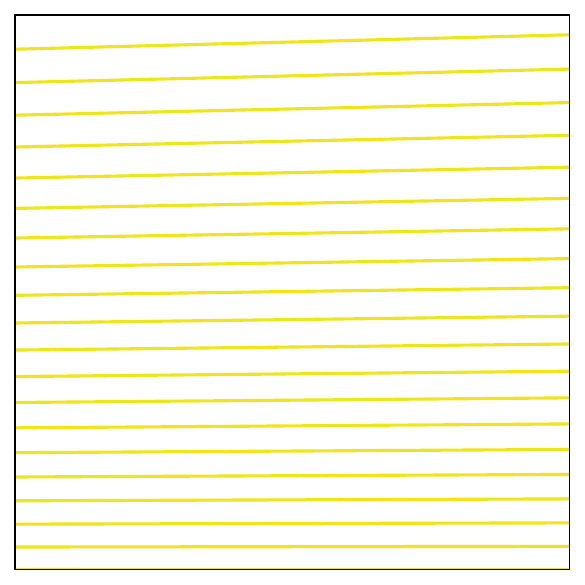} &
        \includegraphics[scale=0.15]{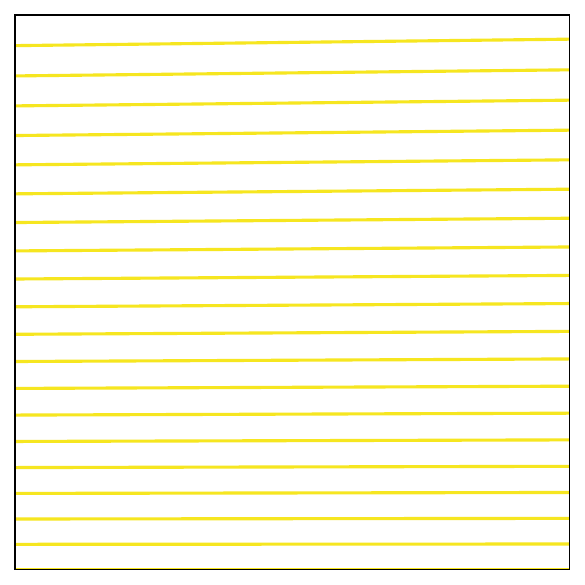} &
        \includegraphics[scale=0.15]{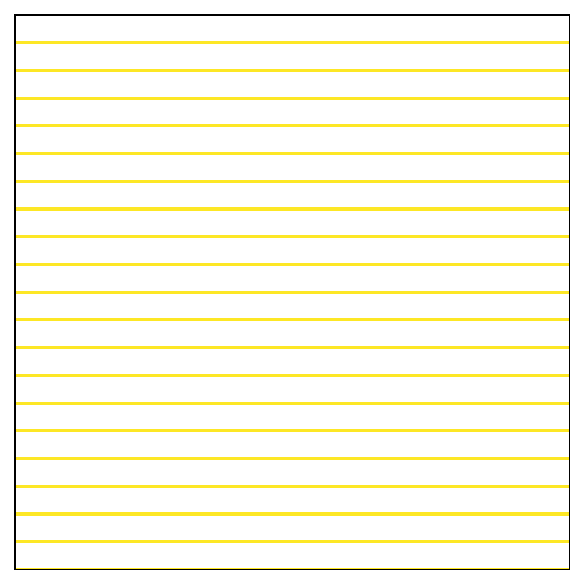} &
        \includegraphics[scale=0.15]{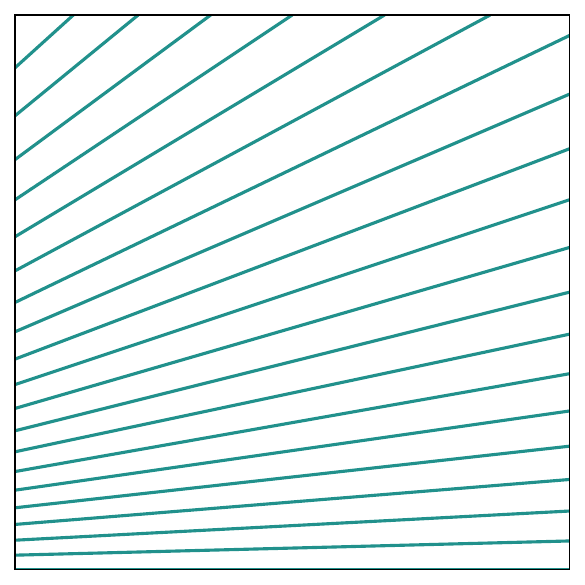} \\
        & 
        $\scoreFBeta[0]=\scorePrecision$ & 
        $\scoreFBeta[\nicefrac{1}{3}]$ & 
        $\scoreFBeta[\nicefrac{1}{2}]$ & 
        $\scoreFBeta[1]$ & 
        $\scoreFBeta[2]$ & 
        $\scoreFBeta[3]$ & 
        $\scoreFBeta[\infty]=\scoreRecall$ & 
        $\scoreSkewInsensitiveVersionFOne$
    \end{tabular}
    \includegraphics[scale=0.7]{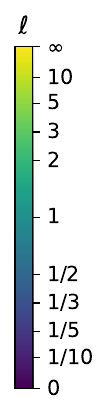}
    \caption{
        Isometrics of $\scoreFBeta$ and $\scoreSkewInsensitiveVersionFOne$ 
        in the \roc space $(\scoreFPR,\scoreTPR)\in[0,1]^2$, for three class priors. An isometric is a locus of equivalent performances, according to the score value. We plotted here those corresponding to the values $0.05, 0.10, 0.15, \ldots, 0.95$. The isometrics of $\scoreFBeta$ form a pencil of lines intersecting at $(\scoreFPR,\scoreTPR)=(-\ell,0)$ with $\ell=\beta^2 \nicefrac{\priorpos}{\priorneg}$. The isometrics of $\scoreSkewInsensitiveVersionFOne$ intersect at $(\scoreFPR,\scoreTPR)=(-1,0)$, no matter what the class priors are. The score $\scoreSkewInsensitiveVersionFOne$ induces the same performance ordering as $\scoreFBeta$ with $\beta^2=\nicefrac{\priorneg}{\priorpos}$.
        \label{fig:roc_isometrics}
    }
\end{figure*}

%% file: sections/3_0_theory.tex
\section{Theory}
\label{sec:theory}

\input{sections/3_1_preliminaries}
\input{sections/3_2_rankings_and_manifold}

\input{sections/3_3_shortest_paths_and_means_of_scores}

\input{sections/3_4_tradeoff}

%% file: sections/3_1_preliminaries.tex
\subsection{Preliminaries}

We use the notations of~\cite{Pierard2025Foundations}. $\allPerformances$ is the set of all possible performances $\aPerformance$ in two-class crisp classification, \ie the set of all probability measures with a sample space $\sampleSpace$ containing the four cases that can arise: $\sampleTN$ (true negative), $\sampleFP$ (false positive), $\sampleFN$ (false negative), and $\sampleTP$ (true positive). The negative and positive class priors are $\priorneg=\aPerformance(\{\sampleTN,\sampleFP\})$ and $\priorpos=\aPerformance(\{\sampleFN,\sampleTP\})$, respectively.

We are interested in the rankings induced by scores. Scores are functions $\aScore:\allPerformances\rightarrow\realNumbers:\aPerformance\mapsto\aScore(\aPerformance)$. %
The unconditional probabilities of a false positive, false negative, and true positive are given by the scores $\scorePFN$, $\scorePFP$, and $\scorePTP$, respectively. The precision ($\scorePrecision$), also called positive predictive value ($\scorePPV$), is defined as 
$%
    \scorePrecision = \scorePPV = \scoreFBeta[0] : \aPerformance\mapsto\aPerformance(\eventTP|\{\sampleFP,\sampleTP\})
$. %
The recall ($\scoreRecall$) or true positive rate ($\scoreTPR$), is 
$%
    \scoreRecall= \scoreTPR = \scoreFBeta[\infty] : \aPerformance\mapsto\aPerformance(\eventTP|\{\sampleFN,\sampleTP\})
$. %
The skew-insensitive version of $\scoreFOne$~\cite{Flach2003TheGeometry} is
$%
    \scoreSkewInsensitiveVersionFOne = \frac{
        2 \scoreTPR
    }{
        \scoreTPR + \scoreFPR + 1
    }
$, %
where $\scoreFPR$ is the false positive rate.

%% file: sections/3_2_rankings_and_manifold.tex
\subsection{The Rankings and Their Manifold}

Without loss of generality, let us consider a finite set $\aSetOfPerformances=\{\aPerformance_1,\ldots,\aPerformance_n\}\subset\allPerformances$ of performances (for infinitely many performances, consider a distribution\footnote{
    Technically, we need to consider a measurable space whose sample space is $\allPerformances$. A performance distribution $\aDistributionOfPerformances$ then corresponds to a probability measure. Scores, defined as real functions on $\allPerformances$, become random variables, and we can rigorously speak of correlation %
    between scores.
} $\aDistributionOfPerformances$ instead of a set $\aSetOfPerformances$). The rankings of $\aSetOfPerformances$ are its permutations, in finite number.

Let us consider a score $\aScore$ that induces meaningful rankings. Assuming that the value of $\aScore$ is defined for all $\aPerformance\in\aSetOfPerformances$, and  that $\aScore$ does not assign the same value to two performances of $\aSetOfPerformances$ (absence of ties), $\aScore$ induces the ranking:
\begin{equation}
    \mathbf{x} 
    = (
        \mathrm{rank}_\aScore(\aPerformance_1), 
        \ldots,
        \mathrm{rank}_\aScore(\aPerformance_n)
    )
    \comma
    \label{eq:ranking-as-point}
\end{equation}
where $\mathrm{rank}_\aScore(\aPerformance)$ denotes the number of performances in $\aSetOfPerformances$ for which $\aScore$ takes a value greater or equal to $\aScore(\aPerformance)$.

A set of scores (\eg all $\scoreFBeta$) corresponds to a set of points as written in \cref{eq:ranking-as-point}. 
The Euclidean distance (\ie along a straight line) between two rankings is their Spearman distance $\distSpearman$, related to Spearman's $\spearman$~\cite{Spearman1904TheProof}. \begin{myImportantResult}The distance measured along the manifold of rankings is Kendall's distance $\distKendall$ (\aka bubble-sort distance), related to Kendall's $\kendall$~\cite{Kendall1938ANewMeasure}.\end{myImportantResult} We can explain it as follows. Let us create the path graph that connects the rankings that are neighbors. Two rankings $\mathbf{x}_a$ and $\mathbf{x}_b$ are neighbors if and only if they differ only by one swap of two classifiers at consecutive ranks. The geodesic distance, \ie the length of the shortest path, between $\mathbf{x}_a$ and $\mathbf{x}_b$, is the minimum number of swaps needed to transform $\mathbf{x}_a$ into $\mathbf{x}_b$, and it is proportional to the Kendall distance $\distKendall$. For this reason, we consider this distance to be adequate for specifying the optimal tradeoff between two rankings.

As we observed in all our case studies that the manifold is curved, as depicted in \cref{fig:graphical_abstract}, we will define the optimal tradeoff in terms of Kendall's distance $\distKendall$ or in terms of Kendall's rank correlation $\kendall$.

%% file: sections/3_3_shortest_paths_and_means_of_scores.tex
\subsection{The Shortest Paths and the Means of Scores}

The score $\scoreFBeta$ is the weighted harmonic mean between precision $\scorePrecision$ and recall $\scoreRecall$. The harmonic mean is the generalized $f$-mean, also called the regular mean~\cite{Kolmogorov1930SurLaNotionDeLaMoyenne}, %
with $f:x\mapsto x^{-1}$. This has an important implication for the performance orderings induced by $\scorePrecision$, $\scoreRecall$, and $\scoreFBeta$.

Consider an interval $\xi\subseteq\realNumbers$, $m$ scores $\aScore_1, \ldots, \aScore_m$ whose images are included in $\xi$, and a continuous strictly monotonic real function $f$ defined on $\xi$. If a performance $\aPerformance_A$ is better than, equivalent to, or worse than a performance $\aPerformance_B$ according to all the scores $\aScore_1, \ldots, \aScore_m$, then so is it according to all their generalized $f$-means $\overline{\aScore}$. This extends to weighted $f$-means. 
So, \begin{myImportantResult}if a performance $\aPerformance_A$ is better than, equivalent to, or worse than a performance $\aPerformance_B$ according to $\scorePrecision$ and to $\scoreRecall$, then so is it according to all the $\scoreFBeta$ scores\end{myImportantResult}.

Let us introduce the indicator $\Delta_{\aScore_{1},\aScore_{2}}^{\aPerformance_{A},\aPerformance_{B}}\in\{0,1\}$ that specifies if the scores $\aScore_1$ and $\aScore_2$ disagree on the relative order between the performances $\aPerformance_A$ and $\aPerformance_B$. 
Given the implication that a mean of scores has on the rankings, we have
\begin{equation}
    \Delta_{\scorePrecision,\scoreRecall}^{\aPerformance_{A},\aPerformance_{B}}
    = \Delta_{\scorePrecision,\scoreFBeta}^{\aPerformance_{A},\aPerformance_{B}}
    + \Delta_{\scoreFBeta,\scoreRecall}^{\aPerformance_{A},\aPerformance_{B}}
     \qquad \forall \beta\ge0
    \point
\end{equation}
And, as \cite{Kendall1938ANewMeasure}
\begin{equation}
    \distKendall(\aScore_1;\aScore_2) = \frac{2}{n(n-1)} \sum_{i<j} \Delta_{\aScore_{1},\aScore_{2}}^{\aPerformance_{i},\aPerformance_{j}}
    \quad
    \in [0,1]
    \comma
    \label{eq:kendall-distance} %
\end{equation}
\begin{equation}
    \distKendall(\scorePrecision;\scoreRecall) = \distKendall(\scorePrecision;\scoreFBeta) + \distKendall(\scoreFBeta;\scoreRecall) \qquad \forall \beta\ge0
    \point
    \label{eq:shortest-path-distances}
\end{equation}
This means that \begin{myImportantResult}the rankings induced by the $\scoreFBeta$ scores form one shortest path (\ie a geodesic) between the rankings induced by $\scorePrecision$ and $\scoreRecall$\end{myImportantResult}. This is a key result for this paper, as it proves meaningful the search for the optimal tradeoff between $\scorePrecision$ and $\scoreRecall$ in the family of $\scoreFBeta$ scores. An example is provided in \cref{fig:cada-rre-main}. As $\kendall=1-2\distKendall$, \cref{eq:shortest-path-distances} can be rewritten in terms of correlations as
\begin{equation}
    1 + \kendall(\scorePrecision;\scoreRecall) = 
    \kendall(\scorePrecision;\scoreFBeta) 
    + \kendall(\scoreFBeta;\scoreRecall) \qquad \forall \beta\ge0
    \point
    \label{eq:shortest-path-correlations}
\end{equation}

\input{figs/cada_rre_main}

%% file: figs/cada_rre_main.tex
\begin{figure*}
    \centering
    \,
    \hfill
    \subfloat[
        The curves $\scoreFBeta$ \wrt $\beta$ for the $16$ performances to rank. %
        The swaps in the ranking are shown by the black dots and the gray vertical lines, which delimit the $44$ different rankings. The ranking with $\scoreFBeta \ge 1.508$ perfectly mimics $\scoreRecall$, ignoring $\scorePrecision$. There is the same amount of black dots on both sides of the optimal compromise (green vertical bar, $\beta=0.426$ by \cref{eq:closed-form-expression}).
        \label{fig:cada-rre-main-fbeta-curves}
    ]{
         \includegraphics[width=0.30\linewidth]{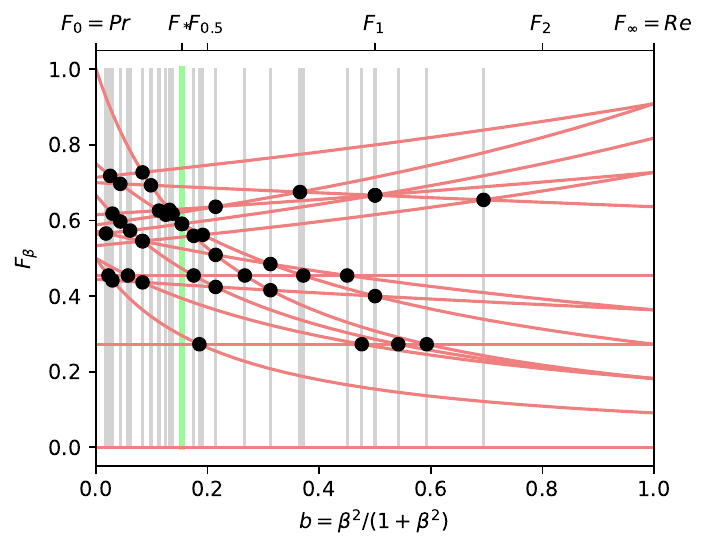}
    }
    \hfill
    \hfill
    \subfloat[
        Visualization of the manifold of rankings inducible by the $\scoreFBeta$ scores with two-components PCA ($93.57\%$ variance explained). %
        Each point corresponds to a different ranking and to a range of $\beta$s. A line segment between consecutive points represents a value of $\beta$ for which there is a swap between classifiers.
        See \cref{fig:graphical_abstract} for a detailed description of such a plot.
        \label{fig:cada-rre-main-pca}
    ]{
         \includegraphics[width=0.30\linewidth]{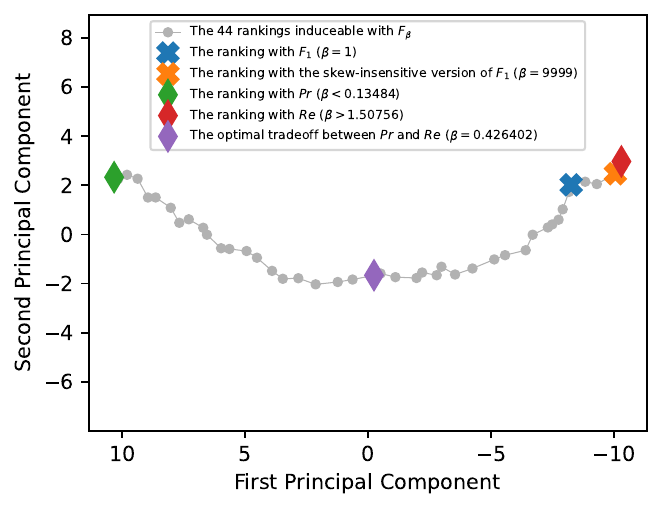}
    }
    \hfill
    \hfill
    \subfloat[
        Fréchet variance $\sigma^2(\beta)$ (\cref{eq:frechet-variance}) \wrt $\beta$. It is a piecewise constant function, the discontinuities occurring at the values for which there is a swap between two classifiers in the ranking. The optimal trade-off(s) between $\scorePrecision$ and $\scoreRecall$ is (are) obtained with the values of $\beta$ for which the Fréchet variance is minimized. The value computed by \cref{eq:choice-optimal} belongs to them (green vertical bar).
        \label{fig:cada-rre-main-frechet}
    ]{
         \includegraphics[width=0.30\linewidth]{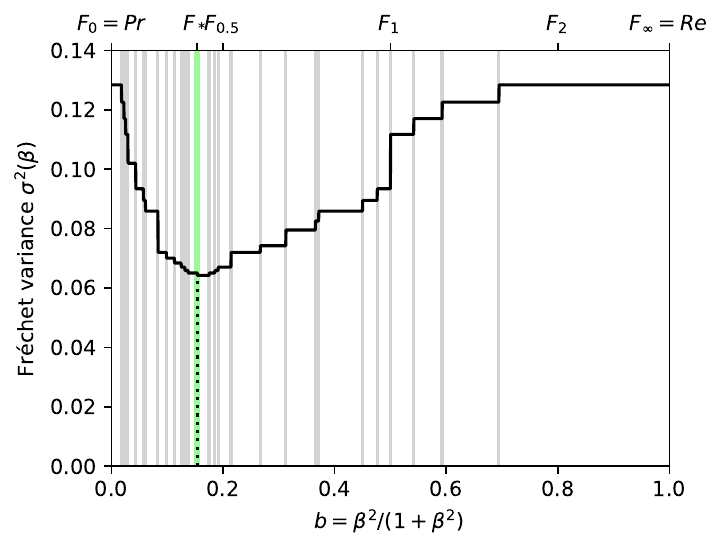}
    }
    \hfill
    \,
    \caption{
        The leaderboard for the medical imaging challenge \linkCADARRE \cite{Ivantsits2021Cerebral}, with $|\aSetOfPerformances|=16$ performances to rank (details provided in supplementary material). Thanks to \cref{eq:shortest-path-distances}, we know that the optimal tradeoff between the rankings induced by the precision $\scorePrecision$ (on the left hand side of each plot) and the recall $\scoreRecall$ (on the right hand side) is induced by some $\scoreFBeta$ score. This figure shows how the rankings change with $\beta$. There are $44$ ($\le |\aSetOfPerformances|^2+1$) different rankings inducible by the $\scoreFBeta$ scores. The optimal trade-off is given by \cref{eq:frechet-variance,eq:closed-form-expression}.
        \label{fig:cada-rre-main}
    }
\end{figure*}

%% file: sections/3_4_tradeoff.tex
\subsection{The Optimal Tradeoff Between $\scorePrecision$ and $\scoreRecall$}

\subsubsection{Optimal Tradeoffs as Karcher Means}
\label{sec:OptimalTradeoffs}

The optimal tradeoffs are the $\scoreFBeta$ scores that minimize the Fréchet variance~\cite{Frechet1948LesElementsAleatoires} (see \cref{fig:cada-rre-main-frechet}):
\begin{equation}
    \sigma^2(\beta) = \distKendall^2(\scorePrecision;\scoreFBeta) + \distKendall^2(\scoreFBeta ; \scoreRecall)
    \point
    \label{eq:frechet-variance}
\end{equation}
The solutions $\scoreFBeta=\scoreOptimalTradeoff$ of \cref{eq:frechet-variance}, known as the Karcher means~\cite{Karcher1977Riemannian} are those that are equidistant of $\scorePrecision$ and $\scoreRecall$ (this is detailed in supplementary material), \ie such that
\begin{align}
 & \distKendall(\scorePrecision;\scoreOptimalTradeoff) = \distKendall(\scoreOptimalTradeoff;\scoreRecall) = \frac{\distKendall(\scorePrecision;\scoreRecall)}{2}\\
\Leftrightarrow \; \; & \kendall(\scorePrecision;\scoreOptimalTradeoff) = \kendall(\scoreOptimalTradeoff;\scoreRecall) = \frac{1+\kendall(\scorePrecision;\scoreRecall)}{2}
\point
\end{align}

Note that, putting aside the perspective of performances and scores, the idea of minimizing either the sum of distances between rankings or the sum of squared distances has been proposed by Kemeny~\cite{Kemeny1959Mathematics}. However, \cref{eq:shortest-path-distances,eq:shortest-path-correlations} show that, in our case, the minimization of the sum of distances is undefined. But the minimization of the sum of squared distances (\cref{eq:frechet-variance}) is well defined.

\subsubsection{A Closed-Form Expression for the Optimal $\beta$}

Let us, by thought, place ourselves at the ranking induced by $\scorePrecision=\scoreFBeta[0]$ and then continuously increase $\beta$, following the path of rankings induced by the $\scoreFBeta$ scores, towards the one induced by $\scoreRecall=\scoreFBeta[\infty]$. With a finite number of performances, we stay at a given ranking during a range of values for $\beta$ and then, suddenly, move to the next ranking. The value of $\beta$ at which such a transition occurs is a value for which two performances are put on an equal footing by $\scoreFBeta$. Let us consider any two performances $\aPerformance_1$ and $\aPerformance_2$. We have $\scoreFBeta ( \aPerformance_1 ) = \scoreFBeta ( \aPerformance_2 )$ if and only if $\beta^2 = \vartheta(\aPerformance_1,\aPerformance_2)$ with
\begin{equation}
    \vartheta(\aPerformance_1,\aPerformance_2) = - \frac{
        \scorePTP(\aPerformance_1) \scorePFP(\aPerformance_2) - \scorePTP(\aPerformance_2) \scorePFP(\aPerformance_1) 
    }{
        \scorePTP(\aPerformance_1) \scorePFN(\aPerformance_2) - \scorePTP(\aPerformance_2) \scorePFN(\aPerformance_1)
    }
    \comma
    \label{eq:beta-sq-swap}
\end{equation}
when this value is positive. Otherwise, there is no score $\scoreFBeta$ that puts the performances $\aPerformance_1$ and $\aPerformance_2$ in equivalence. Kendall's distance between $\scorePrecision$ and $\scoreFBeta$ increases linearly with the number of swaps. This leads us to the conclusion that \begin{myImportantResult}the optimal tradeoff %
is $\scoreOptimalTradeoff=\scoreFBeta$ with
\begin{equation}
    \beta^2 = \mathrm{median} \left( \{ \vartheta(\aPerformance_i,\aPerformance_j) \, \vert \, i \ne j \wedge \vartheta(\aPerformance_i,\aPerformance_j) \ge 0 \} \right)
    \point
    \label{eq:closed-form-expression}
\end{equation}\end{myImportantResult}

\subsubsection{Assessing the Degree of Optimality of Some $\scoreFBeta$}

The choice of $\beta$ comes down to choosing a permutation of the classifiers. Looking at the level of pairwise comparisons, we can identify three mutually exclusive cases: \trivialOrdering, \wrongOrdering, and \correctOrdering.

\trivialOrdering: the pairs of classifiers that are ordered in the same way by $\scorePrecision$ and $\scoreRecall$. As it is also the case for all $\beta$, there is no choice to make. The proportion of such pairs is given by
\begin{align}
    P(\trivialOrdering)
    & = 1 - \distKendall(\scorePrecision;\scoreRecall) \label{eq:no-choice} \\
    & = \nicefrac12 \left[1+\kendall(\scorePrecision;\scoreRecall)\right]
    = \nicefrac12 \left[\kendall(\scorePrecision;\scoreFBeta)+\kendall(\scoreFBeta;\scoreRecall)\right]
    \point
    \nonumber
\end{align}

\wrongOrdering: the pairs of classifiers for which a choice has to be made because $\scorePrecision$ and $\scoreRecall$ contradict each other, and for which the choice is not optimal ($\scoreFBeta$ disagrees with $\scoreOptimalTradeoff$). The proportion of such pairs is given by
\begin{equation}
    P(\wrongOrdering)
    = \distKendall(\scoreFBeta;\scoreOptimalTradeoff)
    = \nicefrac{1}{4} \left| 
        \kendall(\scorePrecision;\scoreFBeta) - \kendall(\scoreFBeta;\scoreRecall)
    \right|
    \point
    \label{eq:choice-not-optimal}
\end{equation}

\correctOrdering: the pairs of classifiers for which a choice has to be made and for which the choice is optimal. The proportion of such pairs is given by 
\begin{equation}
    P(\correctOrdering) 
    = \distKendall(\scorePrecision;\scoreRecall) - \distKendall(\scoreFBeta;\scoreOptimalTradeoff) 
    = 1 - P(\trivialOrdering) - P(\wrongOrdering)
    \point
    \label{eq:choice-optimal}
\end{equation}

It appears from \cref{eq:no-choice,eq:choice-not-optimal} that \begin{myImportantResult}if one uses some $\scoreFBeta$ to rank and reports the values for both $\kendall(\scorePrecision;\scoreFBeta)$ and $\kendall(\scoreFBeta;\scoreRecall)$, then we can determine the degree of optimality for the chosen $\beta$\end{myImportantResult}. We define it as follows:
\begin{equation}
    \optimality 
    = P(\correctOrdering | \overline{\trivialOrdering})
    = \frac{
        P(\correctOrdering)
    }{
        P(\correctOrdering) + P(\wrongOrdering)
    }
    = 1 - \frac{
        P(\wrongOrdering)
    }{
        1 - P(\trivialOrdering)
    }
    \point
    \label{eq:optimality}
\end{equation}
Note that $\optimality=1$ if and only if $\kendall(\scorePrecision;\scoreOptimalTradeoff) = \kendall(\scoreOptimalTradeoff;\scoreRecall)$, which is what is targeted by minimizing the Fréchet variance.%

%% file: sections/4_0_case_studies.tex
\input{tbls/summary}
\input{figs/distributions_in_tetrahedron}

\section{Case Studies}
\label{sec:case-studies}

We now perform case studies for several distributions (see \cref{fig:distributions_in_tetrahedron}) and many sets of performances. The results have been obtained either analytically (all details in supplementary material) or numerically based on Monte Carlo simulations. A summary of the results, emphasizing the degree of optimality and providing the links to results that can be used to determine the optimal tradeoff, is provided in \cref{tbl:summary}.

\input{sections/4_1_distribution_I}
\input{sections/4_2_distribution_II}

\input{sections/4_3_distribution_III}

\input{sections/4_4_distribution_IV}

\input{sections/4_5_distribution_V}

\input{sections/4_6_CDnet}

%% file: tbls/summary.tex
\newcommand{\optimalityStatistics}[3]{$\optimality$: min=#1\%, max=#2\%, mean=#3\%}

\begin{table*}
\begin{centering}
\resizebox{\linewidth}{!}{
\begin{tabular}{|p{4.5cm}|c|p{4.0cm}|p{4.0cm}|p{4.0cm}|p{4.0cm}|}
\hline 
Case study &
$\kendall(\scorePrecision; \scoreRecall)$ &
$\scoreFOne$ ($\beta^2=1$) &
$\scoreSkewInsensitiveVersionFOne$ ($\equiv \beta^2=\frac{\priorneg}{\priorpos}$) &
Heuristic $\beta^2 = \frac{
        \expectedValueSymbol[\scorePFP]
    }{
        \expectedValueSymbol[\scorePFN]
    }$&
Optimal tradeoff, $\optimality = 100\%$\tabularnewline
\hline 
\hline 
uniform distribution over all performances &
$\nicefrac13$ & 
$\optimality = \textcolor{teal}{100\%}$ (optimal) &
\textcolor{red}{NA \newline (meaningless ranking)} &
$\optimality = \textcolor{teal}{100\%}$ (optimal) \newline (it selects $\scoreFOne$) &
\textcolor{teal}{our analytical result} \newline \cf \cref{eq:analytical-solution-set-I} \tabularnewline
\hline 
uniform distributions with fixed probability of true negatives &
$\nicefrac13$ &
$\optimality = \textcolor{teal}{100\%}$ (optimal) &
\textcolor{red}{NA \newline (meaningless ranking)} &
$\optimality = \textcolor{teal}{100\%}$ (optimal) \newline (it selects $\scoreFOne$) &
\textcolor{teal}{our analytical result} \newline \cf \cref{eq:analytical-solution-set-II}  \tabularnewline
\hline 
uniform distributions with fixed class priors &
$\nicefrac12$ &
$\optimality \in [ \textcolor{red}{50\%}, 100\% ]$ \newline optimal only for $\priorpos \simeq 0.381$ &
$\optimality = \log(4) - \nicefrac12 \simeq \textcolor{orange}{88.63 \%}$ &
$\optimality = \log(4) - \nicefrac12 \simeq \textcolor{orange}{88.63 \%}$ \newline (it selects $\scoreSkewInsensitiveVersionFOne$) &
\textcolor{teal}{our analytical result} \newline \cf \cref{eq:analytical-solution-set-III} and \cref{fig:distri_III_adaptation} \tabularnewline
\hline 
uniform distributions with fixed class priors, above no-skill &
$0$ &
$\optimality \in [ \textcolor{red}{50\%}, 100\% ]$ \newline optimal only for $\priorpos \simeq 0.325$ &
$\optimality = \nicefrac56 \simeq \textcolor{orange}{83.33 \%}$ &
$\optimality = \nicefrac56 \simeq \textcolor{orange}{83.33 \%}$ \newline (it selects $\scoreSkewInsensitiveVersionFOne$) &
\textcolor{teal}{our analytical result} \newline \cf \cref{eq:analytical-solution-set-IV} \tabularnewline
\hline 
uniform distributions with fixed class priors, close to oracle &
$\in(0, \nicefrac12)$ &
$\optimality \in [ \textcolor{orange}{87.85\%}, 100\% ]$ \newline optimal only for $\priorpos \rightarrow 1$ &
$\optimality \in [ \textcolor{red}{50\%}, 100\% ]$ \newline optimal only for $\priorpos \simeq 0.561$ &
$\optimality \in [ \textcolor{orange}{87.85\%}, 100\% ]$ \newline (it selects $\scoreFOne$) &
\textcolor{teal}{our numerical result} \newline \cf \cref{fig:distri_V_adaptation} \tabularnewline
\hline 
53 real sets of about 60 performances &
$\in[-0.31, 0.65]$&
$\optimality \in [\textcolor{red}{52.23 \%}, 100.00 \%]$ \newline (mean: $78.44 \%$) &
$\optimality \in [\textcolor{red}{50.00 \%}, 99.13 \%]$ \newline (mean: $56.71 \%$) &
$\optimality \in [\textcolor{orange}{69.69 \%}, 100.00 \%]$ \newline (mean: $89.53 \%$) &
\textcolor{teal}{our closed-form expression for the optimal $\beta$}, \cref{eq:closed-form-expression} \tabularnewline
\hline 
\end{tabular}
}
\par\end{centering}
\caption{Summary of our results emphasizing the degree of optimality $\optimality$ introduced in \cref{eq:optimality}. As shown by the last column, we can always find the optimal ranking score (in green) between $\scorePrecision$ and $\scoreRecall$. Relying on a fixed and arbitrarily chosen score like $\scoreFOne$ or $\scoreSkewInsensitiveVersionFOne$ can lead to catastrophic cases (in red). In our case studies, the simple heuristic introduced in \cref{sec:cdnet} never led to such catastrophic cases, but nevertheless appeared to be suboptimal in some cases (in orange).
\label{tbl:summary}
}
\end{table*}

%% file: figs/distributions_in_tetrahedron.tex
\begin{figure*}
\subfloat[The set $\setI$.\label{fig:setI}]{
    \includegraphics[viewport={11mm 20mm 84mm 96mm},clip,scale=0.45]{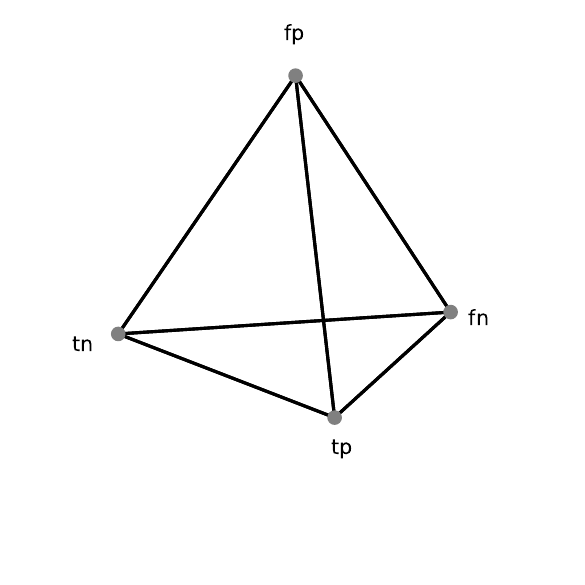}
}
\hfill
\subfloat[The sets $\setII$.\label{fig:setII}]{
    \includegraphics[viewport={11mm 20mm 84mm 96mm},clip,scale=0.45]{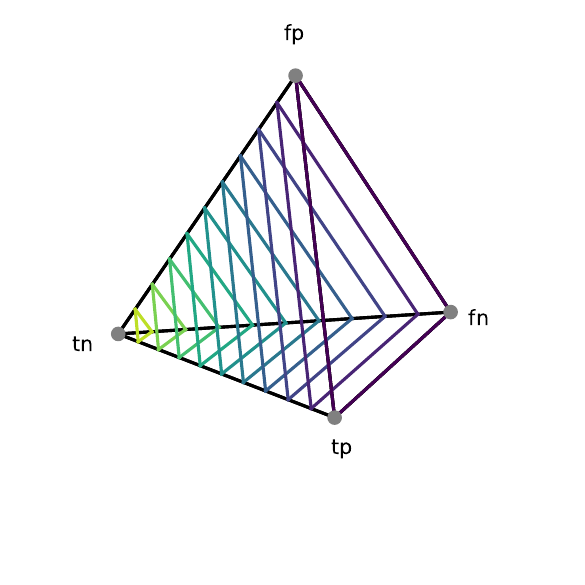}
}
\hfill
\subfloat[The sets $\setIII$.\label{fig:setIII}]{
    \includegraphics[viewport={11mm 20mm 84mm 96mm},clip,scale=0.45]{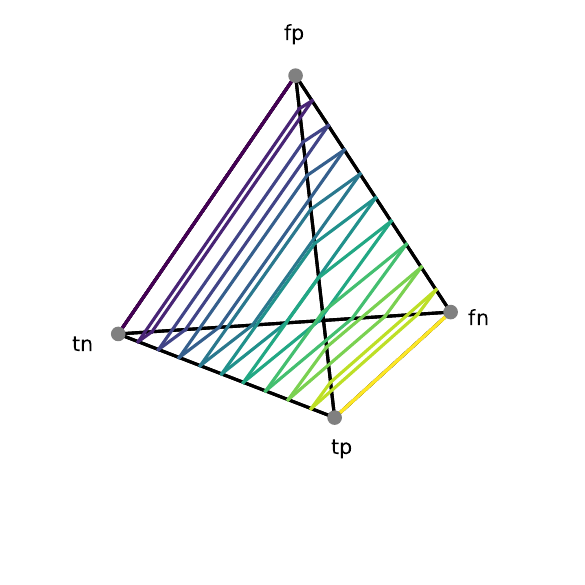}
}
\hfill
\subfloat[The sets $\setIV$.\label{fig:setIV}]{
    \includegraphics[viewport={11mm 20mm 84mm 96mm},clip,scale=0.45]{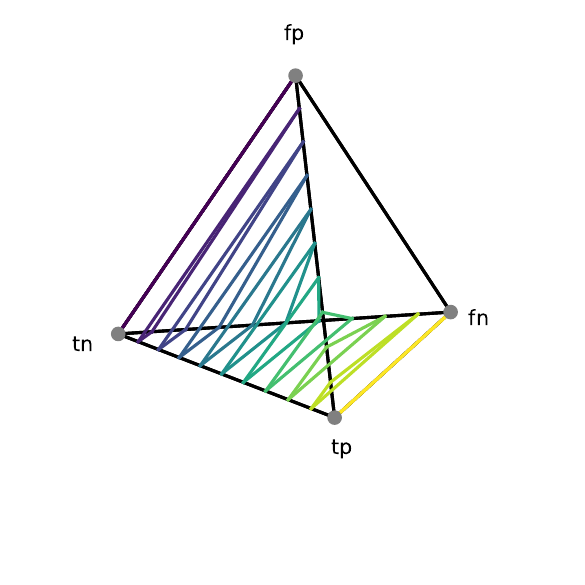}
}
\hfill
\subfloat[The sets $\setV$.\label{fig:setV}]{
    \includegraphics[viewport={11mm 20mm 84mm 96mm},clip,scale=0.45]{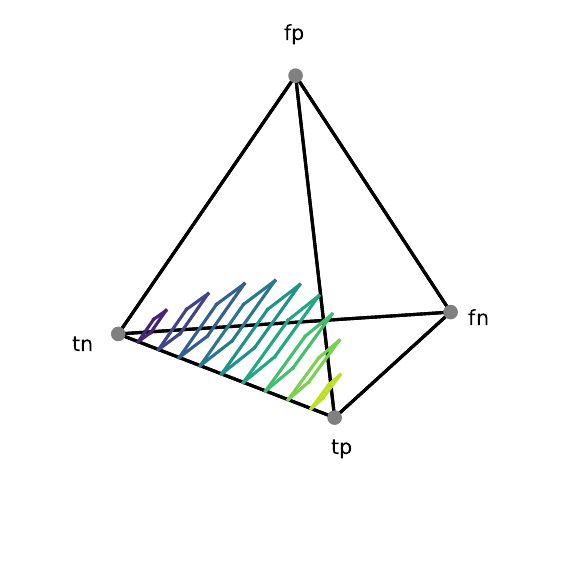}
}

\caption{
    Graphical representation of some sets of two-class crisp classification performances considered in this paper. 
    (a) All possible performances %
    correspond to points in a regular tetrahedron. 
    (b) The performances corresponding to fixed probabilities of true negatives %
    are points in parallel equilateral triangles. 
    (c) The performances corresponding to fixed class priors %
    are points in parallel rectangles~\cite{Lovell2023NeverMind}. 
    (d) The performances corresponding to fixed class priors and ``above'' no-skills %
    are points in parallel right triangles. 
    (e) The performances for given class priors,
    and ``close to the oracle'' %
    are points in parallel squares. The color represents the value of the fixed parameter ($ptn$ or $\priorpos$).
    \label{fig:distributions_in_tetrahedron}
}
\end{figure*}

%% file: sections/4_1_distribution_I.tex
\subsection{Uniform Distribution Over All Performances}

Let us start by considering the uniform distribution over the set $\setI$ of all performances:
$%
    \setI = \allPerformances
$. %
It is the Dirichlet distribution with all concentration parameters set to $1$, and corresponds to a uniform distribution in a tetrahedron (see \cref{fig:setI}). With this distribution, precision and recall are uniformly distributed and positively correlated: $\kendall(\scorePrecision;\scoreRecall)=\nicefrac13$.
The $\scoreFOne$ score has not uniformly distributed values, but is the optimal tradeoff between $\scorePrecision$ and $\scoreRecall$:
\begin{equation}
    \kendall(\scorePrecision;\scoreFOne) 
    = \kendall(\scoreFOne;\scoreRecall)
    = \nicefrac23
    \point
    \label{eq:analytical-solution-set-I}
\end{equation}

The skew-insensitive version of $\scoreFOne$, $\scoreSkewInsensitiveVersionFOne$, does not lead to meaningful rankings as it does not satisfy the \third axiom of the theory of performance-based ranking \cite{Pierard2025Foundations} on $\setI$. Moreover, we computed $\kendall(\scorePrecision;\scoreSkewInsensitiveVersionFOne)\simeq 0.43$ and $\kendall(\scoreSkewInsensitiveVersionFOne;\scoreRecall)\simeq 0.81$, which implies that $\scoreSkewInsensitiveVersionFOne$ it is not located on a shortest path between $\scorePrecision$ and $\scoreRecall$ and is not at equidistance from them.

%% file: sections/4_2_distribution_II.tex
\subsection{With Fixed Probability of True Negatives}

Motivated by the fact that $\scorePrecision$, $\scoreRecall$, and $\scoreFBeta$ give no importance to true negatives, we now consider a second type of distributions, the uniform distribution over the sets 
$%
    \setII = \{ 
        \aPerformance\in\allPerformances : \aPerformance(\eventTN)=ptn
    \}
    \subset \setI
$. %
Note that $\aSetOfPerformances_2(0)$ is the set of performances considered in detection problems. As shown in \cref{fig:setII}, these distributions correspond to uniform distributions over equilateral triangles. $\scorePrecision$ and $\scoreRecall$ have uniformly distributed values and are positively correlated: $\kendall(\scorePrecision;\scoreRecall)=\nicefrac13$.

For $\scoreFOne$, our observations remain the same as what we had without any constraint, \ie it is the optimal tradeoff:%
\begin{equation}
    \kendall(\scorePrecision;\scoreFOne) 
    = \kendall(\scoreFOne;\scoreRecall)
    = \nicefrac23
    \point
    \label{eq:analytical-solution-set-II}
\end{equation}

For $\scoreSkewInsensitiveVersionFOne$, we found that it does not satisfy the \third axiom of the theory of performance-based ranking \cite{Pierard2025Foundations} on $\setII$, unless $ptn=0$. But, to the contrary of what we had without any constraint, $\scoreSkewInsensitiveVersionFOne$ is now located on a shortest path between $\scorePrecision$ and $\scoreRecall$. However, it is still far from being at equidistance from them. For $ptn=0$, we found $\kendall(\scorePrecision;\scoreSkewInsensitiveVersionFOne)=\nicefrac13$ and $\kendall(\scoreSkewInsensitiveVersionFOne;\scoreRecall)=1$.

%% file: sections/4_3_distribution_III.tex
\subsection{With Fixed Class Priors}

We now move to more realistic distributions for classification, and consider that one needs to rank only performances corresponding to some given class priors $(\priorneg,\priorpos)$. For this reason, we consider the uniform distributions over the sets 
$%
    \setIII = \{ 
        \aPerformance\in\allPerformances : \aPerformance(\{\sampleFN,\sampleTP\})=\priorpos
    \}
    \subset \setI
$. %
As shown in \cref{fig:setIII}, these distributions correspond to uniform distributions over rectangles. This classical reference distribution is used, \eg, in \cite{Zhao2022Classifier,Zhao2025Outperformance-arxiv}. %
It turns out that their axes correspond to %
$\scoreFPR\in[0,1]$ and to $\scoreTPR\in[0,1]$. The uniform distributions over $\setIII$ correspond thus to uniform distributions of points in \roc. We found, no matter what the class priors are, that 
$\kendall(\scorePrecision;\scoreRecall)=\nicefrac12$.

If we apply a two-components \pca to the rankings obtained with $\setIII$ for $\priorpos=0.1$, we obtain the curve depicted in \cref{fig:graphical_abstract} ($98.77\%$ of total variance explained).

\input{figs/distri_III_V}
We stress that, to the contrary of what we concluded for the two previous distributions, when the class priors are fixed, $\scoreFOne$ is far from the optimal tradeoff, as shown in \cref{fig:distri_III_f1_is_not_between_ppv_tpr}: unless $\priorpos\simeq 0.381$, $\kendall ( \scorePrecision ; \scoreFOne ) \ne \kendall ( \scoreFOne ; \scoreRecall )$. In the extreme cases for which the prior of the negative (positive) class tends towards $1.0$, the ranking induced by $\scoreFOne$ perfectly mimics the ranking induced by $\scorePrecision$ ($\scoreRecall$), thus totally ignoring the ranking induced by $\scoreRecall$ ($\scorePrecision$).

To find the optimal threshold, we need first to compute either $\kendall ( \scorePrecision ; \scoreFBeta )$ or $\kendall ( \scoreFBeta ; \scoreRecall )$. The other can be found using \cref{eq:shortest-path-correlations}. Analytically, we found
\begin{equation}
    \kendall ( \scoreFBeta ; \scoreRecall ) = \frac12 + \ell - \ell^2 \log\frac{1+\ell}{\ell}\comma 
\end{equation}
with $\ell$ defined as in \cref{eq:ell}. %
Then, we minimized the Fréchet variance, given in \cref{eq:frechet-variance}. 
If one considers some given class priors, then it is a function of $\beta$ (see \cref{fig:distri_III_frechet_variance}) and the minimization leads to the optimal value of $\beta$ for $\scoreOptimalTradeoff=\scoreFBeta$. However, it is also possible to see the Fréchet variance as a function of $\ell$. Its minimization leads to the solution 
$\ell \simeq 0.61585$. 
Injecting this value into \cref{eq:ell}, we obtained the link between the optimal $\beta$ and the class priors:
\begin{equation}
    \beta^2 = \ell \frac{\priorneg}{\priorpos}
    \Rightarrow
    b
    \simeq \frac{
        0.61585 \, (1-\priorpos)
    }{
        \priorpos+ 0.61585 \, (1-\priorpos)
    }
    \label{eq:analytical-solution-set-III}
\end{equation}
It is depicted in \cref{fig:distri_III_adaptation}. We see that the optimal relationship is quite close to that of $\scoreSkewInsensitiveVersionFOne$ (\ie, $b=1-\priorpos)$, but very different from that of $\scoreFOne$ (\ie, $b=\nicefrac{1}{2}$). %

%% file: figs/distri_III_V.tex
\begin{figure*}
    \begin{centering}
        \hfill
        \subfloat[$\scoreFOne$ is not the optimal tradeoff.\label{fig:distri_III_f1_is_not_between_ppv_tpr}]{
            \includegraphics[scale=0.34]{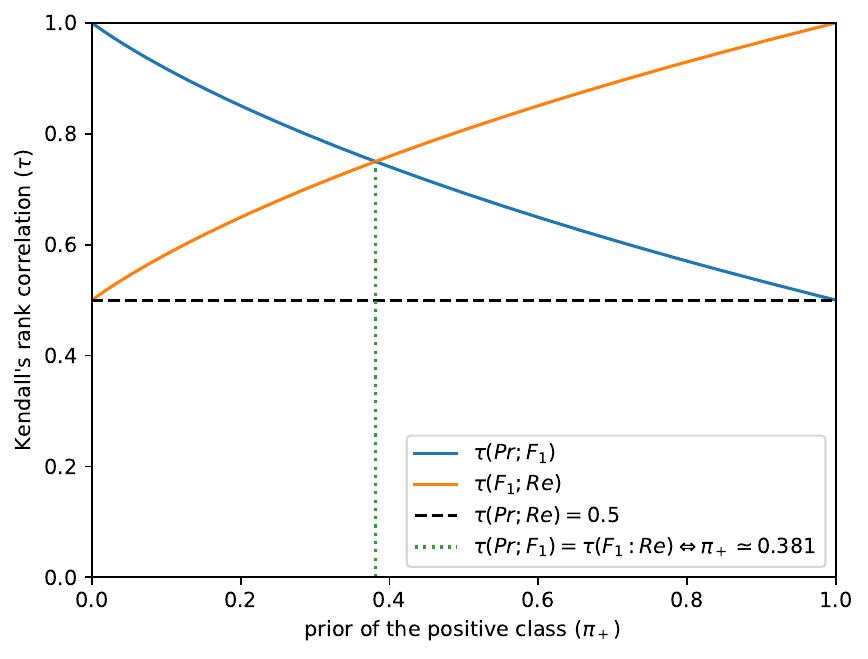}
        }
        \hfill
        \hfill
        \subfloat[Fréchet variance for various priors.\label{fig:distri_III_frechet_variance}]{
            \includegraphics[scale=0.34]{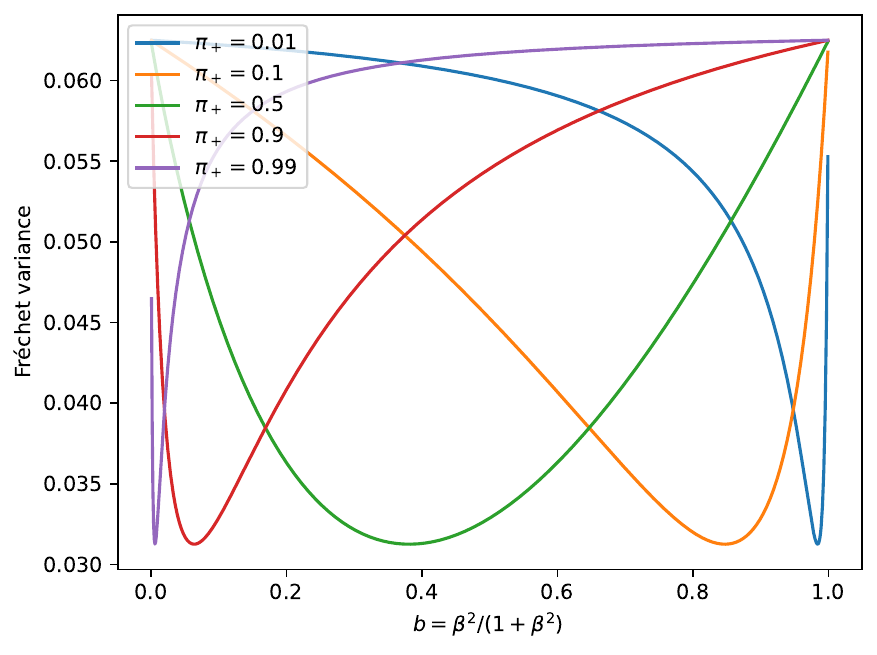}
        }
        \hfill
        \hfill
        \subfloat[Adaptation of $\beta$ \wrt class priors.\label{fig:distri_III_adaptation}]{
            \includegraphics[scale=0.34]{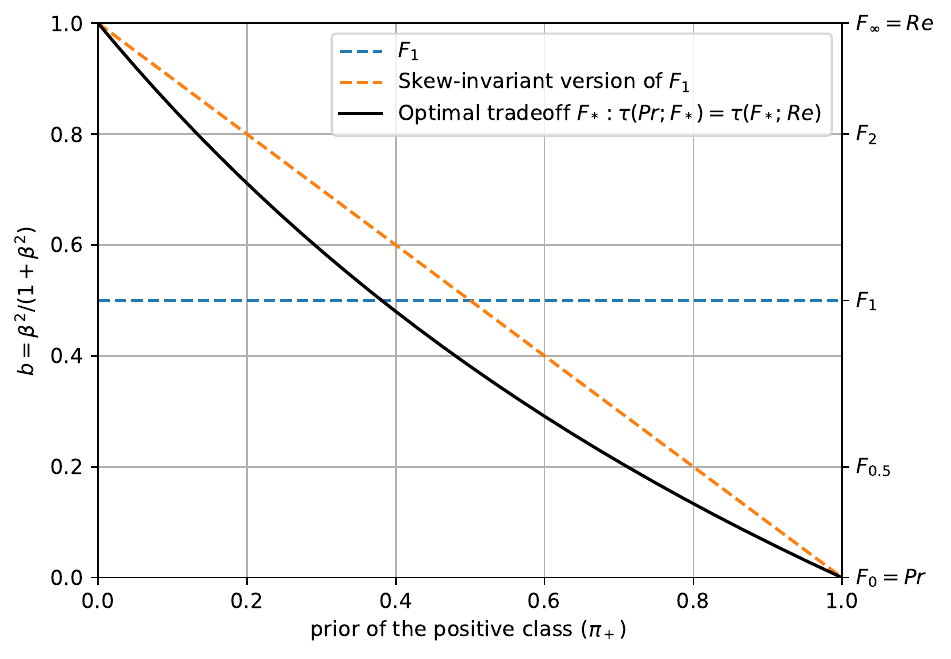}
        }
        \hfill \\
        \hfill
        \subfloat[$\scoreFOne$ is not the optimal tradeoff.\label{fig:distri_V_f1_is_not_between_ppv_tpr}]{
            \includegraphics[scale=0.34]{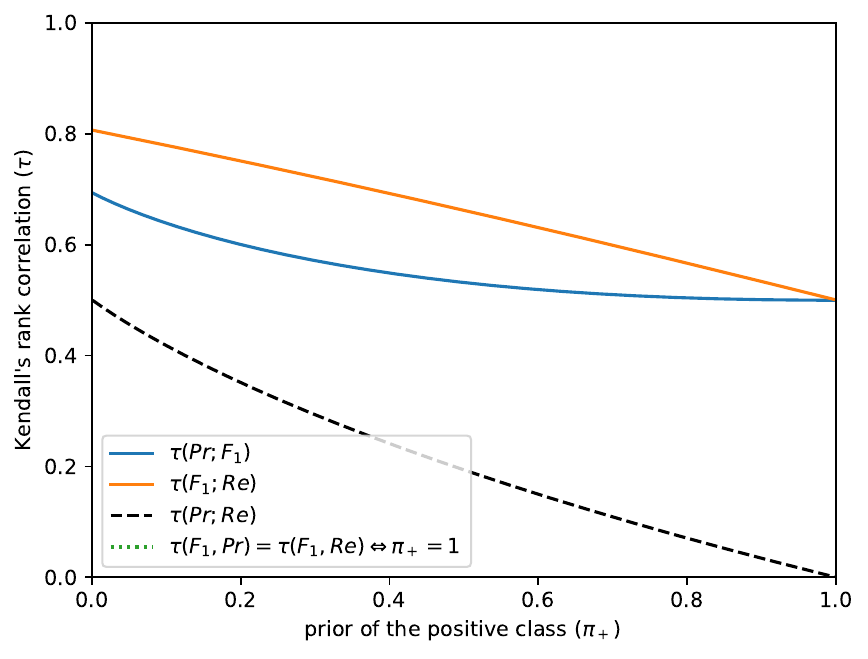}
        }
        \hfill
        \hfill
        \subfloat[Fréchet variance for various priors.\label{fig:distri_V_frechet_variance}]{
            \includegraphics[scale=0.34]{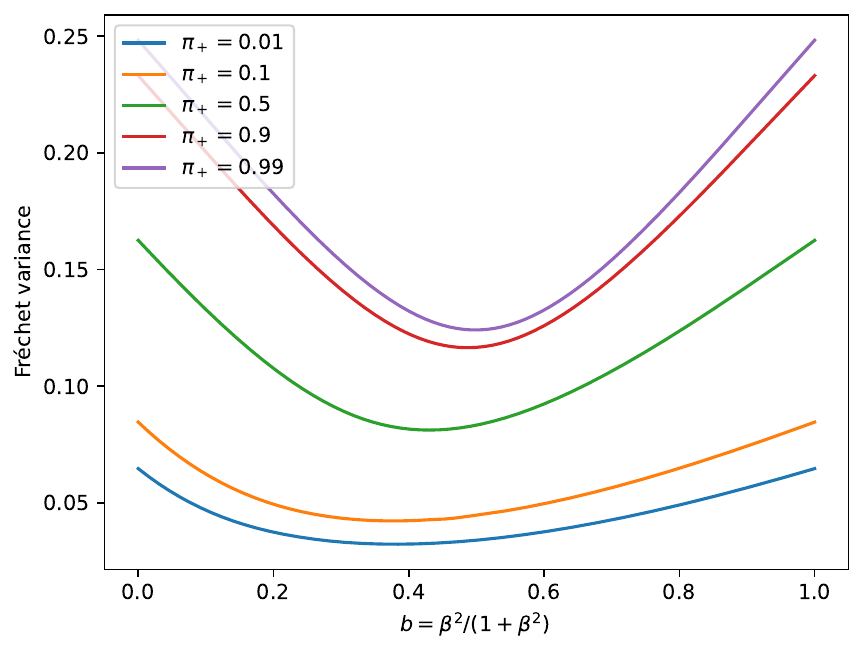}
        }
        \hfill
        \hfill
        \subfloat[Adaptation of $\beta$ \wrt class priors.\label{fig:distri_V_adaptation}]{
            \includegraphics[scale=0.34]{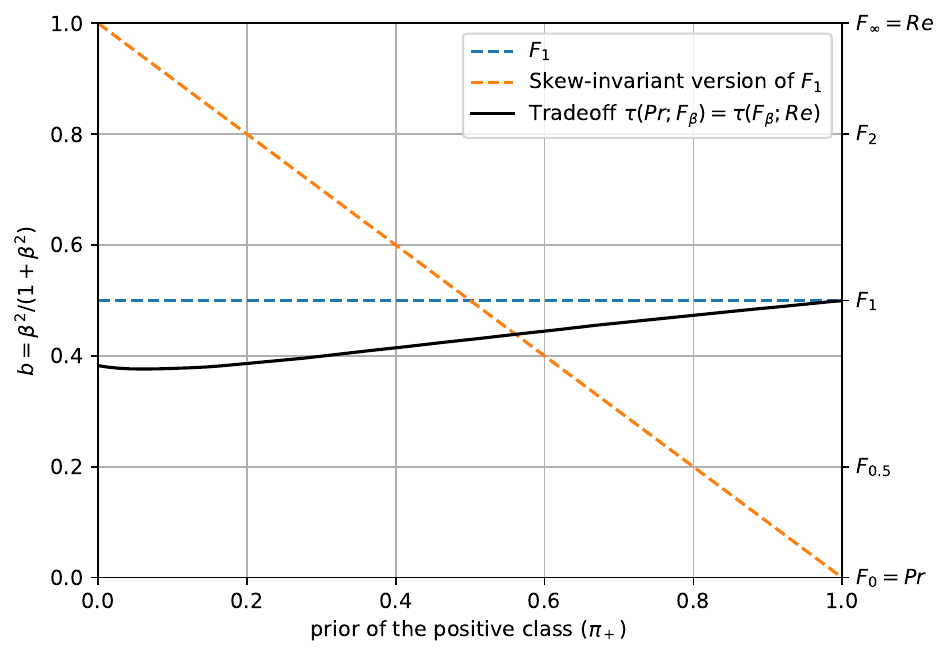}
        }
        \hfill
    \par\end{centering}
        
    \caption{
        Results for uniform distributions with fixed class priors (top), \ie $\setIII$, and close to the oracle (bottom), \ie $\setV$.
        \label{fig:distri_III_V}
    }
\end{figure*}

%% file: sections/4_4_distribution_IV.tex
\subsection{With Fixed Class Priors, Above No-Skill}

Performances below the rising diagonal of ROC, where the no-skill performances lie, are usually of little interest. %
For this reason, we study the case of the uniform distributions over the sets 
$%
    \setIV = \{ 
        \aPerformance\in\allPerformances : \aPerformance(\{\sampleFN,\sampleTP\})=\priorpos \\
        \wedge \scoreTPR(\aPerformance) \ge \scoreFPR(\aPerformance)
    \}
    \subset \setIII
$. %
These sets are depicted in \cref{fig:setIV}. No matter what the class priors are, we obtain 
$\kendall(\scorePrecision;\scoreRecall)=0$. 
All detailed results are 
in supplementary material. In a nutshell, the observations are very close to those reported for the uniform distributions with fixed class priors. The overall shape of the manifold, as observed by \pca, is very similar. %
$\kendall(\scorePrecision;\scoreFOne)$ monotonically increases from $0$ to $1$ (instead of $0.5$ to $1$) and $\kendall(\scoreFOne;\scoreRecall)$ monotonically decreases from $1$ to $0$ (instead of $1$ to $0.5$) when $\priorpos$ sweeps the $[0,1]$ interval. The $\scoreFOne$ score is the optimal tradeoff only for $\priorpos\simeq0.325$. Regarding the adaptation, the curve is slightly more curved and lower than the one we depicted in \cref{fig:distri_III_adaptation}, with the optimal value 
$\ell \simeq 0.48$:
\begin{equation}
    \beta^2 = \ell \frac{\priorneg}{\priorpos}
    \Rightarrow
    b 
    \simeq \frac{
        0.48 \, (1-\priorpos)
    }{
        \priorpos+ 0.48 \, (1-\priorpos)
    }
    \point
    \label{eq:analytical-solution-set-IV}
\end{equation}

%% file: sections/4_5_distribution_V.tex
\subsection{With Fixed Class Priors, Close to Oracle}

In competitions, a setting in which ranking is omnipresent, contenders usually refrain from choosing decision thresholds that would lead either to a high $\scoreTPR$ at the cost of a high $\scoreFPR$ or to a low $\scoreFPR$ at the cost of a low $\scoreTPR$. In our opinion, this means that $\scoreFPR<\priorpos$ and $\scoreTPR>\priorpos$. In \roc, these performances are above and on the left-hand side of the no-skill performance that results from randomly choosing the negative or positive class with respective probabilities equal to $\priorneg$ and $\priorpos$. So, we now discuss the case of uniform distributions over the sets 
$%
    \setV = \{ 
        \aPerformance\in\allPerformances : \aPerformance(\{\sampleFN,\sampleTP\})=\priorpos %
        \wedge \scoreFPR(\aPerformance)<\priorpos
        \wedge \scoreTPR(\aPerformance)>\priorpos
    \}
    \subset \setIV
$. %
As shown in \cref{fig:setV}, these distributions correspond to uniform distributions over squares. Drawing a performance at random can be achieved by drawing independent values uniformly in $[0,\priorpos]$ for $\scoreFPR$ and in $[\priorpos,1]$ for $\scoreTPR$.

The results, shown in \cref{fig:distri_V_f1_is_not_between_ppv_tpr,fig:distri_V_frechet_variance,fig:distri_V_adaptation}, are remarkably different from the previous ones. In comparison to \cref{fig:distri_III_adaptation}, \cref{fig:distri_V_adaptation} shows that the optimal $\beta$ now increases from about $0.8$ to $1$, when $\priorpos$ increases from $0$ to $1$, while it was previously decreasing from $\infty$ to $0$.%

%% file: sections/4_6_CDnet.tex
\subsection{Some Real Sets of Performances}
\label{sec:cdnet}

\input{figs/cdnet_result_on_one_video}
\input{figs/cdnet_prediction_b_tradeoff}
\input{figs/cdnet_2014_probas_fone_vs_tradeoff}

We now report results for performances encountered in existing rankings. They are either performances of some baseline methods or performances that people considered good enough for a competition. We consider a computer vision task known as \emph{background subtraction}, similar to a pixelwise classification between background and foreground \cite{Jodoin2014Overview,Garcia2020Background}. The results of about 60 methods (classifiers), on 53 videos ---a dataset known as \CDnetMMXIV~\cite{Wang2014AnExpanded}--- %
are publicly available on \CDnetPlatform. %
According to~\cite{Goyette2012Changedetection}, the multi-criteria ranking proposed by this platform is well correlated with $\scoreFOne$. On any given video, all performances belong to $\setIII$. Moreover, $99.9\%$ and $99.4\%$ of them belong to $\setIV$ and $\setV$, respectively. Depending on the video, 
$\kendall(\scorePrecision;\scoreRecall)\in [-0.31, 0.65]$.
\Cref{fig:cdnet_result_on_one_video} shows the results for the set of performances reported for one video; more results are in the supplementary material. The optimal $\beta$ is between $0.29$ and $22.79$ depending on the video. It is in the $[0.5, 2]$ range for only $35$ videos among the $53$.

An interesting observation, shown in \cref{fig:cdnet_prediction_b_tradeoff}, is that the class priors $(\priorneg,\priorpos)$ are poor predictors for the optimal value of $\beta$. %
However, it also suggests that 
\begin{equation}
    \beta^2
    =\frac{
        \expectedValueSymbol[\scorePFP]
    }{
        \expectedValueSymbol[\scorePFN]
    }
    =\frac{
        \sum_{\aPerformance \in \aSetOfPerformances} \aPerformance(\eventFP)
    }{
        \sum_{\aPerformance \in \aSetOfPerformances} \aPerformance(\eventFN)
    }
    \label{eq:heuristic}
\end{equation} 
is a simple heuristic that is worth trying as an alternative to the closed-form and optimal solution provided in \cref{eq:closed-form-expression}. The degree of optimality $\optimality$ achieved by this heuristic is provided in \cref{tbl:summary} for all our case studies.

\Cref{fig:cdnet_2014_probas_fone_vs_tradeoff} shows the consequences of using $\scoreFOne$ for ranking instead of the optimal tradeoff $\scoreOptimalTradeoff$. On average, there is $61.98\%$ (\markerNothingToDecide) of chance that a randomly chosen pair of methods is ranked the same by $\scorePrecision$ and $\scoreRecall$. In this case, which tradeoff is chosen among all $\scoreFBeta$ is not relevant. But when $\scorePrecision$ and $\scoreRecall$ disagree, the value of $\beta$ matters. The degree of optimality for $\scoreFOne$ is only about $\optimality\simeq78\%$ $\frac{\textrm{\markerAgreeWithTradeoff}}{\textrm{\markerAgreeWithTradeoff}+\textrm{\markerDisagreeWithTradeoff}}$ on average.

%% file: figs/cdnet_result_on_one_video.tex
\begin{figure*}[t]
    \begin{centering}
        \hfill
        \includegraphics[scale=0.36]{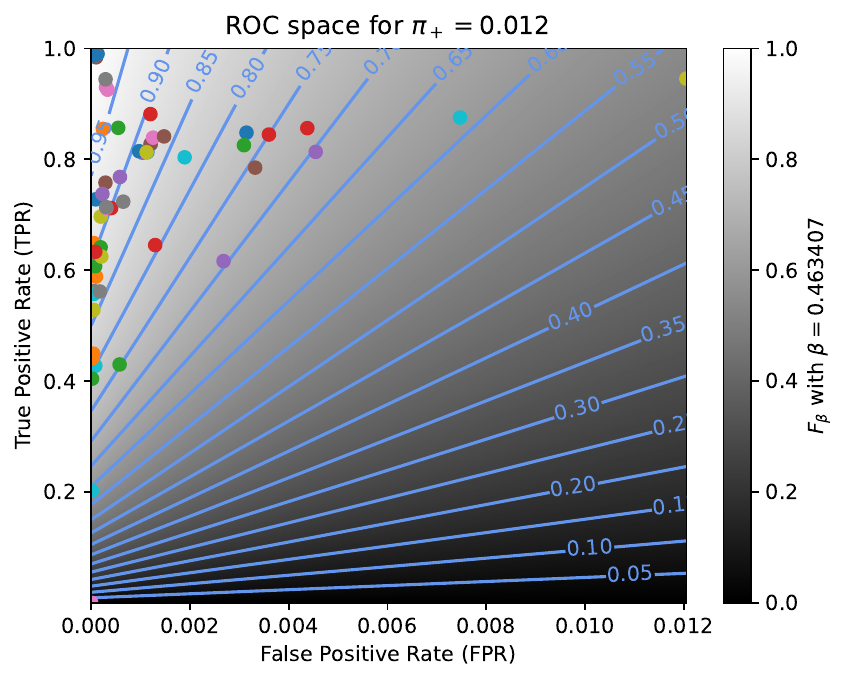}
        \hfill
        \hfill
        \includegraphics[scale=0.36]{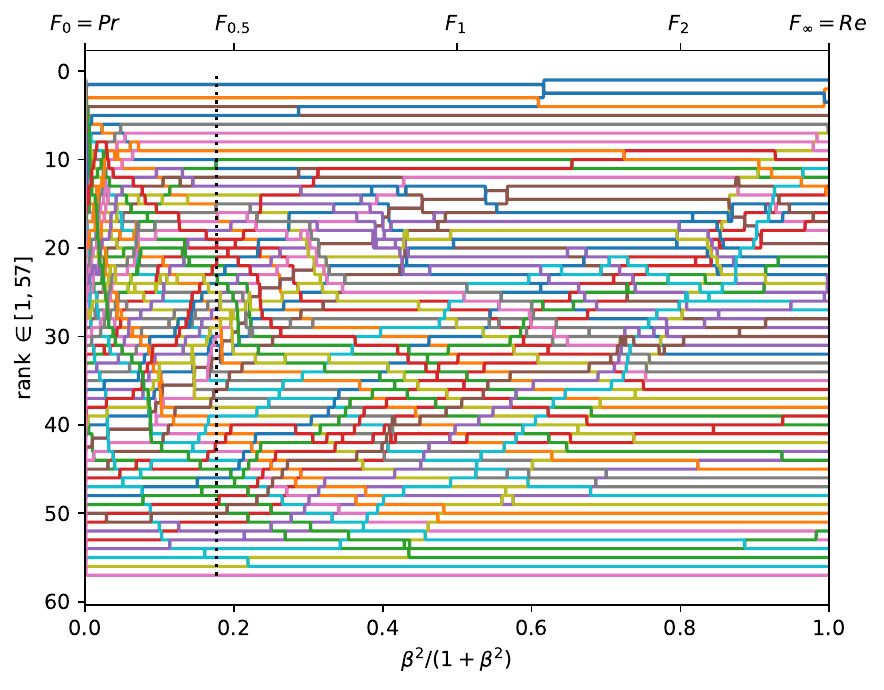}
        \hfill
        \hfill
        \includegraphics[scale=0.36]{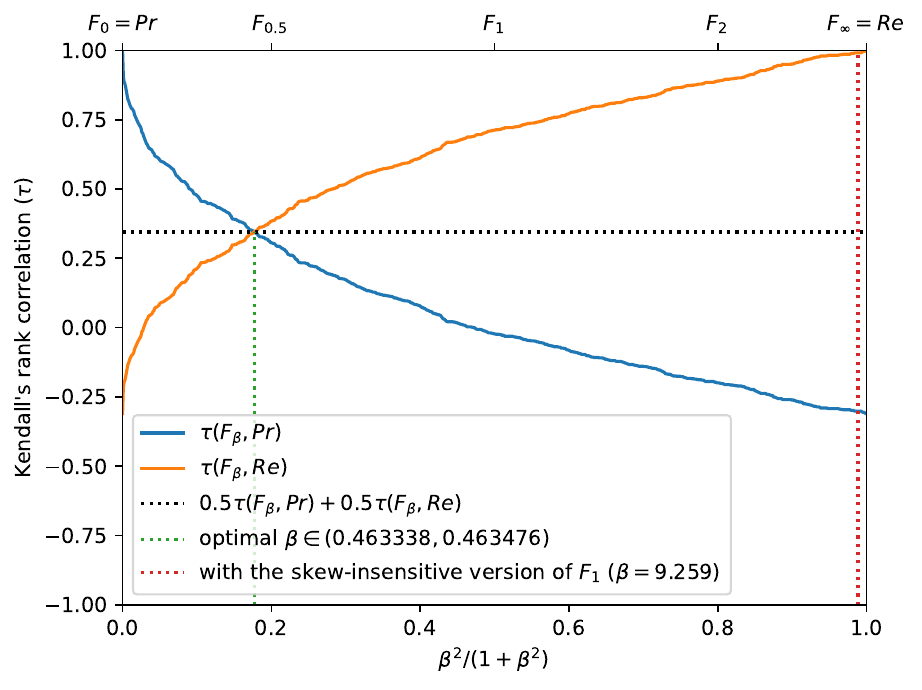}
        \hfill
    \par\end{centering}
    
    \caption{
        What is the optimal tradeoff %
        for the %
        performances of $57$ background subtraction methods (pixelwise classifiers) on the \emph{blizzard} video of \CDnetMMXIV? Left: the $57$ performances to be ranked, depicted in the ROC space, with the isometrics of the optimal tradeoff. Center: the ranks of each classifier, \wrt $\beta$. Right: %
        $\kendall(\scorePrecision;\scoreFBeta)$ and $\kendall(\scoreFBeta;\scoreRecall)$ \wrt $\beta$. The ranking with $\scoreFOne$, at the center, is too far from $\scorePrecision$ and too close to $\scoreRecall$. The optimal tradeoff, for which $\kendall(\scorePrecision;\scoreFBeta)=\kendall(\scoreFBeta;\scoreRecall)$ is at $\beta \simeq 0.463$.
        \label{fig:cdnet_result_on_one_video}
    }
\end{figure*}

%% file: figs/cdnet_prediction_b_tradeoff.tex
\begin{figure}[t]
    \begin{centering}
        \hfill
        \includegraphics[width=0.48\linewidth]{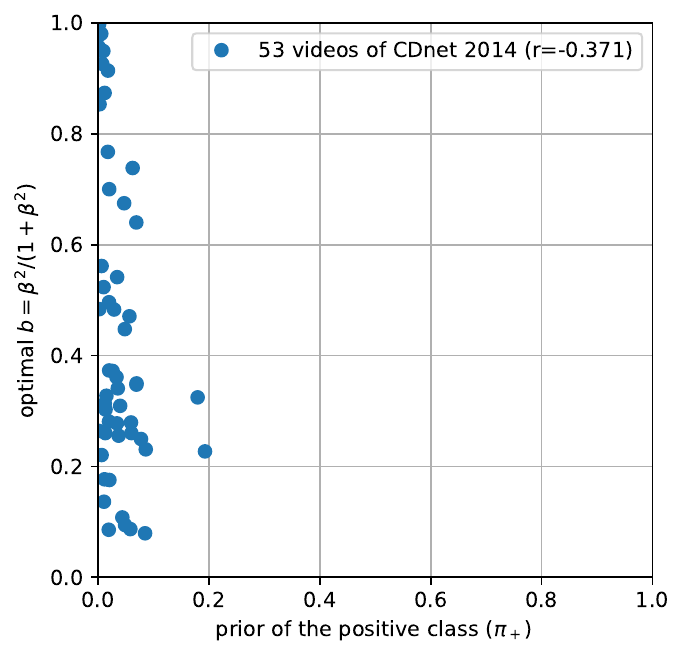}
        \hfill
        \hfill
        \includegraphics[width=0.48\linewidth]{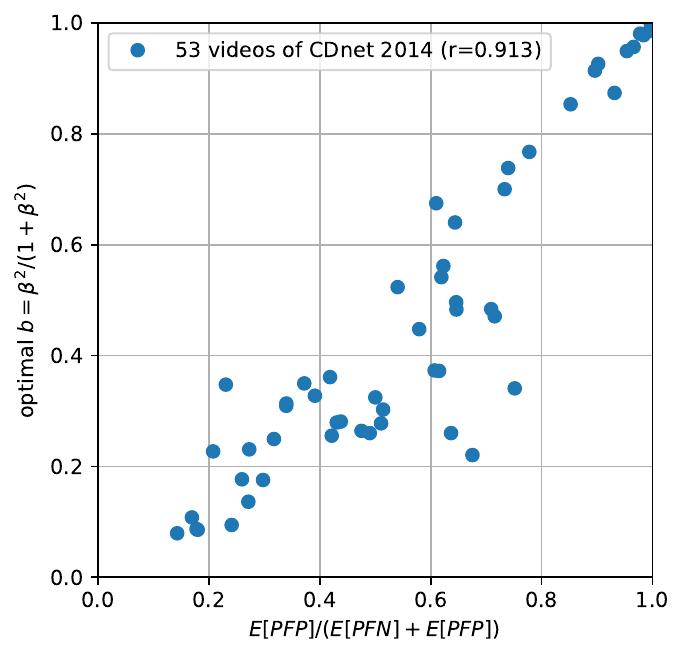}
        \hfill
    \par\end{centering}
    
    \caption{
        Which statistic about the sets of performances can be used to predict the optimal value of $\beta$? These results are based on $53$ sets of performances (those of about $60$ methods on each video of \CDnetMMXIV). Left: the class priors are not suitable. %
        Right: the ratio of the average probability of false positives ($\expectedValueSymbol[\scorePFP]$) to the average probability of errors ($\expectedValueSymbol[\scorePFN]+\expectedValueSymbol[\scorePFP]$) seems to be a good predictor for $\beta$, as Pearson's linear correlation between $\frac{\expectedValueSymbol[\scorePFP]}{\expectedValueSymbol[\scorePFN]+\expectedValueSymbol[\scorePFP]}$ and $\frac{\beta^2}{1+\beta^2}$ is $r\simeq 0.913$.
        \label{fig:cdnet_prediction_b_tradeoff}
    }
\end{figure}

%% file: figs/cdnet_2014_probas_fone_vs_tradeoff.tex
\definecolor{colorSilver}{RGB}{192, 192, 192}
\newcommand{\markerNothingToDecide}{\textcolor{colorSilver}{$\vrectangleblack$}\xspace}
\definecolor{colorLightGreen}{RGB}{166, 236, 153}
\newcommand{\markerAgreeWithTradeoff}{\textcolor{colorLightGreen}{$\vrectangleblack$}\xspace}
\definecolor{colorLightCoral}{RGB}{225, 134, 131}
\newcommand{\markerDisagreeWithTradeoff}{\textcolor{colorLightCoral}{$\vrectangleblack$}\xspace}

\begin{figure}[t]
    \begin{centering}
        \includegraphics[width=\linewidth]{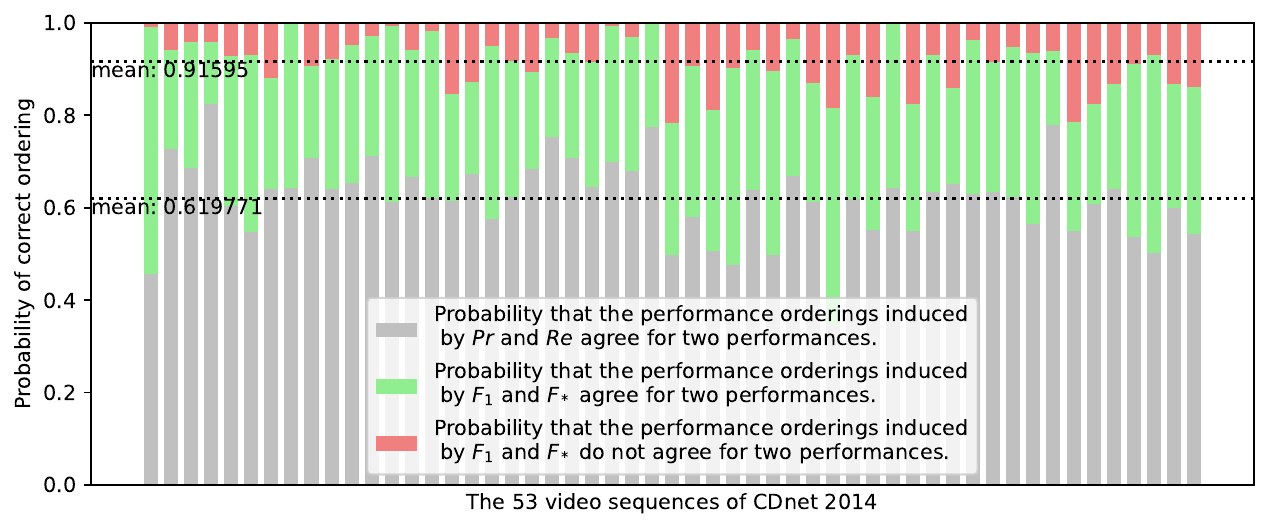}
    \par\end{centering}
    
    \caption{
        How far is $\scoreFOne$ from the optimal tradeoff $\scoreOptimalTradeoff$? %
        Each vertical bar is for a specific set of performances. 
        \markerNothingToDecide: $P(\trivialOrdering)$, \markerAgreeWithTradeoff: $P(\correctOrdering)$, 
        \markerDisagreeWithTradeoff:  $P(\wrongOrdering)$.
        \label{fig:cdnet_2014_probas_fone_vs_tradeoff}
    }
\end{figure}

%% file: sections/5_conclusion.tex
\section{Conclusion}

Nobody would ever choose a classifier solely based on the precision $\scorePrecision$ or the recall $\scoreRecall$. The traditional approach of calculating the harmonic mean $\scoreFOne$ of these two scores and then ranking the classifiers based on it is, unfortunately, nothing but a red herring. The colossal problem, ignored in the literature until now, is that the ranking built upon the balanced average of two scores is generally not a good tradeoff between the rankings obtained with the two base scores. In extreme cases, the ranking obtained with $\scoreFOne$ copies the ranking based on $\scorePrecision$ or $\scoreRecall$ and totally ignores the other.

To solve these issues, we defined and described the manifold of rankings induced by scores (see \cref{fig:graphical_abstract}) and found that the distance along it is linearly related to Kendall's rank correlation $\kendall$. Besides, we showed that all $\scoreFBeta$ scores induce meaningful rankings, even without any constraint on the performances. 
We further proved that the $\scoreFBeta$ scores induce rankings that form a shortest path between the rankings of $\scorePrecision$ and $\scoreRecall$. Also, we provided the theory and methods (including a closed-form expression) for defining and identifying the optimal tradeoffs between the rankings induced by $\scorePrecision$ and $\scoreRecall$. Lastly, we exemplified six case studies, covering numerous distributions and sets of performances. A summary of our results is provided in \cref{tbl:summary}.

\mysection{Acknowledgments.}
S. Pi{\'e}rard is funded by grants 8573 (ReconnAIssance project) and 2010235 (ARIAC by \href{https://www.digitalwallonia.be/en/}{DIGITALWALLONIA4.AI}) of the SPW EER, Wallonia, Belgium; A. Deli{\`e}ge is a \href{https://www.frs-fnrs.be}{F.R.S.-FNRS} postdoc researcher.

%% file: sections/A_0_supplementary.tex
\renewcommand{\thefigure}{\Alph{section}.\arabic{subsection}.\arabic{figure}}
\setcounter{figure}{0} %

\section{Supplementary Material}

This is the supplementary material for paper \emph{\thetitle}.

\section*{Contents} %
\startcontents[mytoc]
\printcontents[mytoc]{}{0}{}

\clearpage
\input{sections/A_1_0_suppl_for_related_work}
\clearpage
\input{sections/A_2_0_suppl_for_theory}
\clearpage
\input{sections/A_3_0_suppl_for_case_studies}

%% file: sections/A_1_0_suppl_for_related_work.tex
\subsection{Supplementary Material About \cref{sec:building-upon-related-work} (\,\,\,\nameref{sec:building-upon-related-work})}

\subsubsection{Wide Use of $\scoreFOne$}

The $\scoreFOne$ score (\aka Dice) is definitely standard in computer vision, serving as ranking criteria in benchmarks and challenges on tasks such as image/video classification in medical imaging (\linkCOVIDCT, \linkPAIP, \linkECDP, \linkTNSCUI), and behavior analysis (\linkABAW), segmentation in video surveillance (\linkIWDD \cite{Bouwmans2026Illegal}), medical imaging (\linkTDTEETHSEG, \linkDRAC, \linkUNICORN), and visual quality control (\linkVAND), detection in medical imaging (\linkMIDOG, \linkUNICORN). Also, \linkCADARRE \cite{Ivantsits2021Cerebral} uses $\scoreFBeta[2]$; we discuss it in \cref{sec:CADA-RRE}. Notably, some are from CVPR'25 (\linkVAND, \linkABAW). More exist in computer vision and beyond.

\subsubsection{Reminder of the Three Axioms of Performance-Based Ranking}

These axioms are from last year CVPR paper~\cite{Pierard2025Foundations}.

\newcommand{\performanceOrdering}{\relWorseOrEquivalent_\aScore}

The first axiom states that a ranking must be established based on a preorder $\relWorseOrEquivalent$ on the performances. Following Theorem~1 of~\cite{Pierard2025Foundations}, in this paper, we consider only preorders $\performanceOrdering$ induced by a score $\aScore$ in the following manner. Regardless of whether the performances $\aPerformance_1$ and $\aPerformance_2$ to be compared belong to the domain of definition of $\aScore$, we decide that $\aPerformance_1 \performanceOrdering \aPerformance_2$ when $\aPerformance_1 = \aPerformance_2$. If $\aPerformance_1$ and $\aPerformance_2$ are both in the domain of definition of $\aScore$, then we decide that $\aPerformance_1 \performanceOrdering \aPerformance_2$ if and only if $\aScore(\aPerformance_1) \le \aScore(\aPerformance_2)$. In all other cases, $\aPerformance_1 \not\performanceOrdering \aPerformance_2$. Such a preorder is called \emph{performance ordering induced by $\aScore$}. As an immediate consequence of this first axiom, the rankings are stable: it is impossible to swap two previously compared methods by inserting or deleting methods in the ranking.

The two other axioms ensure that one can interpret the binary relation $\performanceOrdering$ as \emph{worse or equivalent}. For the sake of consistency, all the other binary relations of interest (\emph{worse than}, \emph{better than}, \emph{equivalent to}, \emph{incomparable with}, \etc) are derived from $\performanceOrdering$. When the two following axioms are satisfied, we say that the performance ordering induced by $\aScore$ satisfy the axioms of performance-based ranking, or more simply that $\aScore$ satisfy the axioms.

The second axiom specifies a condition that must be satisfied for the preorder $\performanceOrdering$ on performances to be compatible with the task. In the case of the two-class crisp classification task studied in this paper, the satisfaction $\randVarSatisfaction$ is binary: we are not at all satisfied with a false positive or a false negative, and we are fully satisfied with a true negative and a true positive. Denoting the accuracy by $\scoreAccuracy$, the second axiom, given in a generic form in~\cite{Pierard2025Foundations}, is in this case equivalent to three conditions: (1)~there is no performance worse than a performance $\aPerformance$ such that $\scoreAccuracy(\aPerformance)=0$; (2) if a performance $\aPerformance_1$ is such that $\scoreAccuracy(\aPerformance_1)=0$ and a performance $\aPerformance_2$ is such that $\scoreAccuracy(\aPerformance_2)=1$, then $\aPerformance_1$ can never be better than $\aPerformance_2$; (3) there exists no performance better than a performance $\aPerformance$ such that $\scoreAccuracy(\aPerformance)=1$. The second axiom thus focuses on the extreme cases of the worst and best performances, but says nothing about intermediate performances. It should be noted that it is very permissive. A score may put performances for which $\scoreAccuracy(\aPerformance)>0$ on the same footing the worst performances, or performances for which $\scoreAccuracy(\aPerformance)<0$ on the same footing the best performances. This is for example the case of $\scoreFOne$: $\scoreAccuracy=0 \Rightarrow \scoreFOne=0$ but $\scoreFOne=0 \not\Rightarrow \scoreAccuracy=0$.

The third axiom specifies a condition that must be satisfied for the preorder $\performanceOrdering$ on performances to be compatible with evaluation (\ie, the mapping from methods to performances). It is this third axiom that constrains the ordering of performances that are neither among the best nor among the worst ones. Consider a \emph{blind} and non-deterministic combination of some base methods to form a hybrid method: each time the hybrid method is used, it begins by randomly selecting one of the base methods, \emph{independently of its input}, and then executes that method. As the combination is \emph{blind}, it would not make sense for the hybrid method to be better or worse than the best or the worst of the combined methods, respectively. %
Note that this axiom is not satisfied by $\scoreSkewInsensitiveVersionFOne$ in general, but is satisfied by it when the priors are fixed.

\subsubsection{All $\scoreFBeta$ Lead to Meaningful Performance Orderings}

It has been shown in \cite{Pierard2025Foundations} that all performance orderings derived from some score of the form
\begin{equation}
    \rankingScore(\aPerformance)
    = \frac {
        \randVarImportance(\sampleTN) \aPerformance(\{\sampleTN\})
        + \randVarImportance(\sampleTP) \aPerformance(\{\sampleTP\})
    }
    {
        \randVarImportance(\sampleTN) \aPerformance(\{\sampleTN\})
        + \randVarImportance(\sampleFP) \aPerformance(\{\sampleFP\})
        + \randVarImportance(\sampleFN) \aPerformance(\{\sampleFN\})
        + \randVarImportance(\sampleTP) \aPerformance(\{\sampleTP\})
    }
    \comma
    \label{eq:ranking-score}
\end{equation}
where $\randVarImportance$ denotes a positive random variable called \emph{Importance}, satisfy all the axioms of the theory of performance-based ranking. These scores are known as \emph{ranking scores}.

In order to prove that all $\scoreFBeta$ lead to meaningful performance orderings, we show here that the $\scoreFBeta$ are particular cases of $\rankingScore$. The proofs for $\beta=\nicefrac12$, $\beta=1$, and $\beta=2$ were already given in \cite{Pierard2025Foundations}, so here we give a generalization. Our proof is based on the well known fact that $\scoreFBeta$ has two nearly equivalent definitions. The first one, that we have used in the introduction (\cref{eq:f-beta-as-harmonic-mean}), is as follows:
\begin{equation}
    \scoreFBeta
    :
    \left\{ \aPerformance \in \allPerformances : \aPerformance(\{\sampleTP\})\ne 0 \right\} \rightarrow [0,1]
    :
    \aPerformance \mapsto \left( \frac{1}{1+\beta^2} \scorePrecision^{-1}(\aPerformance)
    + \frac{\beta^2}{1+\beta^2} \scoreRecall^{-1}(\aPerformance) \right)^{-1}
    \point
    \label{eq:f-beta-definition-1}
\end{equation}
The second one is
\begin{equation}
    \scoreFBeta
    :
    \left\{ \aPerformance \in \allPerformances : \aPerformance(\{\sampleTN\})\ne 1 \right\} \rightarrow [0,1]
    :
    \aPerformance \mapsto \frac {
        (1+\beta^2) \, \aPerformance(\{\sampleTP\})
    }
    {
        1 \, \aPerformance(\{\sampleFP\})
        + \beta^2 \, \aPerformance(\{\sampleFN\})
        + (1+\beta^2) \, \aPerformance(\{\sampleTP\})
    }
    \point
    \label{eq:f-beta-definition-2}
\end{equation}
It turns out that the first definition is a restriction of the second one: $\aPerformance(\{\sampleTP\})\ne 0 \Rightarrow \aPerformance(\{\sampleTN\})\ne 1$ and both $\scoreFBeta$ are equal on $\left\{ \aPerformance \in \allPerformances : \aPerformance(\{\sampleTP\})\ne 0 \right\}$. It is now possible to compare \cref{eq:f-beta-definition-2} with \cref{eq:ranking-score} and to see that, for any given $\beta$, $\scoreFBeta$ is a particular case of ranking score with
\begin{equation}
    \left(
        \randVarImportance(\sampleTN),
        \randVarImportance(\sampleFP),
        \randVarImportance(\sampleFN),
        \randVarImportance(\sampleTP)
    \right)
    \propto
    \left(
        0,
        1,
        \beta^2,
        1+\beta^2
    \right)
    \comma
    \label{eq:importance-fbeta}
\end{equation}
or equivalently
\begin{equation}
    \left(
        \randVarImportance(\sampleTN),
        \randVarImportance(\sampleFP),
        \randVarImportance(\sampleFN),
        \randVarImportance(\sampleTP)
    \right)
    \propto
    \left(
        0,
        1-b,
        b,
        1
    \right)
    \comma
\end{equation}
with $b=\nicefrac{\beta^2}{1+\beta^2}$.

\subsubsection{$\scoreSkewInsensitiveVersionFOne$ Leads to a Meaningful Performance Ordering When the Class Priors Are Fixed}

The skew-insensitive version of $\scoreFOne$ as been defined in \cite{Flach2003TheGeometry} as
\begin{equation}
    \scoreSkewInsensitiveVersionFOne = \frac{
        2 \scoreTPR
    }{
        \scoreTPR + \scoreFPR + 1
    }
    \point
\end{equation}
When the class priors are fixed and such that $\priorneg \ne 0$ and $\priorpos \ne 0$,
\begin{multline}
    \scoreSkewInsensitiveVersionFOne(\aPerformance) = \frac{
        2 \frac{\aPerformance(\{\sampleTP\})}{\priorpos}
    }{
        \frac{\aPerformance(\{\sampleTP\})}{\priorpos} + \frac{\aPerformance(\{\sampleFP\})}{\priorneg} + 1
    }
    = \frac{
        2 \, \priorneg \, \aPerformance(\{\sampleTP\})
    }{
        \priorneg \, \aPerformance(\{\sampleTP\}) + \priorpos \, \aPerformance(\{\sampleFP\}) + \priorneg \, \priorpos
    } \\
    = \frac{
        2 \, \priorneg \, \aPerformance(\{\sampleTP\})
    }{
        \priorneg \, \aPerformance(\{\sampleTP\}) + \priorpos \, \aPerformance(\{\sampleFP\}) + \priorneg \, ( \aPerformance(\{\sampleFN\}) + \aPerformance(\{\sampleTP\}) )
    }
    = \frac{
        2 \, \priorneg \, \aPerformance(\{\sampleTP\})
    }{
        \priorpos \, \aPerformance(\{\sampleFP\})
        + \priorneg \, \aPerformance(\{\sampleFN\})
        + 2 \, \priorneg \, \aPerformance(\{\sampleTP\})
    }
    \point
    \label{eq:sivf-as-ranking-score}
\end{multline}
The comparison of \cref{eq:sivf-as-ranking-score} with \cref{eq:ranking-score} shows that $\scoreSkewInsensitiveVersionFOne$ is a particular case of ranking score with
\begin{equation}
    \left(
        \randVarImportance(\sampleTN),
        \randVarImportance(\sampleFP),
        \randVarImportance(\sampleFN),
        \randVarImportance(\sampleTP)
    \right)
    \propto
    \left(
        0,
        \priorpos,
        \priorneg,
        2 \, \priorneg
    \right)
    \point
    \label{eq:importance-sivf}
\end{equation}

\subsubsection{$\scoreSkewInsensitiveVersionFOne$ Leads to the Same Ranking as $\scoreFBeta$ with $\beta^2=\frac{\priorneg}{\priorpos}$}

\paragraph{A first demonstration, based on the relationship between $\scoreSkewInsensitiveVersionFOne$ and $\scoreFBeta$.}

From \cite{Flach2003TheGeometry}, it is already known that $\scoreSkewInsensitiveVersionFOne=\scoreFOne$ when $\priorneg=\priorpos=\nicefrac12$. 
Here, we give a generalization for $\priorneg \ne 0$ and $\priorpos \ne 0$. 
We noticed that
\begin{equation}
    \beta^2=\frac{\priorneg}{\priorpos}
    \Rightarrow
    \scoreFBeta = \frac{\scoreSkewInsensitiveVersionFOne}{(\priorpos-\priorneg) \, \scoreSkewInsensitiveVersionFOne + 2 \priorneg}
    \point
\end{equation}
When $\priorneg \ne 0$, we have
\begin{equation}
    \beta^2=\frac{\priorneg}{\priorpos}
    \Rightarrow
    \frac{\partial \scoreFBeta}{\partial \scoreSkewInsensitiveVersionFOne} > 0
    \point
\end{equation}
Therefore, when the priors are fixed, $\scoreSkewInsensitiveVersionFOne$ leads to the same performance ordering as $\scoreFBeta$ with $\beta^2=\frac{\priorneg}{\priorpos}$. %

\paragraph{A second demonstration, based on the importance values.}

Property 4 of \cite{Pierard2025Foundations}, when particularized for two-class classification, states that the performance ordering induced by a ranking score is insensitive to the uniform scaling of the importance given to the unsatisfying samples ($\sampleFP$ and $\sampleFN$) or to the satisfying ($\sampleTN$ and $\sampleTP$) samples. The comparison between \cref{eq:importance-fbeta} and \cref{eq:importance-sivf} shows that:
\begin{itemize}
    \item for the unsatisfying samples, we have $\left(
        \bullet,
        1,
        \beta^2,
        \bullet
    \right)
    \propto
    \left(
        \bullet,
        \priorpos,
        \priorneg,
        \bullet
    \right)$ when $\beta^2=\frac{\priorneg}{\priorpos}$;
    \item for the satisfying samples, we have always $\left(
        0,
        \bullet,
        \bullet,
        1+\beta^2
    \right)
    \propto
    \left(
        0,
        \bullet,
        \bullet,
        2 \, \priorneg
    \right)$.
\end{itemize}
Therefore, when the priors are fixed, $\scoreSkewInsensitiveVersionFOne$ leads to the same performance ordering as $\scoreFBeta$ with $\beta^2=\frac{\priorneg}{\priorpos}$.

%% file: sections/A_2_0_suppl_for_theory.tex
\subsection{Supplementary Material About \cref{sec:theory} (\,\,\,\nameref{sec:theory})}

\subsubsection{On the Minimization of Fréchet Variance (\cref{eq:frechet-variance})}

In \cref{sec:OptimalTradeoffs}, we say that the optimal tradeoffs are the $\scoreFBeta$ scores that minimize the Fréchet variance~\cite{Frechet1948LesElementsAleatoires}:
\begin{equation*}
    \sigma^2(\beta) = \distKendall^2(\scorePrecision;\scoreFBeta) + \distKendall^2(\scoreFBeta ; \scoreRecall)
    \point
\end{equation*}
We also say that the solutions, known as the Karcher means~\cite{Karcher1977Riemannian} are those that are equidistant of $\scorePrecision$ and $\scoreRecall$, \ie such that
\begin{align*}
 & \distKendall(\scorePrecision;\scoreOptimalTradeoff) = \distKendall(\scoreOptimalTradeoff;\scoreRecall) = \frac{\distKendall(\scorePrecision;\scoreRecall)}{2}\\
\Leftrightarrow \; \; & \kendall(\scorePrecision;\scoreOptimalTradeoff) = \kendall(\scoreOptimalTradeoff;\scoreRecall) = \frac{1+\kendall(\scorePrecision;\scoreRecall)}{2}
\point
\end{align*}

Hereafter, we discuss here the case of a continuum of rankings, as shown in \cref{fig:optimal_tradeoff_continuous}. The case of a finite amount of rankings is depicted in \cref{fig:optimal_tradeoff_odd} and \cref{fig:optimal_tradeoff_even} for, respectively, an odd and an even amount.

By restricting to the family of $\scoreFBeta$ scores, the solution is actually unique and is thus a Fréchet mean. Indeed, since~\cref{eq:shortest-path-distances} states that 
\begin{equation*}
    \distKendall(\scorePrecision;\scoreRecall) = \distKendall(\scorePrecision;\scoreFBeta) + \distKendall(\scoreFBeta;\scoreRecall) \qquad \forall \beta\ge0
\end{equation*}
we have
\begin{align}
    \sigma^2(\beta) & = \distKendall^2(\scorePrecision;\scoreFBeta) + (\distKendall(\scorePrecision;\scoreRecall)-\distKendall(\scorePrecision;\scoreFBeta))^2 \\
    & = 2 \distKendall^2(\scorePrecision;\scoreFBeta) - 2 \distKendall(\scorePrecision;\scoreRecall) \distKendall(\scorePrecision;\scoreFBeta) + \distKendall^2(\scorePrecision;\scoreRecall)
\end{align}
which is quadratic in $\distKendall(\scorePrecision;\scoreFBeta)$ and thus uniquely minimized for $\beta$ such that
\begin{equation}
\distKendall(\scorePrecision;\scoreFBeta) = \frac{\distKendall(\scorePrecision;\scoreRecall)}{2}
\point
\end{equation}
This incidentally yields the same equality for $\distKendall(\scoreFBeta;\scoreRecall)$. The result in terms of correlation stems from the relation $\kendall=1-2\distKendall$.

\input{figs/optimal_tradeoff}

\subsubsection{On the Close-Form Expression for the Optimal $\beta$ (\cref{eq:beta-sq-swap,eq:closed-form-expression})}
\label{sec:suppl-optimal-beta}

\paragraph{Complexity/scalability.}
After evaluating $n$ classifiers, a straightforward computation of the optimal $\beta$ with our formulas requires indeed $O(n^2)$ uses of \cref{eq:beta-sq-swap} (same as for Kendall's $\tau$). Even for $n=1000$ (hardly met in practice), this takes less than one second on a modern laptop.

\paragraph{An alternative to \cref{eq:beta-sq-swap}.}
We thank Peter Flach for pointing out that \cref{eq:beta-sq-swap} can be rewritten in terms of only precision and recall values as follows:
\begin{equation}
    \vartheta(\aPerformance_1,\aPerformance_2) = - \frac{
        \scorePrecision^{-1}(\aPerformance_1) - \scorePrecision^{-1}(\aPerformance_2)
    }{
        \scoreRecall^{-1}(\aPerformance_1) - \scoreRecall^{-1}(\aPerformance_2)
    }
    \point
\end{equation}
This formula is Eq. (6) in~\cite{Flach2015Precision}.

\paragraph{Downstream objective.}
We can generalize \cref{eq:frechet-variance} by weighting $\distKendall^2$ and \cref{eq:closed-form-expression} by replacing the median by a quantile. To ensure that the $0\%$ quantile corresponds to $\scorePrecision$ and that the $100\%$ quantile corresponds to $\scoreRecall$, the quantile is taken over $\{0,\infty\} \cup \{ \vartheta(\aPerformance_i,\aPerformance_j) \, \vert \, i \ne j \wedge \vartheta(\aPerformance_i,\aPerformance_j) \ge 0 \}$. For instance, to weight four times more $\scoreRecall$ than $\scorePrecision$, from the \emph{point of view of values} one uses $b=0.8$ in \cref{eq:f-beta-as-harmonic-mean} (\ie, $\beta=2$,  as in \linkCADARRE). But from the \emph{point of view of ranks}, we take the $80^{th}$ percentile of $\vartheta$ values. This leads to $\beta=0.914$ in the example of \linkCADARRE (\cref{sec:CADA-RRE}). The relationship between the downstream objective (\ie, the chosen quantile) and the optimal value for $\beta$ is shown in \cref{fig:cada-rre-quantiles}.

\input{figs/cada_rre_quantiles}

\subsubsection{On the Formulas for Degree of Optimality of Some $\scoreFBeta$}

We provide here a few explanations about how \cref{eq:no-choice,eq:choice-not-optimal,eq:choice-optimal} cen be derived. For that, let us remind three interpretations for Kendall's distance $\distKendall(\aScore_1;\aScore_2)\in[0,1]$.

\begin{itemize}
    \item First, \cref{eq:kendall-distance} shows that $\distKendall(\aScore_1;\aScore_2)$ is equal to the proportion of pairs of elements for which $\aScore_1$ and $\aScore_2$ do not agree on the relative order. This interpretation is the cornerstone for probabilistic reasoning.
    \item Second, $\distKendall(\aScore_1;\aScore_2)$ is equal to the minimum number of swaps of consecutive elements that are needed to transform the ranking induced by $\aScore_1$ into the ranking induced by $\aScore_2$ and vice versa. This interpretation is the cornerstone for geometric reasoning, as it makes the connection with the distance along the manifold (or path graph) of rankings.
    \item Third, $\distKendall(\aScore_1;\aScore_2)$ is linearly related to the rank correlation $\kendall(\aScore_1;\aScore_2)$ as $\distKendall = \frac{1-\kendall}{2}$.
\end{itemize}

Keeping these three interpretations in mind is helpful to derive \cref{eq:no-choice,eq:choice-not-optimal,eq:choice-optimal}.

Let us consider \cref{eq:choice-not-optimal}, for example. We are interested in the probability $P(\wrongOrdering)$. The first interpretation allows us to express it as $\distKendall(\scoreFBeta;\scoreOptimalTradeoff)$. Then, it is possible to perform some 
geometric reasoning based on the second interpretation. As shown in \cref{fig:optimality_formulas_explanation}, $\distKendall(\scoreFBeta;\scoreOptimalTradeoff)=\nicefrac12 \left|\distKendall(\scoreFBeta;\scoreRecall)-\distKendall(\scorePrecision;\scoreFBeta)\right|$. Finally, using the third interpretation, we find that $\nicefrac12 \left|\distKendall(\scoreFBeta;\scoreRecall)-\distKendall(\scorePrecision;\scoreFBeta)\right| = \nicefrac14 \left|\kendall(\scorePrecision;\scoreFBeta)-\kendall(\scoreFBeta;\scoreRecall)\right|$.

\input{figs/optimality-formulas-explanation}

\subsubsection{On the CADA-RRE Example Used to Illustrate the Theory}
\label{sec:CADA-RRE}

Among the various challenges hosted on \emph{Grand Challenge}, we found that the third task of the \emph{Cerebral Aneurysm Detection and Analysis} (CADA) challenge \cite{Ivantsits2021Cerebral} is particularly interesting. This task is known as CADA-RRE for \emph{Cerebral Aneurysm Rupture Risk Estimation} and is hosted at \url{https://cada-rre.grand-challenge.org/}. It consists in the automatic classification of known ruptured and unruptured aneurysms based on rotational X-ray angiographic images.

Even if the ranking chosen by the organizers is only based on $\scoreFBeta[2]$, the logs contain for each entry the values of five ranking scores~\cite{Pierard2025Foundations}: the precision $\scorePrecision$, recall $\scoreRecall$, accuracy $\scoreAccuracy$, $\scoreFBeta[1]$, and indeed $\scoreFBeta[2]$. This information allows one to recover the complete (normalized) confusion matrix with the probabilities of a true negative ($\scorePTN$), false positive ($\scorePFP$), false negative ($\scorePFN$), and true positive ($\scorePTP$). From these confusion matrices, we observed that the class priors are fixed and given by $\priorneg=\nicefrac{19}{30}$ and $\priorpos=\nicefrac{11}{30}$. The values of $\scorePFP$, $\scorePFN$, and $\scorePTP$ can be used in \cref{eq:beta-sq-swap}, which gives the values used in \cref{eq:closed-form-expression} to compute an optimal $\beta$. The values of $\scorePFP$ and $\scorePFN$ can be used to derive $\expectedValueSymbol[\scorePFP]$ and $\expectedValueSymbol[\scorePFN]$, which in turn can be used to recommend a value for $\beta$ with our heuristic \cref{eq:heuristic}.

At the time of writing these lines, the \href{https://cada-rre.grand-challenge.org/evaluation/challenge/leaderboard/}{challenge leaderboard} contains $29$ entries. However, there are only $16$ unique pairs of precision and recall values. This can be explained by the low number of cases on which the methods are tested. In our example based on CADA-RRE, we decided to rank the classifiers corresponding to these $16$ unique pairs. This stands in contrast with the official ranking of CADA-RRE as, for each participant, the organizers only considered the first three submissions with $\scoreFBeta[2] \ge 0.1$. On the one hand, according to \cite{Ivantsits2021Cerebral}, the official ranking only contains three methods as only three teams submitted their solution: the \first place is for \cite{Ivantsits2021Intracranial} with $\scoreFBeta[2]=0.702$, the \second place is for \cite{Liu2021Cerebral} with $\scoreFBeta[2]=0.678$, and the \third place is for an unpublished method with $\scoreFBeta[2]=0.377$. On the other hand, the $\scoreFBeta[2]$ score is between $0.0$ and $0.862$ in the $16$ entries that we rank. The results that we have obtained with the $16$ performances to rank are provided in \cref{fig:cada-rre-main,fig:cada-rre-quantiles,fig:cada-rre-heuristic}.

What is particularly interesting with this example is that the organizers chose to rank the participants according to the $\scoreFBeta[2]$ score, as ``the identification of aneurysms at risk is considered more important than the avoidance of false-positive risk classification''. As shown in the plots of \cref{fig:cada-rre-main}, the ranking induced by $\scoreFBeta[2]$ perfectly mimics the ranking induced by $\scoreRecall$, totally ignoring the ranking induced by $\scorePrecision$. This is certainly not what is expected when one chooses to rank according to some $\scoreFBeta$. As shown in \cref{fig:cada-rre-quantiles} and already discussed in \cref{sec:suppl-optimal-beta}, $\beta=2$ is the value to use if one wants to weight four times more $\scoreRecall$ than $\scorePrecision$ from the \emph{point of view of values}, but $\beta=0.914$ is the value to use to weight four times more $\scoreRecall$ than $\scorePrecision$ from the \emph{point of view of ranks}.

%% file: figs/optimal_tradeoff.tex
\begin{figure*}[p]
    \begin{centering}
        \subfloat[
            \textbf{Case in which there is a bijection between the values of $\beta$ and the rankings induced by $\scoreFBeta$}. 
            The rankings form a continuum (manifold) and the minimization of Fréchet variance leads to the midpoint where $\distKendall(\scorePrecision;\scoreOptimalTradeoff) = \distKendall(\scoreOptimalTradeoff;\scoreRecall)$.
            \label{fig:optimal_tradeoff_continuous}
        ]{
            \includegraphics[width=0.6\linewidth]{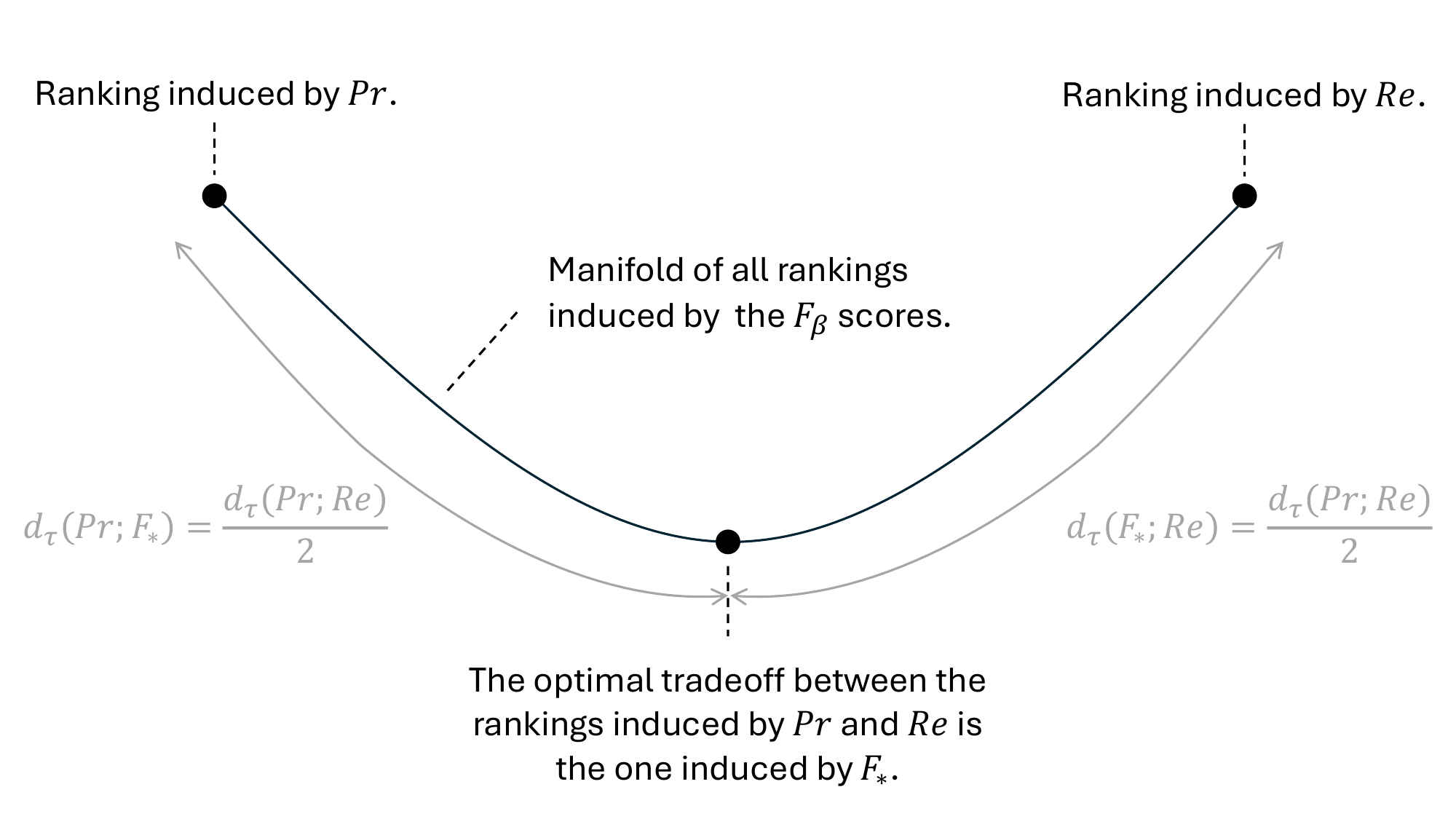}
        }
        
        \subfloat[
            \textbf{Case in which the number of rankings induced by the $\scoreFBeta$ scores is odd.} 
            The rankings form a path graph. The nodes correspond to the rankings, each of them being induced by a range of values for $\beta$. The edges correspond to the swaps that occur between the consecutive rankings, at some given values $\vartheta$ for the $\beta$ parameter. Assuming there is no group of at least three co-aligned performances, the median of all these $\vartheta$ values belongs to the range of $\beta$s for which the ranking is at the midpoint where $\distKendall(\scorePrecision;\scoreOptimalTradeoff) = \distKendall(\scoreOptimalTradeoff;\scoreRecall)$.
            \label{fig:optimal_tradeoff_odd}
        ]{
            \includegraphics[width=0.6\linewidth]{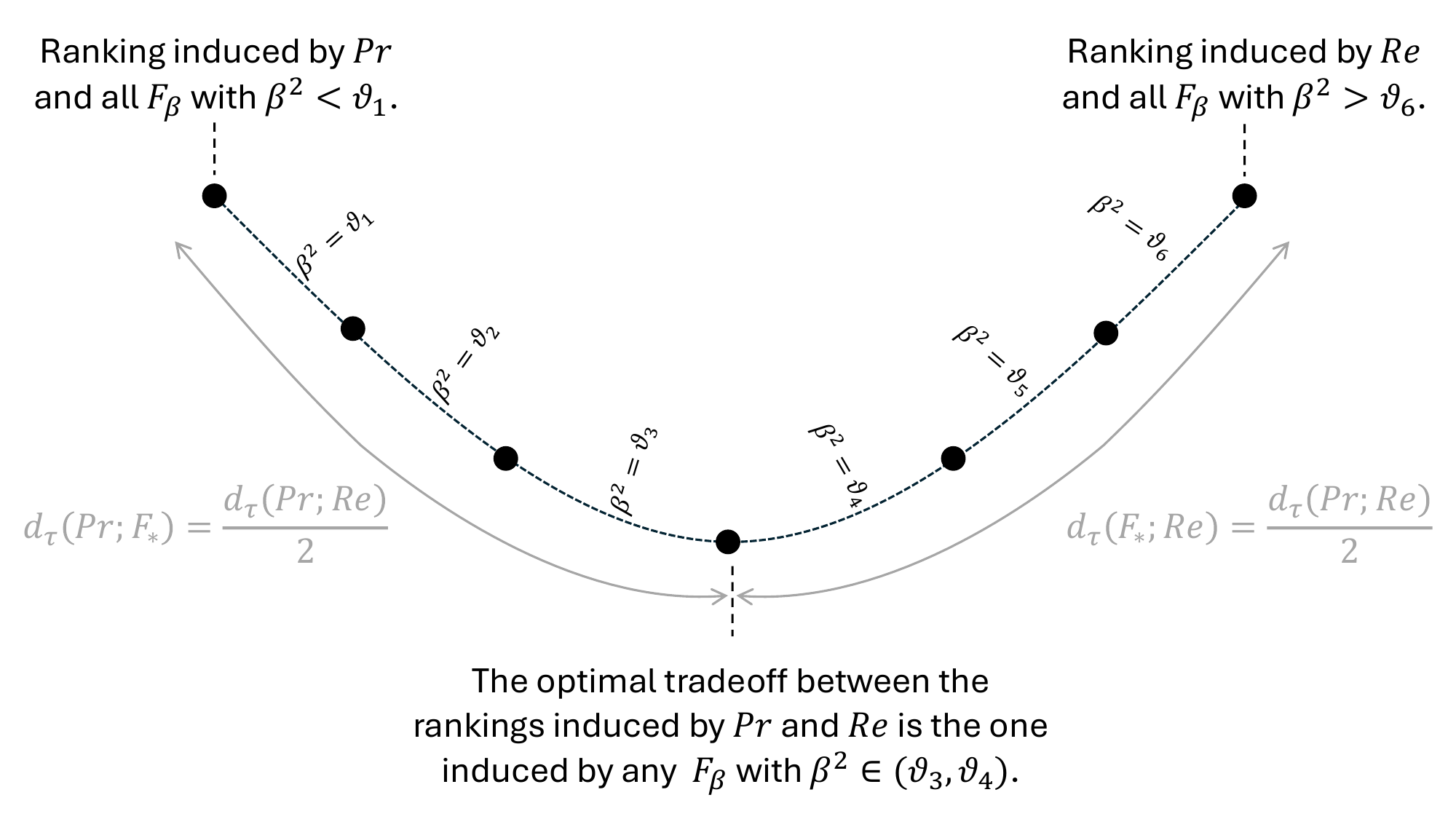}
        }
        
        \subfloat[
            \textbf{Case in which the number of rankings induced by the $\scoreFBeta$ scores is even.} 
            Assuming there is no group of at least three co-aligned performances, the median of all the $\vartheta$ values corresponds to the swap between the two rankings minimizing Fréchet variance.
            \label{fig:optimal_tradeoff_even}
        ]{
            \includegraphics[width=0.6\linewidth]{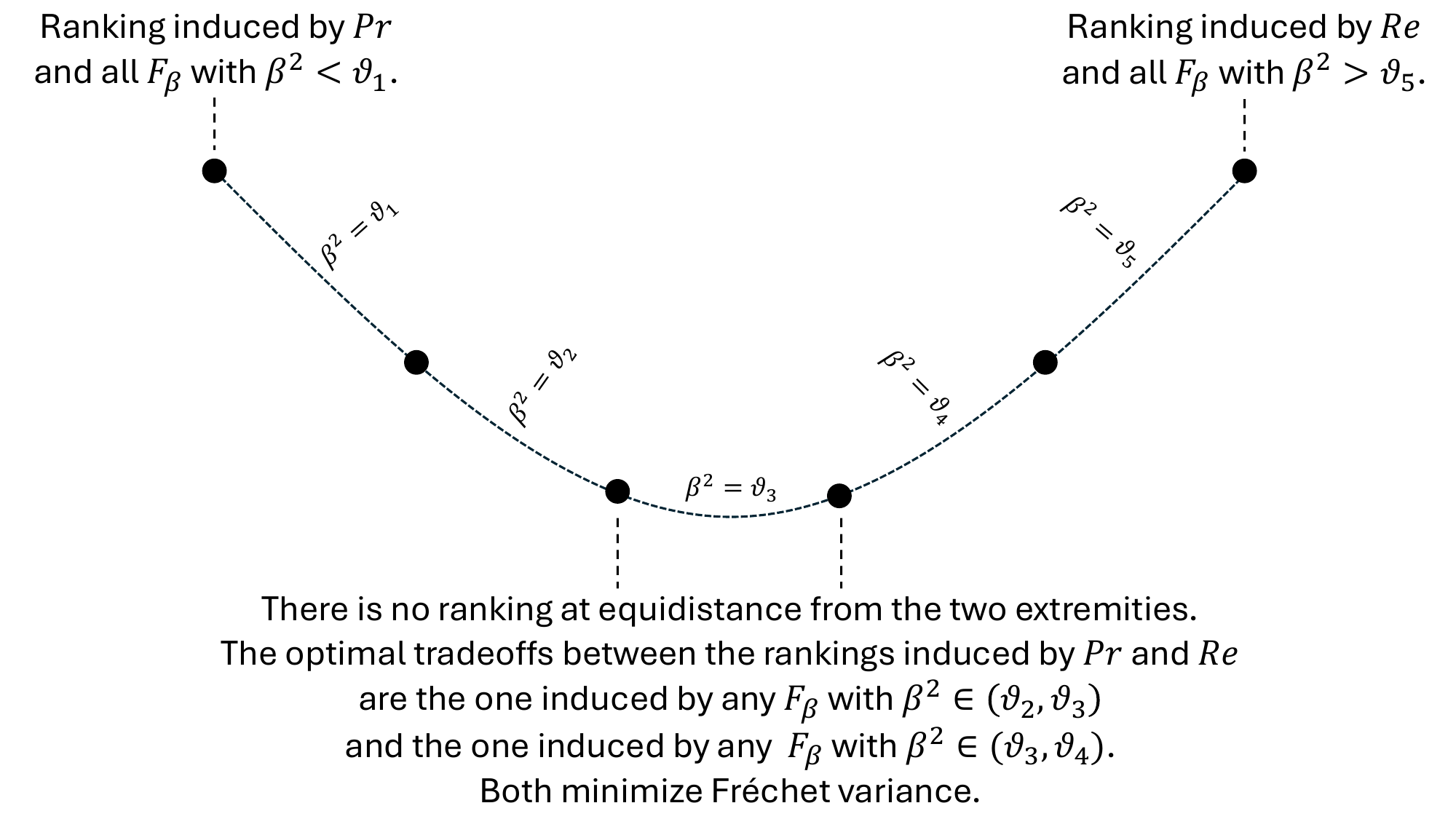}
        }
    \par\end{centering}
        
    \caption{
        The optimal tradeoff as the solution of the minimization of Fréchet variance $\sigma^2(\beta)$.
        \label{optimal_tradeoff}
    }
\end{figure*}

%% file: figs/cada_rre_quantiles.tex
\begin{figure}
    \centering
    \includegraphics[width=0.8\linewidth]{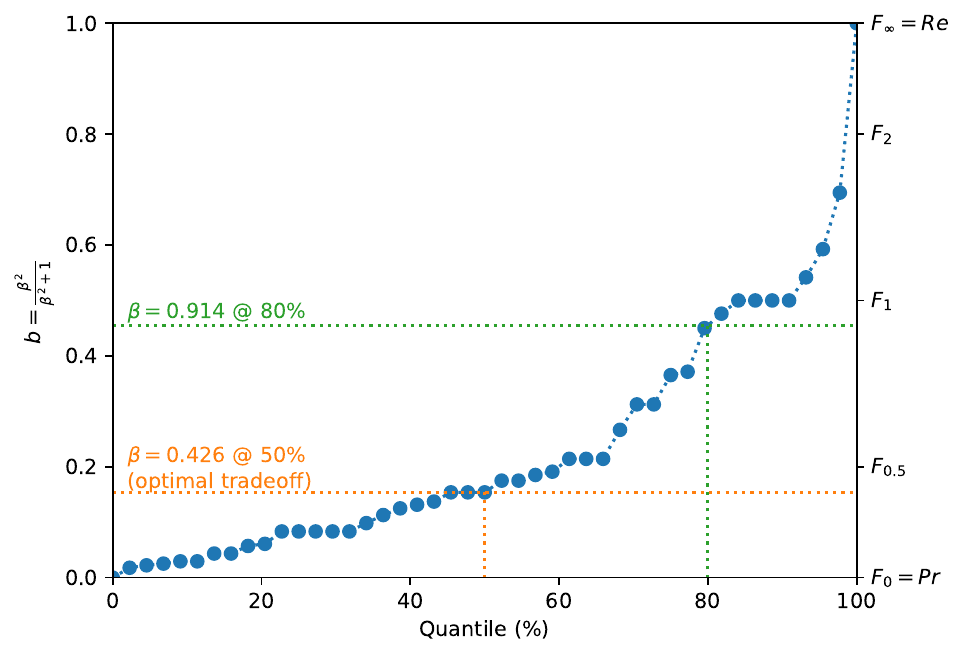}
    \caption{
        The relationship between the downstream objective (\ie, the chosen quantile) and the optimal value for $\beta$, for the \linkCADARRE example described in \cref{sec:CADA-RRE}. The points correspond to the various values of $\beta$ for which a swap between two consecutive classifiers occurs in the ranking. Taking the median (\ie, the $50\%$ quantile) leads to the optimal, balanced, tradeoff studied in this paper.
        \label{fig:cada-rre-quantiles}
    }
\end{figure}

%% file: figs/optimality-formulas-explanation.tex
\begin{figure*}%
    \begin{centering}
        \subfloat[
            Case in which the ranking induced by $\scoreFBeta$ is closer to the ranking induced by $\scoreRecall$ than to the ranking induced by $\scorePrecision$. 
            We see that $\distKendall(\scoreFBeta;\scoreOptimalTradeoff)=\frac{\distKendall(\scorePrecision;\scoreFBeta)-\distKendall(\scoreFBeta;\scoreRecall)}{2}$.
        ]{
            \includegraphics[width=0.6\linewidth]{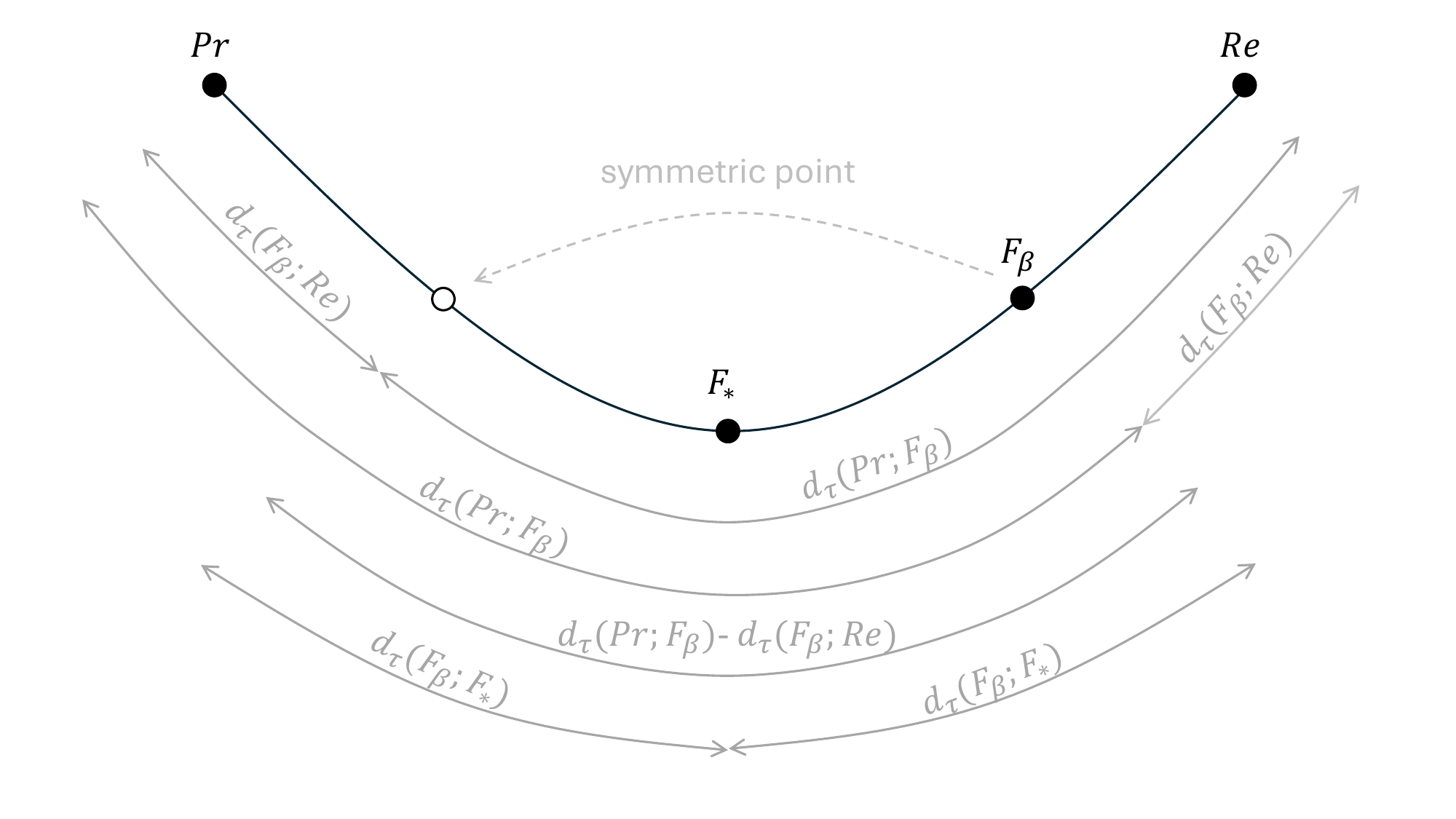}
        }

        \subfloat[
            Case in which the ranking induced by $\scoreFBeta$ is closer to the ranking induced by $\scorePrecision$ than to the ranking induced by $\scoreRecall$. 
            We see that $\distKendall(\scoreFBeta;\scoreOptimalTradeoff)=\frac{\distKendall(\scoreFBeta;\scoreRecall)-\distKendall(\scorePrecision;\scoreFBeta)}{2}$.
        ]{
            \includegraphics[width=0.6\linewidth]{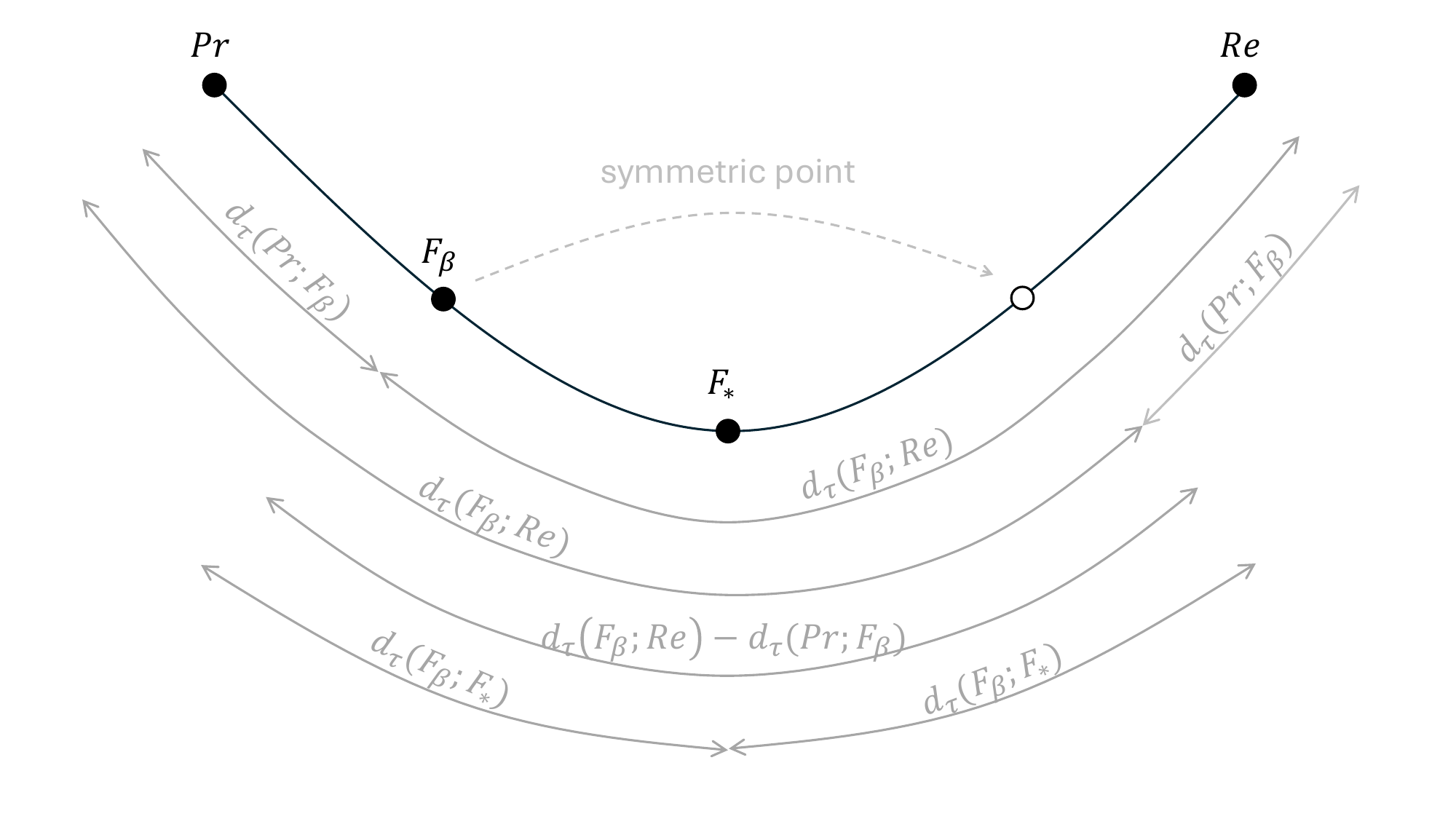}
        }
    \par\end{centering}
        
    \caption{
        Kendall distances on the manifold of rankings induced by the $\scoreFBeta$ scores. We see that $\distKendall(\scoreFBeta;\scoreOptimalTradeoff)=\frac{\left|\distKendall(\scoreFBeta;\scoreRecall)-\distKendall(\scorePrecision;\scoreFBeta)\right|}{2}$.
        \label{fig:optimality_formulas_explanation}
    }
\end{figure*}

%% file: sections/A_3_0_suppl_for_case_studies.tex
\subsection{Supplementary Material About \cref{sec:case-studies} (\,\,\,\nameref{sec:case-studies})}

\input{sections/A_3_P_principle_analytic_solution}
\clearpage
\input{sections/A_3_CS1_detailed_distribution_I}
\clearpage
\input{sections/A_3_CS2_detailed_distribution_II}
\clearpage
\input{sections/A_3_CS3_detailed_distribution_III}

\clearpage
\input{sections/A_3_CS4_detailed_distribution_IV}

\clearpage
\input{sections/A_3_CS5_detailed_distribution_V}

\clearpage
\input{sections/A_3_CS6_all_results_CDnet}
\clearpage
\input{sections/A_3_H_our_heuristic}

%% file: sections/A_3_P_principle_analytic_solution.tex
\subsubsection{Principle of Analytical Computations for Kendall's rank correlations $\kendall$ with Distributions of Performances}

Let us consider two scores $\aScore_1$, $\aScore_2$. We denote the probability for them to order differently two (different) performances $\aPerformance_A$ and $\aPerformance_B$ drawn at random, independently, by $\distKendall(\aScore_1$; $\aScore_2)$. Kendall's rank correlation is given by $\kendall(\aScore_1$; $\aScore_2)=1-2 \distKendall(\aScore_1$; $\aScore_2)$. When we consider a distribution of performances $\aDistributionOfPerformances$, instead of a finite set $\aSetOfPerformances$, we cannot compute $\distKendall(\aScore_1; \aScore_2)$ with \cref{eq:kendall-distance}. Instead, we can compute analytically the following probabilities:
\begin{equation}
    \mathrm{Proba}\left[
        \aScore_1(\aPerformance_A)<\aScore_1(\aPerformance_B),
        \aScore_2(\aPerformance_A)<\aScore_2(\aPerformance_B)
    \right] = 
    \int_{\aPerformance_A\in\allPerformances}
    \int_{\aPerformance_B\in\allPerformances}
    \indicatorSymbol_{
        \aScore_1(\aPerformance_A)<\aScore_1(\aPerformance_B),
        \aScore_2(\aPerformance_A)<\aScore_2(\aPerformance_B)
    }
    f_\aDistributionOfPerformances(\aPerformance_A)
    f_\aDistributionOfPerformances(\aPerformance_B)
    d\aPerformance_A
    d\aPerformance_B
    \comma
\end{equation}
\begin{equation}
    \mathrm{Proba}\left[
        \aScore_1(\aPerformance_A)<\aScore_1(\aPerformance_B),
        \aScore_2(\aPerformance_A)>\aScore_2(\aPerformance_B)
    \right] = 
    \int_{\aPerformance_A\in\allPerformances}
    \int_{\aPerformance_B\in\allPerformances}
    \indicatorSymbol_{
        \aScore_1(\aPerformance_A)<\aScore_1(\aPerformance_B),
        \aScore_2(\aPerformance_A)>\aScore_2(\aPerformance_B)
    }
    f_\aDistributionOfPerformances(\aPerformance_A)
    f_\aDistributionOfPerformances(\aPerformance_B)
    d\aPerformance_A
    d\aPerformance_B
    \comma
\end{equation}
\begin{equation}
    \mathrm{Proba}\left[
        \aScore_1(\aPerformance_A)>\aScore_1(\aPerformance_B),
        \aScore_2(\aPerformance_A)<\aScore_2(\aPerformance_B)
    \right] = 
    \int_{\aPerformance_A\in\allPerformances}
    \int_{\aPerformance_B\in\allPerformances}
    \indicatorSymbol_{
        \aScore_1(\aPerformance_A)>\aScore_1(\aPerformance_B),
        \aScore_2(\aPerformance_A)<\aScore_2(\aPerformance_B)
    }
    f_\aDistributionOfPerformances(\aPerformance_A)
    f_\aDistributionOfPerformances(\aPerformance_B)
    d\aPerformance_A
    d\aPerformance_B
    \comma
\end{equation}
\begin{equation}
    \mathrm{Proba}\left[
        \aScore_1(\aPerformance_A)>\aScore_1(\aPerformance_B),
        \aScore_2(\aPerformance_A)>\aScore_2(\aPerformance_B)
    \right] = 
    \int_{\aPerformance_A\in\allPerformances}
    \int_{\aPerformance_B\in\allPerformances}
    \indicatorSymbol_{
        \aScore_1(\aPerformance_A)>\aScore_1(\aPerformance_B),
        \aScore_2(\aPerformance_A)>\aScore_2(\aPerformance_B)
    }
    f_\aDistributionOfPerformances(\aPerformance_A)
    f_\aDistributionOfPerformances(\aPerformance_B)
    d\aPerformance_A
    d\aPerformance_B
    \comma
\end{equation}
where the symbol $\indicatorSymbol$ denotes the indicator and $f_\aDistributionOfPerformances$ the probability density function related to the distribution $\aDistributionOfPerformances$.

Kendall's rank correlation $\kendall$ is then given by
\begin{equation}
    \kendall(\aScore_1; \aScore_2) = 1 - 2 \left(
        \mathrm{Proba}\left[
            \aScore_1(\aPerformance_A)<\aScore_1(\aPerformance_B),
            \aScore_2(\aPerformance_A)>\aScore_2(\aPerformance_B)
        \right]
        +
        \mathrm{Proba}\left[
            \aScore_1(\aPerformance_A)>\aScore_1(\aPerformance_B),
            \aScore_2(\aPerformance_A)<\aScore_2(\aPerformance_B)
        \right]
    \right)
    \point
\end{equation}
Note that $\aPerformance_A$ and $\aPerformance_B$ are exchangeable in these equalities since they are drawn independently from the same distribution. Thus, by symmetry, we have
\begin{equation}
    \mathrm{Proba}\left[
        \aScore_1(\aPerformance_A)<\aScore_1(\aPerformance_B),
        \aScore_2(\aPerformance_A)<\aScore_2(\aPerformance_B)
    \right]
    =
    \mathrm{Proba}\left[
        \aScore_1(\aPerformance_A)>\aScore_1(\aPerformance_B),
        \aScore_2(\aPerformance_A)>\aScore_2(\aPerformance_B)
    \right]
    \comma
\end{equation}
and
\begin{equation}
    \mathrm{Proba}\left[
        \aScore_1(\aPerformance_A)<\aScore_1(\aPerformance_B),
        \aScore_2(\aPerformance_A)>\aScore_2(\aPerformance_B)
    \right]
    =
    \mathrm{Proba}\left[
        \aScore_1(\aPerformance_A)>\aScore_1(\aPerformance_B),
        \aScore_2(\aPerformance_A)<\aScore_2(\aPerformance_B)
    \right]
    \point
\end{equation}
Moreover, the four probabilities sum to one. Thus, computing one of these four probabilities suffices to be able to determine $\kendall$, and if we know $\kendall$, then we are able to recover the four probabilities:
\begin{equation}
    \kendall(\aScore_1; \aScore_2) = 4 \, \mathrm{Proba}\left[
        \aScore_1(\aPerformance_A)<\aScore_1(\aPerformance_B),
        \aScore_2(\aPerformance_A)<\aScore_2(\aPerformance_B)
    \right]
    - 1
    \comma
    \label{eq:tau-from-pLtLt}
\end{equation}
\begin{equation}
    \kendall(\aScore_1; \aScore_2) = 1 - 4 \, \mathrm{Proba}\left[
        \aScore_1(\aPerformance_A)<\aScore_1(\aPerformance_B),
        \aScore_2(\aPerformance_A)>\aScore_2(\aPerformance_B)
    \right]
    \comma
    \label{eq:tau-from-pLtGt}
\end{equation}
\begin{equation}
    \kendall(\aScore_1; \aScore_2) = 1 - 4 \, \mathrm{Proba}\left[
        \aScore_1(\aPerformance_A)>\aScore_1(\aPerformance_B),
        \aScore_2(\aPerformance_A)<\aScore_2(\aPerformance_B)
    \right]
    \comma
    \label{eq:tau-from-pGtLt}
\end{equation}
\begin{equation}
    \kendall(\aScore_1; \aScore_2) = 4 \, \mathrm{Proba}\left[
        \aScore_1(\aPerformance_A)>\aScore_1(\aPerformance_B),
        \aScore_2(\aPerformance_A)>\aScore_2(\aPerformance_B)
    \right]
    - 1
    \point
    \label{eq:tau-from-pGtGt}
\end{equation}

The source codes for the analytical results provided hereafter are for \emph{Wolfram 14.2} (\aka \emph{Mathematica}).

%% file: sections/A_3_CS1_detailed_distribution_I.tex
\subsubsection{Detailed Results for the Uniform Distribution Over All Performances}

To find the probabilities necessary to compute $\kendall$, we integrate over the tetrahedron. We start with a few initializations.

\begin{lstlisting}[backgroundcolor = \color{lightgray},language=Mathematica,numbers=left,numberstyle={\small},breaklines=true]
Tetra = Simplex[{{1, 0, 0, 0}, {0, 1, 0, 0}, {0, 0, 1, 0}, {0, 0, 0, 1}}]
Ppv[ptn_, pfp_, pfn_, ptp_] := ptp/(pfp + ptp)
Tpr[ptn_, pfp_, pfn_, ptp_] := ptp/(pfn + ptp)
Fone[ptn_, pfp_, pfn_, ptp_] := 2*ptp/(pfp + pfn + 2*ptp)

tot = Integrate[
    1,
    {ptn1, pfp1, pfn1, ptp1} \[Element] Tetra,
    {ptn2, pfp2, pfn2, ptp2} \[Element] Tetra
]
\end{lstlisting}

We have 
\begin{equation}
    \kendall(\scorePrecision; \scoreRecall) = \frac{1}{3}
    \comma
\end{equation}
as shown with the code:
\begin{lstlisting}[backgroundcolor = \color{lightgray},language=Mathematica,numbers=left,numberstyle={\small},breaklines=true]
pLtLt = Integrate[
    Boole[
        Ppv[ptn1, pfp1, pfn1, ptp1] < Ppv[ptn2, pfp2, pfn2, ptp2] &&
        Tpr[ptn1, pfp1, pfn1, ptp1] < Tpr[ptn2, pfp2, pfn2, ptp2]
    ],
    {ptn1, pfp1, pfn1, ptp1} \[Element] Tetra,
    {ptn2, pfp2, pfn2, ptp2} \[Element] Tetra
] / tot
tau = 4 * pLtLt - 1
\end{lstlisting}

We have
\begin{equation}
    \kendall(\scorePrecision; \scoreFOne) = \frac{2}{3}
    \comma
\end{equation}
as shown with the code:
\begin{lstlisting}[backgroundcolor = \color{lightgray},language=Mathematica,numbers=left,numberstyle={\small},breaklines=true]
pLtGt = Integrate[
    Boole[
        Ppv[ptn1, pfp1, pfn1, ptp1] < Ppv[ptn2, pfp2, pfn2, ptp2] && 
        Fone[ptn1, pfp1, pfn1, ptp1] > Fone[ptn2, pfp2, pfn2, ptp2]
    ],
    {ptn1, pfp1, pfn1, ptp1} \[Element] Tetra, 
    {ptn2, pfp2, pfn2, ptp2} \[Element] Tetra
] / tot
tau = 1 - 4 * pLtGt
\end{lstlisting}

And we have
\begin{equation}
    \kendall(\scoreFOne; \scoreRecall) = \frac{2}{3}
    \comma
\end{equation}
as shown with the code:
\begin{lstlisting}[backgroundcolor = \color{lightgray},language=Mathematica,numbers=left,numberstyle={\small},breaklines=true]
pLtGt = Integrate[
    Boole[
        Fone[ptn1, pfp1, pfn1, ptp1] < Fone[ptn2, pfp2, pfn2, ptp2] &&
        Tpr[ptn1, pfp1, pfn1, ptp1] > Tpr[ptn2, pfp2, pfn2, ptp2]
    ],
    {ptn1, pfp1, pfn1, ptp1} \[Element] Tetra,
    {ptn2, pfp2, pfn2, ptp2} \[Element] Tetra
] / tot
tau = 1 - 4 * pLtGt
\end{lstlisting}

%% file: sections/A_3_CS2_detailed_distribution_II.tex
\subsubsection{Detailed Results for the Uniform Distributions With Fixed Probability of True Negatives}

The demonstration could be provided similarly as we did for the previous case study, taking a simplex in two dimensions instead of a simplex in three dimensions. However, we would like to show a geometrical reasoning instead.

The equilateral triangles depicted in \cref{fig:setII} form a "triangular space" in which the performances are assumed to be uniformly distributed. This "space" is parameterized as follows.

\begin{center}
    \scalebox{0.4}{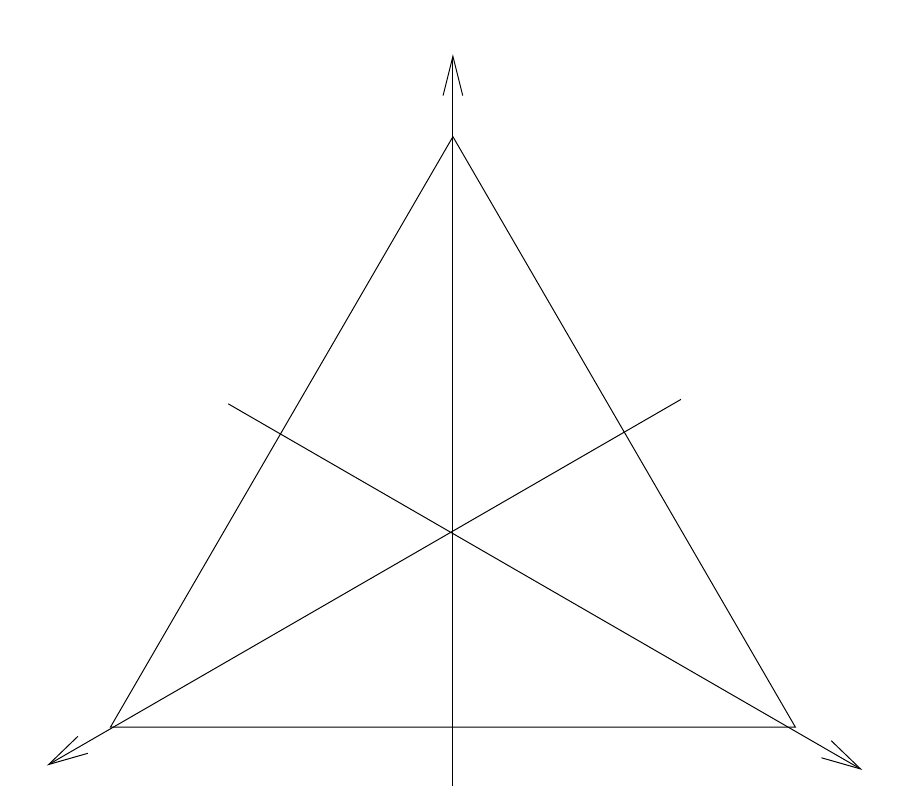}
    \vspace{10mm}
\end{center}

\global\long\def\scoreIoU{IoU}%

In fact, in the drawing shown here-above, the vertical axis corresponds to a score known as the \emph{Intersection over Union} $\scoreIoU$ or \emph{Jaccard Coefficient} $\scoreJaccardPos$. It is monotonically increasing with $\scoreFOne$:
\begin{equation}
    \scoreFOne = \frac{2 \scoreIoU}{1+ \scoreIoU}
    \Rightarrow
    \frac{\partial \scoreFOne}{\partial \scoreIoU} > 0
\end{equation}

The isometrics of $\scorePrecision$ form a pencil of lines whose vertex is located at the bottom-right corner. The isometrics of $\scoreFOne$ are horizontal lines. The isometrics of $\scoreRecall$ form a pencil of lines whose vertex is located at the bottom-left corner.

\begin{center}
    \scalebox{0.4}{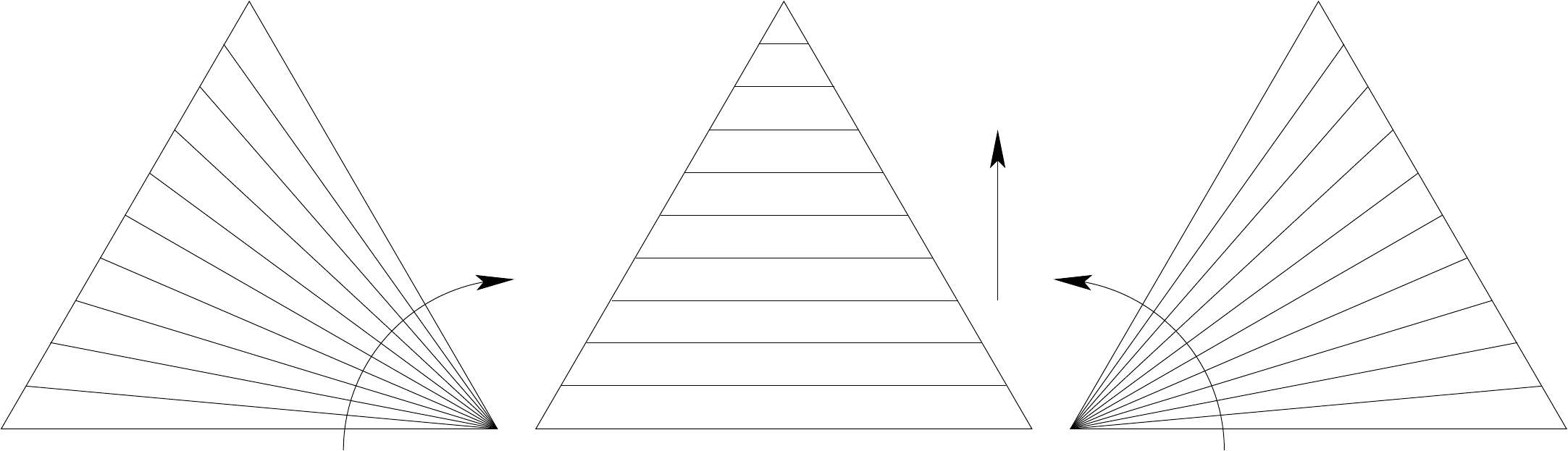}
\end{center}

We now show that $\kendall(\scorePrecision;\scoreRecall)=\nicefrac13$. In the following drawings, we consider a performance $\aPerformance_1$ that can be anywhere in the "triangular space" and we choose one particular performance $\aPerformance_2$, shown as a black point. The pink area represents the probability that $\scorePrecision(\aPerformance_1)<\scorePrecision(\aPerformance_2)$ and $\scoreRecall(\aPerformance_1)>\scoreRecall(\aPerformance_2)$. The six cases are symmetrical: one can obtain them by mirroring or rotating.

\begin{center}
    \scalebox{0.4}{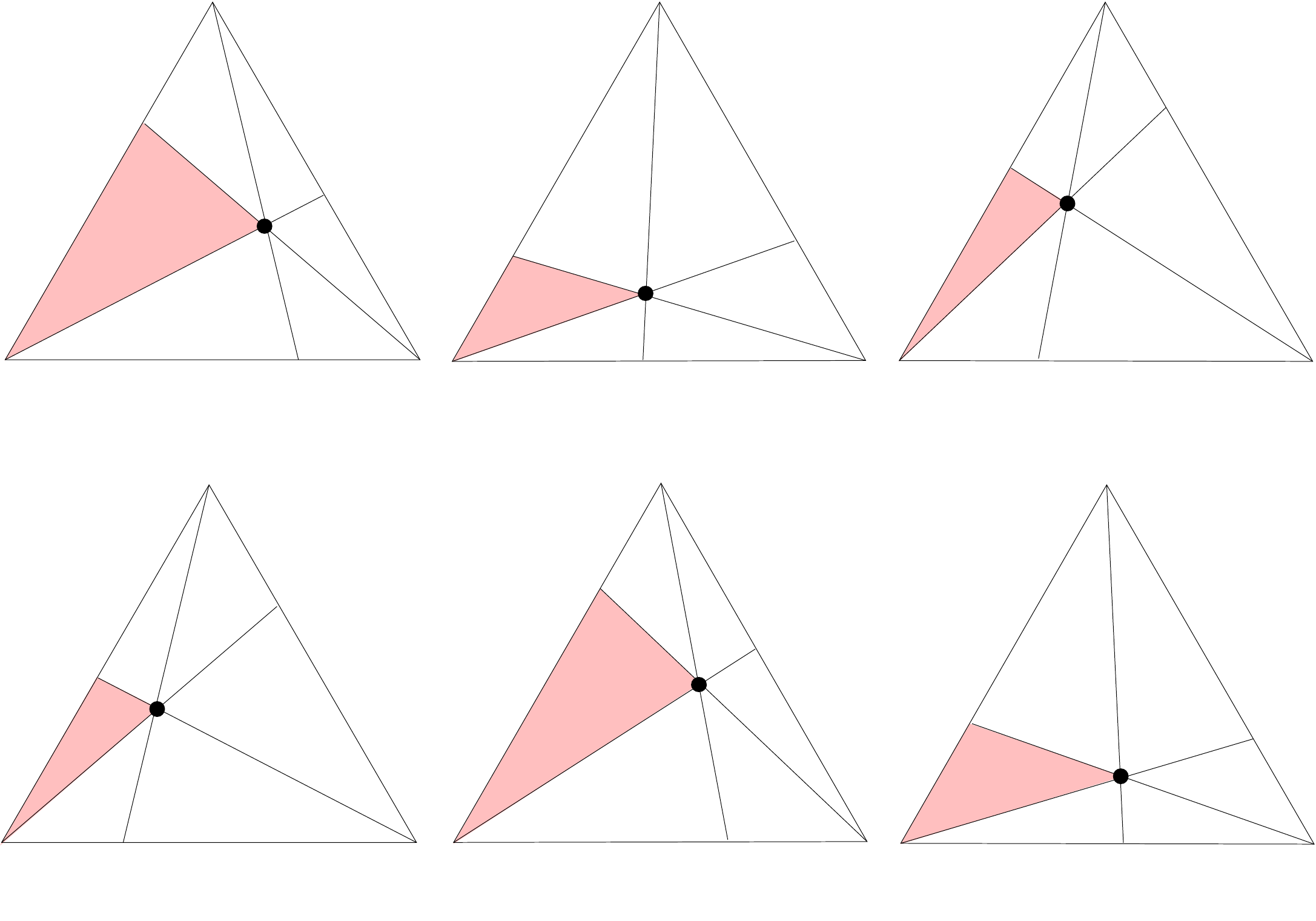}
\end{center}

We see that, in average (that is, if we let $\aPerformance_2$ be anywhere in the "triangular space"), the probability that $\scorePrecision(\aPerformance_1)<\scorePrecision(\aPerformance_2)$ and $\scoreRecall(\aPerformance_1)>\scoreRecall(\aPerformance_2)$ is $\nicefrac{1}{6}$. Using \cref{eq:tau-from-pLtGt}, we find the announced result:
\begin{equation}
    \kendall(\scorePrecision;\scoreRecall)=\nicefrac13
\end{equation}

We now show that $\kendall(\scorePrecision;\scoreFOne) = \kendall(\scoreRecall;\scoreFOne)$. Using \cref{eq:tau-from-pLtGt}, we know that it is the case if and only if the probability that $\scorePrecision(\aPerformance_1)<\scorePrecision(\aPerformance_2)$ and $\scoreFOne(\aPerformance_1)>\scoreFOne(\aPerformance_2)$ is equal to the probability that $\scoreRecall(\aPerformance_1)<\scoreRecall(\aPerformance_2)$ and $\scoreFOne(\aPerformance_1)>\scoreFOne(\aPerformance_2)$. These two probabilities are shown as pink areas, for some arbitrarily fixed $\aPerformance_2$ (the black dot), in the following drawings.

\begin{center}
    \scalebox{0.4}{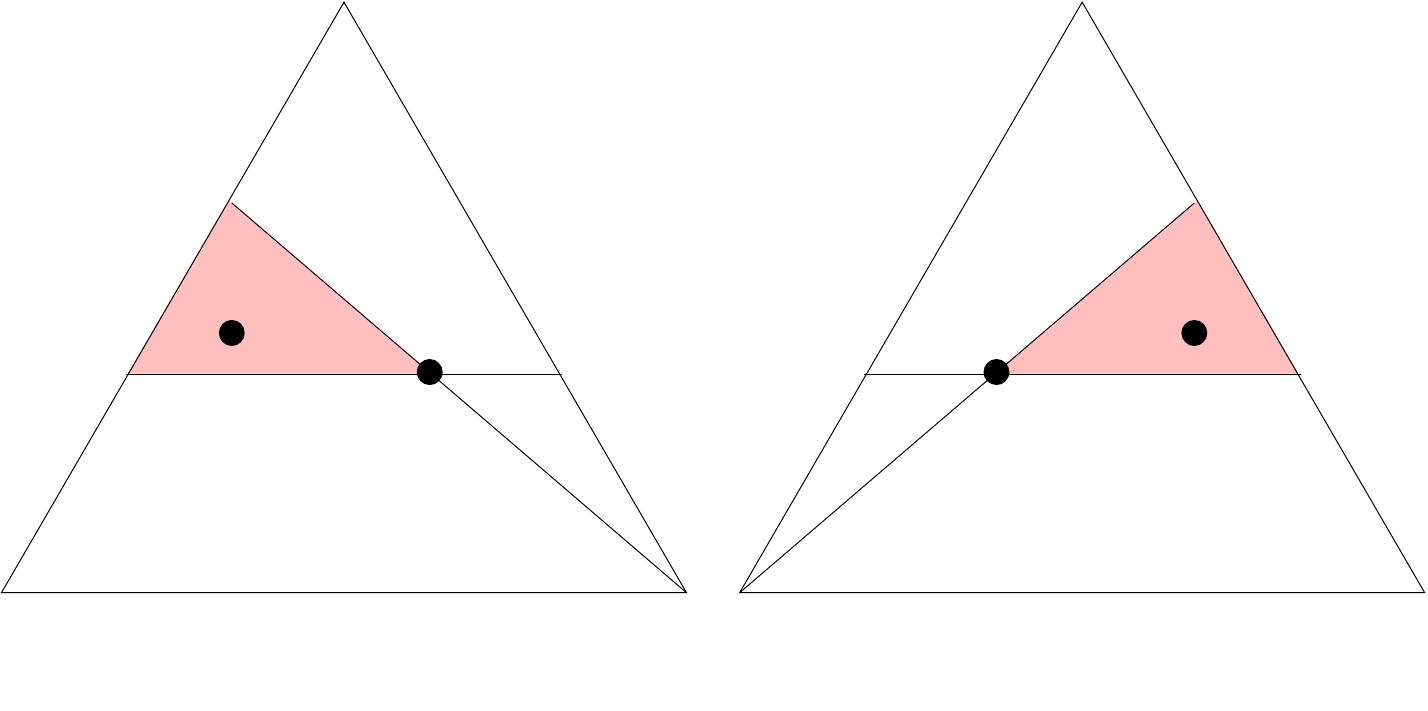}
\end{center}

We see that, by symmetry, the two probabilities are equal. If, furthermore, we do not fix anymore the performance $\aPerformance_2$ but let it be anywhere in the "triangular space", the two probabilities remain equal. Using \cref{eq:tau-from-pLtGt}, we find the announced result:
\begin{equation}
    \kendall(\scorePrecision;\scoreFOne) = \kendall(\scoreRecall;\scoreFOne)
\end{equation}

%% file: drawings/distri_2_space.pdf_t
\begin{picture}(0,0)%
\includegraphics{distri_2_space.pdf}%
\end{picture}%
\setlength{\unitlength}{4144sp}%
\begin{picture}(6930,6079)(-3451,-4199)
\put( 91,-4021){\makebox(0,0)[lb]{\smash{\fontsize{18}{21.6}\usefont{T1}{ptm}{m}{n}{\color[rgb]{0,0,0}$0$}%
}}}
\put( 91,929){\makebox(0,0)[lb]{\smash{\fontsize{18}{21.6}\usefont{T1}{ptm}{m}{n}{\color[rgb]{0,0,0}$1$}%
}}}
\put(  1,1649){\makebox(0,0)[b]{\smash{\fontsize{18}{21.6}\usefont{T1}{ptm}{m}{n}{\color[rgb]{0,0,0}$P(\{tp\}|\{fp,fn,tp\})$}%
}}}
\put(1616,-1130){\rotatebox{120.0}{\makebox(0,0)[lb]{\smash{\fontsize{18}{21.6}\usefont{T1}{ptm}{m}{n}{\color[rgb]{0,0,0}$0$}%
}}}}
\put(-2671,-3605){\rotatebox{120.0}{\makebox(0,0)[lb]{\smash{\fontsize{18}{21.6}\usefont{T1}{ptm}{m}{n}{\color[rgb]{0,0,0}$1$}%
}}}}
\put(-3249,-4043){\rotatebox{120.0}{\makebox(0,0)[b]{\smash{\fontsize{18}{21.6}\usefont{T1}{ptm}{m}{n}{\color[rgb]{0,0,0}$P(\{fp\}|\{fp,fn,tp\})$}%
}}}}
\put(-1678,-1320){\rotatebox{240.0}{\makebox(0,0)[lb]{\smash{\fontsize{18}{21.6}\usefont{T1}{ptm}{m}{n}{\color[rgb]{0,0,0}$0$}%
}}}}
\put(2609,-3795){\rotatebox{240.0}{\makebox(0,0)[lb]{\smash{\fontsize{18}{21.6}\usefont{T1}{ptm}{m}{n}{\color[rgb]{0,0,0}$1$}%
}}}}
\put(3277,-4077){\rotatebox{240.0}{\makebox(0,0)[b]{\smash{\fontsize{18}{21.6}\usefont{T1}{ptm}{m}{n}{\color[rgb]{0,0,0}$P(\{fn\}|\{fp,fn,tp\})$}%
}}}}
\end{picture}%

%% file: drawings/distri_2_isometrics.pdf_t
\begin{picture}(0,0)%
\includegraphics{distri_2_isometrics.pdf}%
\end{picture}%
\setlength{\unitlength}{4144sp}%
\begin{picture}(16494,4745)(-2621,-3894)
\put(7876, 74){\makebox(0,0)[b]{\smash{\fontsize{16}{19.2}\usefont{T1}{ptm}{m}{n}{\color[rgb]{0,0,0}$\scoreIoU$}%
}}}
\put(2791,-1861){\makebox(0,0)[rb]{\smash{\fontsize{16}{19.2}\usefont{T1}{ptm}{m}{n}{\color[rgb]{0,0,0}$\scorePrecision$}%
}}}
\put(8461,-1861){\rotatebox{360.0}{\makebox(0,0)[lb]{\smash{\fontsize{16}{19.2}\usefont{T1}{ptm}{m}{n}{\color[rgb]{0,0,0}$\scoreRecall$}%
}}}}
\put(7876,-286){\makebox(0,0)[b]{\smash{\fontsize{16}{19.2}\usefont{T1}{ptm}{m}{n}{\color[rgb]{0,0,0}$\scoreFOne$}%
}}}
\put(7876,389){\makebox(0,0)[b]{\smash{\fontsize{16}{19.2}\usefont{T1}{ptm}{m}{n}{\color[rgb]{0,0,0}$\scoreJaccardPos$}%
}}}
\end{picture}%

%% file: drawings/distri_2_tau_Pr_Re_Lt_Gt.pdf_t
\begin{picture}(0,0)%
\includegraphics{distri_2_tau_Pr_Re_Lt_Gt.pdf}%
\end{picture}%
\setlength{\unitlength}{4144sp}%
\begin{picture}(16547,11574)(2951,-10723)
\put(4891,-2041){\makebox(0,0)[lb]{\smash{\fontsize{16}{19.2}\usefont{T1}{ptm}{m}{n}{\color[rgb]{0,0,0}\Romanbar{1}}%
}}}
\put(5543,-893){\makebox(0,0)[lb]{\smash{\fontsize{16}{19.2}\usefont{T1}{ptm}{m}{n}{\color[rgb]{0,0,0}\Romanbar{2}}%
}}}
\put(6368,-1209){\makebox(0,0)[lb]{\smash{\fontsize{16}{19.2}\usefont{T1}{ptm}{m}{n}{\color[rgb]{0,0,0}\Romanbar{3}}%
}}}
\put(6795,-2288){\makebox(0,0)[lb]{\smash{\fontsize{16}{19.2}\usefont{T1}{ptm}{m}{n}{\color[rgb]{0,0,0}\Romanbar{4}}%
}}}
\put(6758,-3174){\makebox(0,0)[lb]{\smash{\fontsize{16}{19.2}\usefont{T1}{ptm}{m}{n}{\color[rgb]{0,0,0}\Romanbar{5}}%
}}}
\put(5618,-2963){\makebox(0,0)[lb]{\smash{\fontsize{16}{19.2}\usefont{T1}{ptm}{m}{n}{\color[rgb]{0,0,0}\Romanbar{6}}%
}}}
\put(6316,-8116){\makebox(0,0)[rb]{\smash{\fontsize{16}{19.2}\usefont{T1}{ptm}{m}{n}{\color[rgb]{0,0,0}\Romanbar{1}}%
}}}
\put(5664,-6968){\makebox(0,0)[rb]{\smash{\fontsize{16}{19.2}\usefont{T1}{ptm}{m}{n}{\color[rgb]{0,0,0}\Romanbar{2}}%
}}}
\put(4839,-7284){\makebox(0,0)[rb]{\smash{\fontsize{16}{19.2}\usefont{T1}{ptm}{m}{n}{\color[rgb]{0,0,0}\Romanbar{3}}%
}}}
\put(4412,-8363){\makebox(0,0)[rb]{\smash{\fontsize{16}{19.2}\usefont{T1}{ptm}{m}{n}{\color[rgb]{0,0,0}\Romanbar{4}}%
}}}
\put(4449,-9249){\makebox(0,0)[rb]{\smash{\fontsize{16}{19.2}\usefont{T1}{ptm}{m}{n}{\color[rgb]{0,0,0}\Romanbar{5}}%
}}}
\put(5589,-9038){\makebox(0,0)[rb]{\smash{\fontsize{16}{19.2}\usefont{T1}{ptm}{m}{n}{\color[rgb]{0,0,0}\Romanbar{6}}%
}}}
\put(16469,-7719){\makebox(0,0)[rb]{\smash{\fontsize{16}{19.2}\usefont{T1}{ptm}{m}{n}{\color[rgb]{0,0,0}\Romanbar{1}}%
}}}
\put(17585,-7936){\makebox(0,0)[rb]{\smash{\fontsize{16}{19.2}\usefont{T1}{ptm}{m}{n}{\color[rgb]{0,0,0}\Romanbar{6}}%
}}}
\put(18479,-8863){\makebox(0,0)[rb]{\smash{\fontsize{16}{19.2}\usefont{T1}{ptm}{m}{n}{\color[rgb]{0,0,0}\Romanbar{5}}%
}}}
\put(17694,-9489){\makebox(0,0)[rb]{\smash{\fontsize{16}{19.2}\usefont{T1}{ptm}{m}{n}{\color[rgb]{0,0,0}\Romanbar{4}}%
}}}
\put(16764,-9644){\makebox(0,0)[rb]{\smash{\fontsize{16}{19.2}\usefont{T1}{ptm}{m}{n}{\color[rgb]{0,0,0}\Romanbar{3}}%
}}}
\put(15760,-9123){\makebox(0,0)[rb]{\smash{\fontsize{16}{19.2}\usefont{T1}{ptm}{m}{n}{\color[rgb]{0,0,0}\Romanbar{2}}%
}}}
\put(15565,-2964){\makebox(0,0)[lb]{\smash{\fontsize{16}{19.2}\usefont{T1}{ptm}{m}{n}{\color[rgb]{0,0,0}\Romanbar{2}}%
}}}
\put(15475,-2009){\makebox(0,0)[lb]{\smash{\fontsize{16}{19.2}\usefont{T1}{ptm}{m}{n}{\color[rgb]{0,0,0}\Romanbar{3}}%
}}}
\put(16530,-2548){\makebox(0,0)[lb]{\smash{\fontsize{16}{19.2}\usefont{T1}{ptm}{m}{n}{\color[rgb]{0,0,0}\Romanbar{1}}%
}}}
\put(17142,-1557){\makebox(0,0)[lb]{\smash{\fontsize{16}{19.2}\usefont{T1}{ptm}{m}{n}{\color[rgb]{0,0,0}\Romanbar{6}}%
}}}
\put(16846,-527){\makebox(0,0)[lb]{\smash{\fontsize{16}{19.2}\usefont{T1}{ptm}{m}{n}{\color[rgb]{0,0,0}\Romanbar{5}}%
}}}
\put(16018,-951){\makebox(0,0)[lb]{\smash{\fontsize{16}{19.2}\usefont{T1}{ptm}{m}{n}{\color[rgb]{0,0,0}\Romanbar{4}}%
}}}
\put(11300,-1817){\makebox(0,0)[lb]{\smash{\fontsize{16}{19.2}\usefont{T1}{ptm}{m}{n}{\color[rgb]{0,0,0}\Romanbar{1}}%
}}}
\put(10463,-1997){\makebox(0,0)[lb]{\smash{\fontsize{16}{19.2}\usefont{T1}{ptm}{m}{n}{\color[rgb]{0,0,0}\Romanbar{6}}%
}}}
\put(9637,-2962){\makebox(0,0)[lb]{\smash{\fontsize{16}{19.2}\usefont{T1}{ptm}{m}{n}{\color[rgb]{0,0,0}\Romanbar{5}}%
}}}
\put(10432,-3414){\makebox(0,0)[lb]{\smash{\fontsize{16}{19.2}\usefont{T1}{ptm}{m}{n}{\color[rgb]{0,0,0}\Romanbar{4}}%
}}}
\put(11402,-3501){\makebox(0,0)[lb]{\smash{\fontsize{16}{19.2}\usefont{T1}{ptm}{m}{n}{\color[rgb]{0,0,0}\Romanbar{3}}%
}}}
\put(12308,-2835){\makebox(0,0)[lb]{\smash{\fontsize{16}{19.2}\usefont{T1}{ptm}{m}{n}{\color[rgb]{0,0,0}\Romanbar{2}}%
}}}
\put(12622,-8916){\makebox(0,0)[rb]{\smash{\fontsize{16}{19.2}\usefont{T1}{ptm}{m}{n}{\color[rgb]{0,0,0}\Romanbar{2}}%
}}}
\put(10979,-8915){\makebox(0,0)[rb]{\smash{\fontsize{16}{19.2}\usefont{T1}{ptm}{m}{n}{\color[rgb]{0,0,0}\Romanbar{1}}%
}}}
\put(10860,-7501){\makebox(0,0)[rb]{\smash{\fontsize{16}{19.2}\usefont{T1}{ptm}{m}{n}{\color[rgb]{0,0,0}\Romanbar{6}}%
}}}
\put(11206,-6692){\makebox(0,0)[rb]{\smash{\fontsize{16}{19.2}\usefont{T1}{ptm}{m}{n}{\color[rgb]{0,0,0}\Romanbar{5}}%
}}}
\put(11955,-6923){\makebox(0,0)[rb]{\smash{\fontsize{16}{19.2}\usefont{T1}{ptm}{m}{n}{\color[rgb]{0,0,0}\Romanbar{4}}%
}}}
\put(12556,-8071){\makebox(0,0)[rb]{\smash{\fontsize{16}{19.2}\usefont{T1}{ptm}{m}{n}{\color[rgb]{0,0,0}\Romanbar{3}}%
}}}
\put(5626,-4561){\makebox(0,0)[b]{\smash{\fontsize{15}{18}\usefont{T1}{ptm}{m}{n}{\color[rgb]{0,0,0}$Proba=\frac{\Romanbar{1}}{\Romanbar{1}+\Romanbar{2}+\Romanbar{3}+\Romanbar{4}+\Romanbar{5}+\Romanbar{6}}$}%
}}}
\put(16876,-4561){\makebox(0,0)[b]{\smash{\fontsize{15}{18}\usefont{T1}{ptm}{m}{n}{\color[rgb]{0,0,0}$Proba=\frac{\Romanbar{3}}{\Romanbar{1}+\Romanbar{2}+\Romanbar{3}+\Romanbar{4}+\Romanbar{5}+\Romanbar{6}}$}%
}}}
\put(17101,-10636){\makebox(0,0)[b]{\smash{\fontsize{15}{18}\usefont{T1}{ptm}{m}{n}{\color[rgb]{0,0,0}$Proba=\frac{\Romanbar{2}}{\Romanbar{1}+\Romanbar{2}+\Romanbar{3}+\Romanbar{4}+\Romanbar{5}+\Romanbar{6}}$}%
}}}
\put(5626,-10636){\makebox(0,0)[b]{\smash{\fontsize{15}{18}\usefont{T1}{ptm}{m}{n}{\color[rgb]{0,0,0}$Proba=\frac{\Romanbar{4}}{\Romanbar{1}+\Romanbar{2}+\Romanbar{3}+\Romanbar{4}+\Romanbar{5}+\Romanbar{6}}$}%
}}}
\put(11251,-4561){\makebox(0,0)[b]{\smash{\fontsize{15}{18}\usefont{T1}{ptm}{m}{n}{\color[rgb]{0,0,0}$Proba=\frac{\Romanbar{5}}{\Romanbar{1}+\Romanbar{2}+\Romanbar{3}+\Romanbar{4}+\Romanbar{5}+\Romanbar{6}}$}%
}}}
\put(11251,-10636){\makebox(0,0)[b]{\smash{\fontsize{15}{18}\usefont{T1}{ptm}{m}{n}{\color[rgb]{0,0,0}$Proba=\frac{\Romanbar{6}}{\Romanbar{1}+\Romanbar{2}+\Romanbar{3}+\Romanbar{4}+\Romanbar{5}+\Romanbar{6}}$}%
}}}
\end{picture}%

%% file: drawings/distri_2_Fone_optimal.pdf_t
\begin{picture}(0,0)%
\includegraphics{distri_2_Fone_optimal.pdf}%
\end{picture}%
\setlength{\unitlength}{4144sp}%
\begin{picture}(10869,5409)(1834,-4738)
\put(5221,-1951){\makebox(0,0)[lb]{\smash{\fontsize{16}{19.2}\usefont{T1}{ptm}{m}{n}{\color[rgb]{0,0,0}$\aPerformance_2$}%
}}}
\put(3511,-1681){\makebox(0,0)[rb]{\smash{\fontsize{16}{19.2}\usefont{T1}{ptm}{m}{n}{\color[rgb]{0,0,0}$\aPerformance_1$}%
}}}
\put(9316,-1951){\rotatebox{360.0}{\makebox(0,0)[rb]{\smash{\fontsize{16}{19.2}\usefont{T1}{ptm}{m}{n}{\color[rgb]{0,0,0}$\aPerformance_2$}%
}}}}
\put(11026,-1681){\rotatebox{360.0}{\makebox(0,0)[lb]{\smash{\fontsize{16}{19.2}\usefont{T1}{ptm}{m}{n}{\color[rgb]{0,0,0}$\aPerformance_1$}%
}}}}
\put(4501,-4291){\makebox(0,0)[b]{\smash{\fontsize{16}{19.2}\usefont{T1}{ptm}{m}{n}{\color[rgb]{0,0,0}$\scorePrecision(\aPerformance_1)<\scorePrecision(\aPerformance_2)$}%
}}}
\put(4501,-4651){\makebox(0,0)[b]{\smash{\fontsize{16}{19.2}\usefont{T1}{ptm}{m}{n}{\color[rgb]{0,0,0}$\scoreFOne(\aPerformance_1)>\scoreFOne(\aPerformance_2)$}%
}}}
\put(10171,-4291){\makebox(0,0)[b]{\smash{\fontsize{16}{19.2}\usefont{T1}{ptm}{m}{n}{\color[rgb]{0,0,0}$\scoreRecall(\aPerformance_1)<\scoreRecall(\aPerformance_2)$}%
}}}
\put(10171,-4651){\makebox(0,0)[b]{\smash{\fontsize{16}{19.2}\usefont{T1}{ptm}{m}{n}{\color[rgb]{0,0,0}$\scoreFOne(\aPerformance_1)>\scoreFOne(\aPerformance_2)$}%
}}}
\end{picture}%

%% file: sections/A_3_CS3_detailed_distribution_III.tex
\subsubsection{Detailed Results for the Uniform Distributions With Fixed Class Priors}

To find the probabilities necessary to compute $\kendall$, we integrate over the whole \roc space. Let us denote by $(\scoreFPR,\scoreTPR)=(x,y)$ the coordinates of a performance in this space. We have
\begin{equation}
    \scorePrecision(\aPerformance_1) < \scorePrecision(\aPerformance_2) \Leftrightarrow \frac{y_1}{x_1} < \frac{y_2}{x_2}
\end{equation}
\begin{equation}
    \scoreRecall(\aPerformance_1) < \scoreRecall(\aPerformance_2) \Leftrightarrow y_1 < y_2
\end{equation}
\begin{equation}
    \scoreFBeta(\aPerformance_1) < \scoreFBeta(\aPerformance_2) \Leftrightarrow \frac{y_1}{x_1+\ell} < \frac{y_2}{x_2+\ell}
\end{equation}

We start with a few initializations.

\begin{lstlisting}[backgroundcolor = \color{lightgray},language=Mathematica,numbers=left,numberstyle={\small},breaklines=true]
tot = Integrate[
    1, 
    {x1, 0, 1}, {y1, 0, 1}, 
    {x2, 0, 1}, {y2, 0, 1}
]
\end{lstlisting}

We have
\begin{equation}
    \kendall(\scorePrecision; \scoreRecall) = \frac{1}{2}
    \comma
\end{equation}
as shown with the code (see \cref{fig:distri_3_tau_Pr_Re_Lt_Gt}):
\begin{lstlisting}[backgroundcolor = \color{lightgray},language=Mathematica,numbers=left,numberstyle={\small},breaklines=true]
pLtGt = Integrate[
	Boole[(y1 x2 < y2 x1) && (y1 > y2)], 
	{x1, 0, 1}, {y1, 0, 1}, 
	{x2, 0, 1}, {y2, 0, 1}
] / tot
tau = 1 - 4 * pLtGt
\end{lstlisting}

We have
\begin{equation}
    \kendall(\scorePrecision; \scoreFBeta) = 1-\ell \left(\ell \log \left(\frac{\ell}{\ell+1}\right)+1\right)
    \comma
\end{equation}
as shown with the code (see \cref{fig:distri_3_tau_Pr_Fbeta_Lt_Gt}):
\begin{lstlisting}[backgroundcolor = \color{lightgray},language=Mathematica,numbers=left,numberstyle={\small},breaklines=true]
pLtGt = Integrate[
	Boole[(y1 x2 < y2 x1) && (y1 (x2 + l) > y2 (x1 + l))],
	{x1, 0, 1}, {y1, 0, 1},
	{x2, 0, 1}, {y2, 0, 1}, 
	Assumptions -> l > 0
] / tot
pLtGt = FullSimplify[pLtGt, Assumptions -> l > 0]
tau = 1 - 4 * pLtGt
\end{lstlisting}

And we have
\begin{equation}
    \kendall(\scoreFBeta; \scoreRecall) = \ell^2 \left(-\log \left(\frac{1}{\ell}+1\right)\right)+\ell+\frac{1}{2}
    \comma
\end{equation}
as shown with the code (see \cref{fig:distri_3_tau_Fbeta_Re_Lt_Gt}):
\begin{lstlisting}[backgroundcolor = \color{lightgray},language=Mathematica,numbers=left,numberstyle={\small},breaklines=true]
pLtGt = Integrate[
	Boole[(y1 (x2 + l) < y2 (x1 + l)) && (y1 > y2)], 
	{x1, 0, 1}, {y1, 0, 1}, 
	{x2, 0, 1}, {y2, 0, 1}, 
	Assumptions -> l > 0
] / tot
tau = 1 - 4 * pLtGt
tau = FullSimplify[tau, Assumptions -> l > 0]
\end{lstlisting}

\input{figs/distri_III_proof}

\input{figs/detailed_results_distri_III}
The plots are provided in \cref{fig:detailed_results_distri_III}.

%% file: figs/distri_III_proof.tex
\begin{figure*}[p]
\begin{centering}
    \subfloat[To compute $\kendall(\scorePrecision; \scoreRecall)$ with \cref{{eq:tau-from-pLtGt}}, we determine the probability that $\scorePrecision(\aPerformance_1)<\scorePrecision(\aPerformance_2)$ (the performance $\aPerformance_1$ is under the blue line) and $\scoreRecall(\aPerformance_1)>\scoreRecall(\aPerformance_2)$ (the performance $\aPerformance_1$ is above the green line), which is the pink area averaged for all positions of $\aPerformance_2$.\label{fig:distri_3_tau_Pr_Re_Lt_Gt}]{
        \resizebox{0.8\linewidth}{!}{
            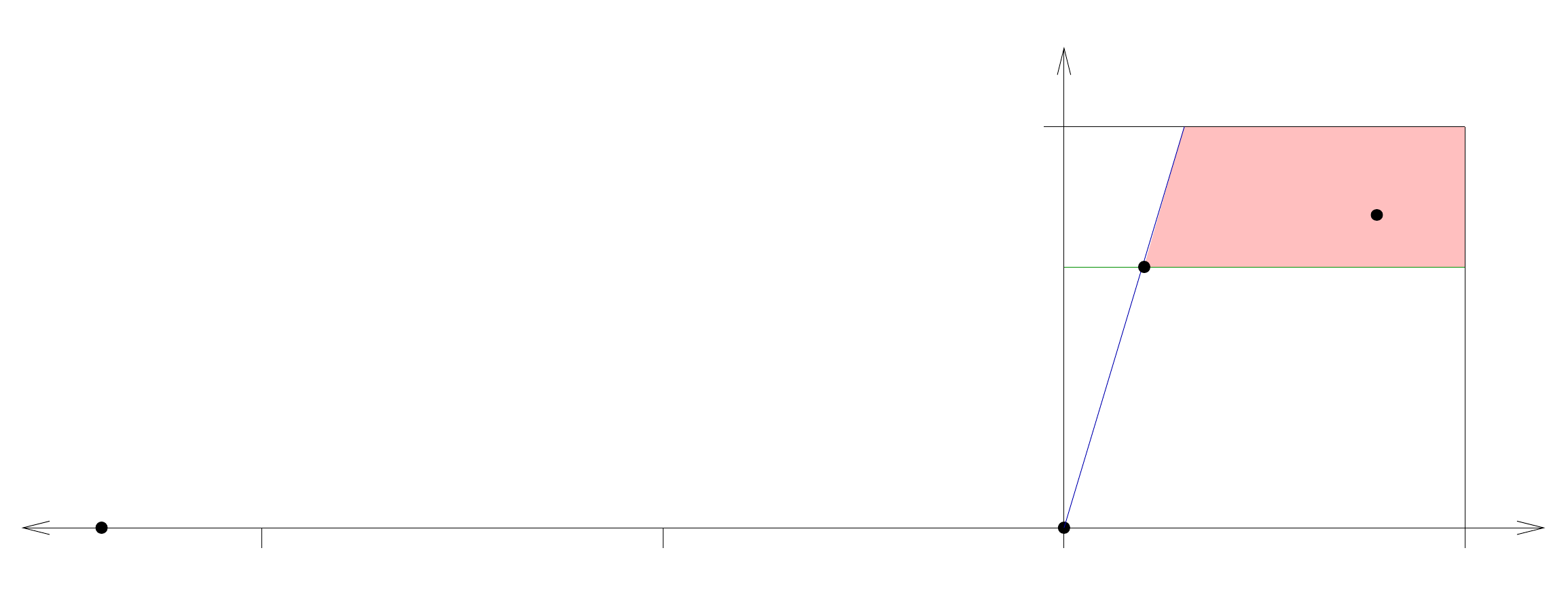
        }
    }
\par\end{centering}
\begin{centering}
    \subfloat[To compute $\kendall(\scorePrecision; \scoreFBeta)$ with \cref{{eq:tau-from-pLtGt}}, we determine the probability that $\scorePrecision(\aPerformance_1)<\scorePrecision(\aPerformance_2)$ (the performance $\aPerformance_1$ is under the blue line) and $\scoreFBeta(\aPerformance_1)>\scoreFBeta(\aPerformance_2)$ (the performance $\aPerformance_1$ is above the red line), which is the pink area averaged for all positions of $\aPerformance_2$.\label{fig:distri_3_tau_Pr_Fbeta_Lt_Gt}]{
        \resizebox{0.8\linewidth}{!}{
            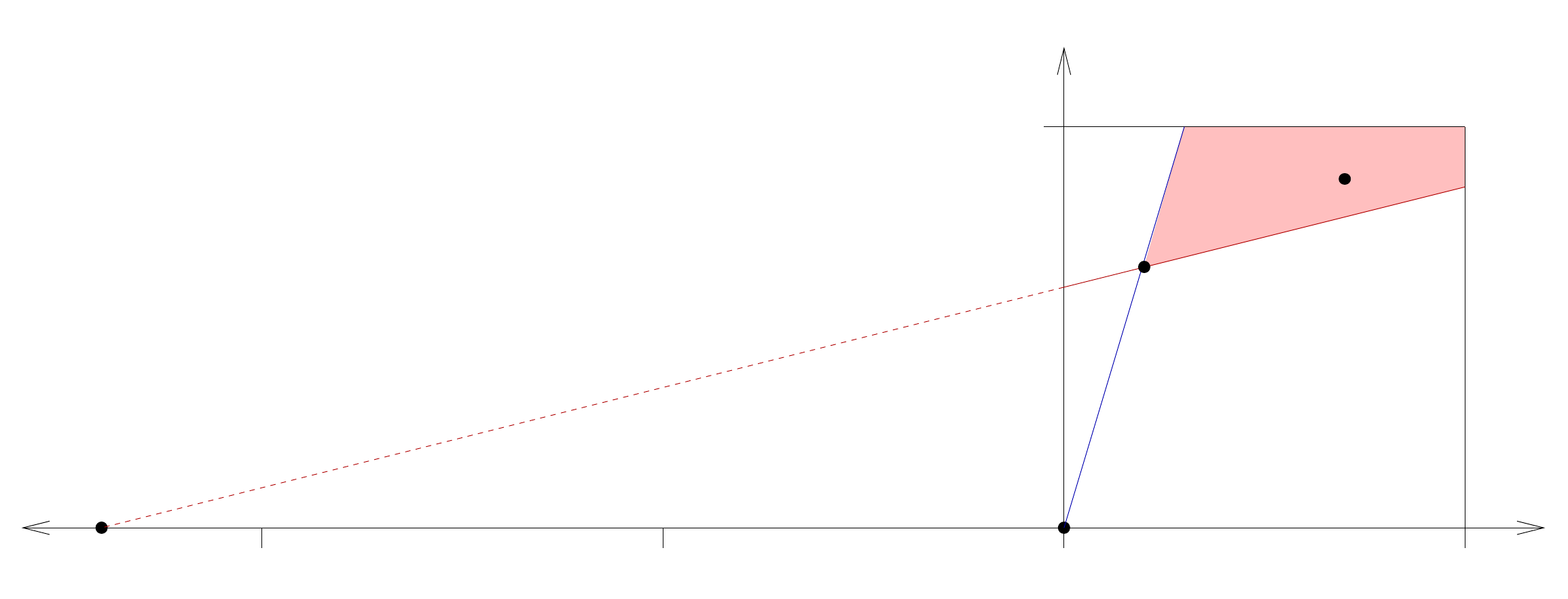
        }
    }
\par\end{centering}
\begin{centering}
    \subfloat[To compute $\kendall(\scoreFBeta; \scoreRecall)$ with \cref{{eq:tau-from-pLtGt}}, we determine the probability that $\scoreFBeta(\aPerformance_1)<\scoreFBeta(\aPerformance_2)$ (the performance $\aPerformance_1$ is under the red line) and $\scoreRecall(\aPerformance_1)>\scoreRecall(\aPerformance_2)$ (the performance $\aPerformance_1$ is above the green line), which is the pink area averaged for all positions of $\aPerformance_2$.\label{fig:distri_3_tau_Fbeta_Re_Lt_Gt}]{
        \resizebox{0.8\linewidth}{!}{
            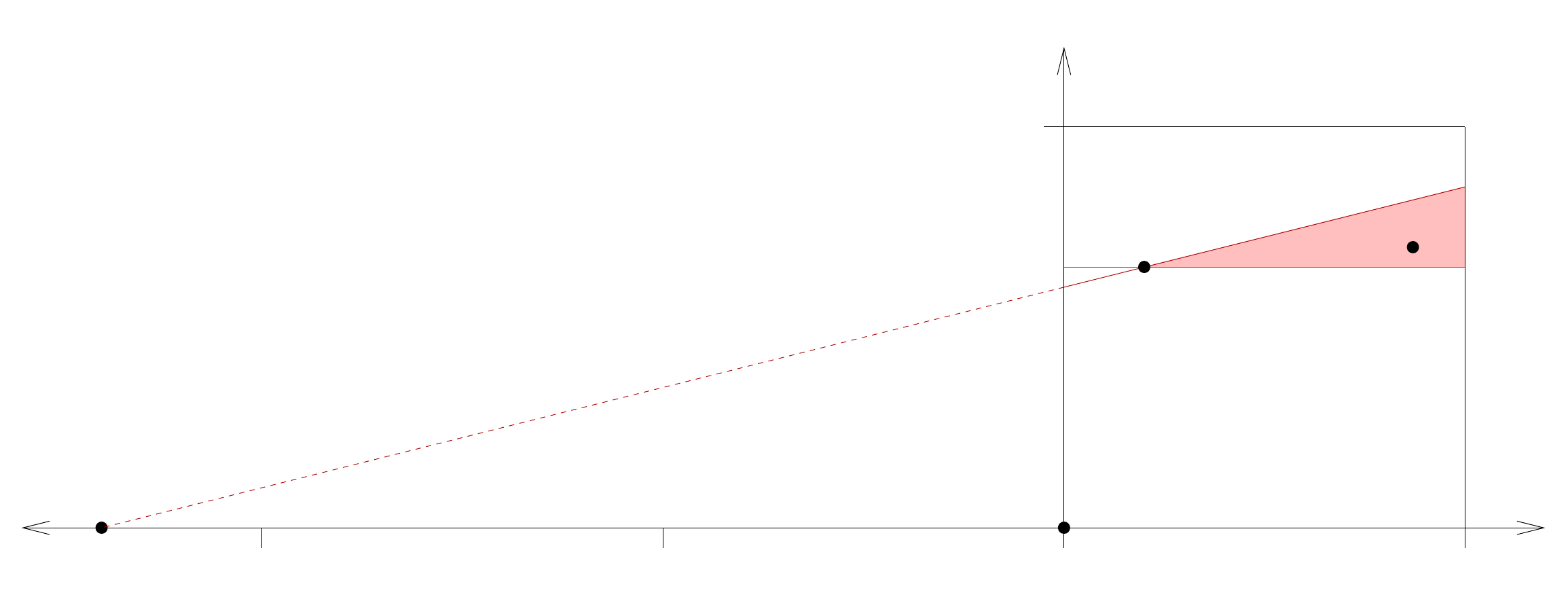
        }
    }
\par\end{centering}
\caption{Graphical representation, in and around the \roc space, of the principle that we use to derive the analytical expression for the optimal tradeoff between precision and recall for the purpose of ranking, in the case of uniform distributions with fixed class priors. Note that the \roc space is a linear mapping of the rectangles depicted in \cref{fig:setIII}.}

\end{figure*}

%% file: drawings/distri_3_tau_Pr_Re_Lt_Gt.pdf_t
\begin{picture}(0,0)%
\includegraphics{distri_3_tau_Pr_Re_Lt_Gt.pdf}%
\end{picture}%
\setlength{\unitlength}{4144sp}%
\begin{picture}(17580,6709)(-11939,-4454)
\put(676,-511){\makebox(0,0)[rb]{\smash{\fontsize{24}{28.8}\usefont{T1}{ptm}{m}{n}{\color[rgb]{0,0,0}$P_2$}%
}}}
\put(3286, 74){\makebox(0,0)[rb]{\smash{\fontsize{24}{28.8}\usefont{T1}{ptm}{m}{n}{\color[rgb]{0,0,0}$P_1$}%
}}}
\put(5626,-3661){\makebox(0,0)[lb]{\smash{\fontsize{24}{28.8}\usefont{T1}{ptm}{m}{n}{\color[rgb]{0,0,0}$x$}%
}}}
\put(2251,1064){\makebox(0,0)[b]{\smash{\fontsize{24}{28.8}\usefont{T1}{ptm}{m}{n}{\color[rgb]{0,0,0}ROC space}%
}}}
\put(  1,1964){\makebox(0,0)[b]{\smash{\fontsize{24}{28.8}\usefont{T1}{ptm}{m}{n}{\color[rgb]{0,0,0}$y$}%
}}}
\put(-11924,-3661){\makebox(0,0)[rb]{\smash{\fontsize{24}{28.8}\usefont{T1}{ptm}{m}{n}{\color[rgb]{0,0,0}$\ell$}%
}}}
\put(  1,-4336){\makebox(0,0)[b]{\smash{\fontsize{24}{28.8}\usefont{T1}{ptm}{m}{n}{\color[rgb]{0,0,0}$(0,0,0)$}%
}}}
\put(-449,839){\makebox(0,0)[rb]{\smash{\fontsize{24}{28.8}\usefont{T1}{ptm}{m}{n}{\color[rgb]{0,0,0}$1$}%
}}}
\put(4501,-4336){\makebox(0,0)[b]{\smash{\fontsize{24}{28.8}\usefont{T1}{ptm}{m}{n}{\color[rgb]{0,0,0}$1$}%
}}}
\put(-4499,-4336){\makebox(0,0)[b]{\smash{\fontsize{24}{28.8}\usefont{T1}{ptm}{m}{n}{\color[rgb]{0,0,0}$1$}%
}}}
\put(-8999,-4336){\makebox(0,0)[b]{\smash{\fontsize{24}{28.8}\usefont{T1}{ptm}{m}{n}{\color[rgb]{0,0,0}$2$}%
}}}
\put(-10799,-4336){\makebox(0,0)[b]{\smash{\fontsize{24}{28.8}\usefont{T1}{ptm}{m}{n}{\color[rgb]{0,0,0}$\beta^2 \frac{\pi_-}{\pi_+}$}%
}}}
\end{picture}%

%% file: drawings/distri_3_tau_Pr_Fbeta_Lt_Gt.pdf_t
\begin{picture}(0,0)%
\includegraphics{distri_3_tau_Pr_Fbeta_Lt_Gt.pdf}%
\end{picture}%
\setlength{\unitlength}{4144sp}%
\begin{picture}(17580,6709)(-11939,-4454)
\put(676,-511){\makebox(0,0)[rb]{\smash{\fontsize{24}{28.8}\usefont{T1}{ptm}{m}{n}{\color[rgb]{0,0,0}$P_2$}%
}}}
\put(2926,479){\makebox(0,0)[rb]{\smash{\fontsize{24}{28.8}\usefont{T1}{ptm}{m}{n}{\color[rgb]{0,0,0}$P_1$}%
}}}
\put(5626,-3661){\makebox(0,0)[lb]{\smash{\fontsize{24}{28.8}\usefont{T1}{ptm}{m}{n}{\color[rgb]{0,0,0}$x$}%
}}}
\put(2251,1064){\makebox(0,0)[b]{\smash{\fontsize{24}{28.8}\usefont{T1}{ptm}{m}{n}{\color[rgb]{0,0,0}ROC space}%
}}}
\put(  1,1964){\makebox(0,0)[b]{\smash{\fontsize{24}{28.8}\usefont{T1}{ptm}{m}{n}{\color[rgb]{0,0,0}$y$}%
}}}
\put(-11924,-3661){\makebox(0,0)[rb]{\smash{\fontsize{24}{28.8}\usefont{T1}{ptm}{m}{n}{\color[rgb]{0,0,0}$\ell$}%
}}}
\put(  1,-4336){\makebox(0,0)[b]{\smash{\fontsize{24}{28.8}\usefont{T1}{ptm}{m}{n}{\color[rgb]{0,0,0}$(0,0,0)$}%
}}}
\put(-449,839){\makebox(0,0)[rb]{\smash{\fontsize{24}{28.8}\usefont{T1}{ptm}{m}{n}{\color[rgb]{0,0,0}$1$}%
}}}
\put(4501,-4336){\makebox(0,0)[b]{\smash{\fontsize{24}{28.8}\usefont{T1}{ptm}{m}{n}{\color[rgb]{0,0,0}$1$}%
}}}
\put(-4499,-4336){\makebox(0,0)[b]{\smash{\fontsize{24}{28.8}\usefont{T1}{ptm}{m}{n}{\color[rgb]{0,0,0}$1$}%
}}}
\put(-8999,-4336){\makebox(0,0)[b]{\smash{\fontsize{24}{28.8}\usefont{T1}{ptm}{m}{n}{\color[rgb]{0,0,0}$2$}%
}}}
\put(-10799,-4336){\makebox(0,0)[b]{\smash{\fontsize{24}{28.8}\usefont{T1}{ptm}{m}{n}{\color[rgb]{0,0,0}$\beta^2 \frac{\pi_-}{\pi_+}$}%
}}}
\end{picture}%

%% file: drawings/distri_3_tau_Fbeta_Re_Lt_Gt.pdf_t
\begin{picture}(0,0)%
\includegraphics{distri_3_tau_Fbeta_Re_Lt_Gt.pdf}%
\end{picture}%
\setlength{\unitlength}{4144sp}%
\begin{picture}(17580,6709)(-11939,-4454)
\put(676,-511){\makebox(0,0)[rb]{\smash{\fontsize{24}{28.8}\usefont{T1}{ptm}{m}{n}{\color[rgb]{0,0,0}$P_2$}%
}}}
\put(3691,-286){\makebox(0,0)[rb]{\smash{\fontsize{24}{28.8}\usefont{T1}{ptm}{m}{n}{\color[rgb]{0,0,0}$P_1$}%
}}}
\put(5626,-3661){\makebox(0,0)[lb]{\smash{\fontsize{24}{28.8}\usefont{T1}{ptm}{m}{n}{\color[rgb]{0,0,0}$x$}%
}}}
\put(2251,1064){\makebox(0,0)[b]{\smash{\fontsize{24}{28.8}\usefont{T1}{ptm}{m}{n}{\color[rgb]{0,0,0}ROC space}%
}}}
\put(  1,1964){\makebox(0,0)[b]{\smash{\fontsize{24}{28.8}\usefont{T1}{ptm}{m}{n}{\color[rgb]{0,0,0}$y$}%
}}}
\put(-11924,-3661){\makebox(0,0)[rb]{\smash{\fontsize{24}{28.8}\usefont{T1}{ptm}{m}{n}{\color[rgb]{0,0,0}$\ell$}%
}}}
\put(  1,-4336){\makebox(0,0)[b]{\smash{\fontsize{24}{28.8}\usefont{T1}{ptm}{m}{n}{\color[rgb]{0,0,0}$(0,0,0)$}%
}}}
\put(-449,839){\makebox(0,0)[rb]{\smash{\fontsize{24}{28.8}\usefont{T1}{ptm}{m}{n}{\color[rgb]{0,0,0}$1$}%
}}}
\put(4501,-4336){\makebox(0,0)[b]{\smash{\fontsize{24}{28.8}\usefont{T1}{ptm}{m}{n}{\color[rgb]{0,0,0}$1$}%
}}}
\put(-4499,-4336){\makebox(0,0)[b]{\smash{\fontsize{24}{28.8}\usefont{T1}{ptm}{m}{n}{\color[rgb]{0,0,0}$1$}%
}}}
\put(-8999,-4336){\makebox(0,0)[b]{\smash{\fontsize{24}{28.8}\usefont{T1}{ptm}{m}{n}{\color[rgb]{0,0,0}$2$}%
}}}
\put(-10799,-4336){\makebox(0,0)[b]{\smash{\fontsize{24}{28.8}\usefont{T1}{ptm}{m}{n}{\color[rgb]{0,0,0}$\beta^2 \frac{\pi_-}{\pi_+}$}%
}}}
\end{picture}%

%% file: figs/detailed_results_distri_III.tex
\begin{figure*}
    \begin{centering}
        \hfill
        \subfloat[$\scoreFOne$, the traditional (balanced) F-score, is not the optimal tradeoff: $\kendall(\scorePrecision; \scoreFOne) \ne \kendall(\scoreFOne; \scoreRecall)$.]{
            \includegraphics[scale=0.5]{images/distri_3_f1_is_not_between_ppv_tpr_analytical.pdf}
        }
        \hfill
        \hfill
        \subfloat[$\scoreSkewInsensitiveVersionFOne$, the skew-insensitive version of $\scoreFOne$~\cite{Flach2003TheGeometry}, is not the optimal tradeoff: $\kendall(\scorePrecision; \scoreSkewInsensitiveVersionFOne) \ne \kendall(\scoreSkewInsensitiveVersionFOne; \scoreRecall)$.]{
            \includegraphics[scale=0.5]{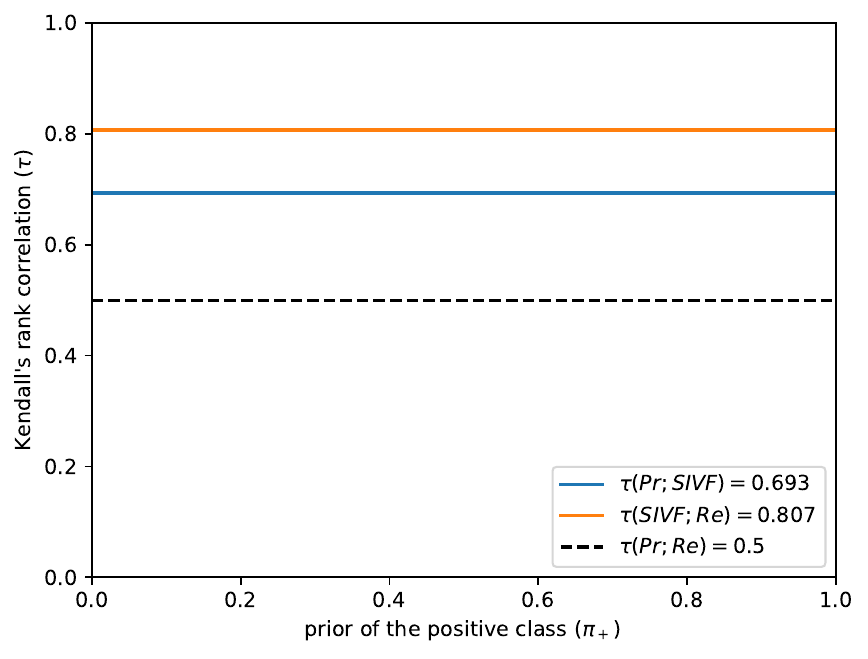}
        }
        \hfill
    \par\end{centering}
    
    \begin{centering}
        \hfill
        \subfloat[Fréchet variance for various priors. It is defined at \cref{eq:frechet-variance} and should be minimized to obtain the optimal tradeoff.]{
            \includegraphics[scale=0.5]{images/distri_3_frechet_variance_analytical.pdf}
        }
        \hfill
        \hfill
        \subfloat[$\scoreOptimalTradeoff$ if the optimal tradeoff: $\kendall(\scorePrecision; \scoreOptimalTradeoff) = \kendall(\scoreOptimalTradeoff; \scoreRecall)$.]{
            \includegraphics[scale=0.5]{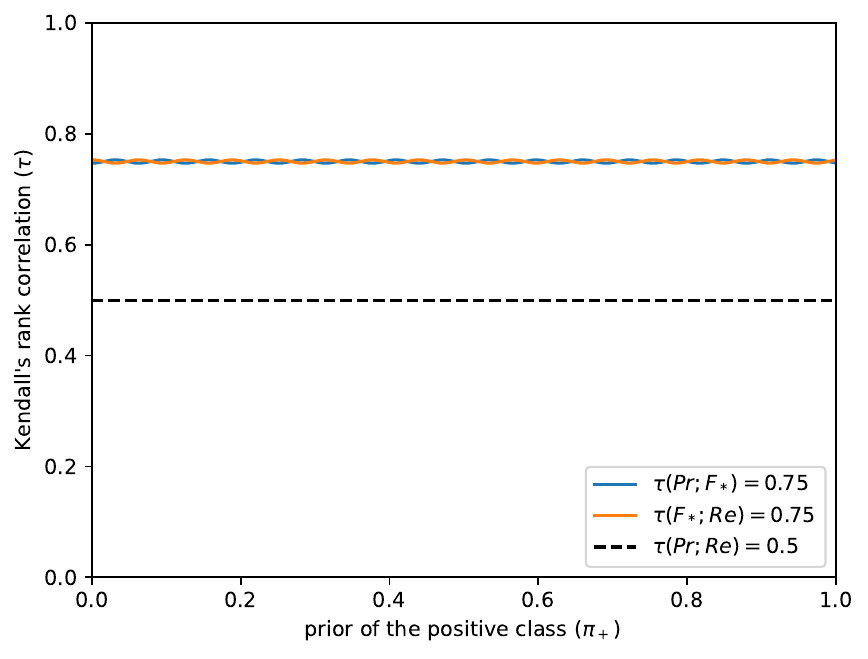}
        }
        \hfill
    \par\end{centering}
    
    \begin{centering}
        \hfill
        \subfloat[Adaptation of $\beta$ \wrt class priors.]{
            \includegraphics[scale=0.5]{images/distri_3_adaptation_analytical.pdf}
        }
        \hfill
        \hfill
        \subfloat[\pca of the manifold (for $\priorpos=0.1$).]{
            \includegraphics[scale=0.5]{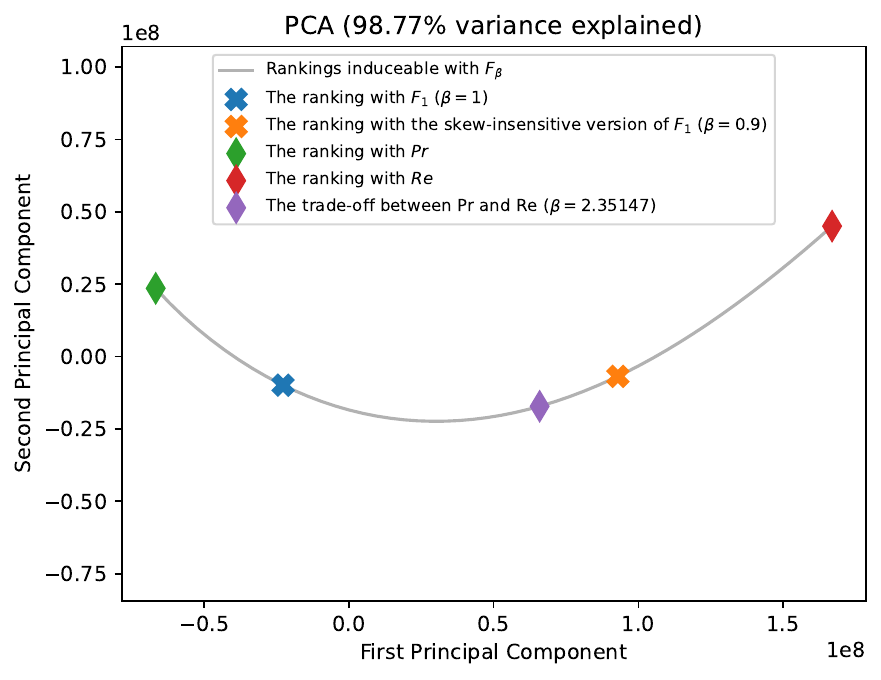}
        }
        \hfill
    \par\end{centering}
        
    \caption{
        Results for uniform distributions over the performances with fixed class priors, \ie $\setIII$.
        \label{fig:detailed_results_distri_III}
    }
\end{figure*}

%% file: sections/A_3_CS4_detailed_distribution_IV.tex
\subsubsection{Detailed Results for the Uniform Distributions With Fixed Class Priors, Above No-Skill}

To find the probabilities necessary to compute $\kendall$, we integrate over half of the \roc space, the half above the raising diagonal. Let us denote by $(\scoreFPR,\scoreTPR)=(x,y)$ the coordinates of a performance in this space. We have
\begin{equation}
    \scorePrecision(\aPerformance_1) < \scorePrecision(\aPerformance_2) \Leftrightarrow \frac{y_1}{x_1} < \frac{y_2}{x_2}
\end{equation}
\begin{equation}
    \scoreRecall(\aPerformance_1) < \scoreRecall(\aPerformance_2) \Leftrightarrow y_1 < y_2
\end{equation}
\begin{equation}
    \scoreFBeta(\aPerformance_1) < \scoreFBeta(\aPerformance_2) \Leftrightarrow \frac{y_1}{x_1+\ell} < \frac{y_2}{x_2+\ell}
\end{equation}

We start with a few initializations.

\begin{lstlisting}[backgroundcolor = \color{lightgray},language=Mathematica,numbers=left,numberstyle={\small},breaklines=true]
tot = Integrate[
    1, 
    {x1, 0, 1}, {y1, x1, 1}, 
    {x2, 0, 1}, {y2, x2, 1}
]
\end{lstlisting}

We have
\begin{equation}
    \kendall(\scorePrecision; \scoreRecall) = 0
    \comma
\end{equation}
as shown with the code (see \cref{fig:distri_4_tau_Pr_Re_Lt_Gt}):
\begin{lstlisting}[backgroundcolor = \color{lightgray},language=Mathematica,numbers=left,numberstyle={\small},breaklines=true]
pLtGt = Integrate[
	Boole[(y1 x2 < y2 x1) && (y1 > y2)], 
    {x1, 0, 1}, {y1, x1, 1}, 
    {x2, 0, 1}, {y2, x2, 1}
] / tot
tau = 1 - 4 * pLtGt
\end{lstlisting}

We have
\begin{equation}
    \kendall(\scorePrecision; \scoreFBeta) = 1-\frac{2}{3} \ell \left(-6 \ell^2-6 \left(\ell^2-1\right) \ell \log \left(\frac{\ell}{\ell+1}\right)+3 \ell+4\right)
    \comma
\end{equation}
as shown with the code (see \cref{fig:distri_4_tau_Pr_Fbeta_Lt_Gt}):
\begin{lstlisting}[backgroundcolor = \color{lightgray},language=Mathematica,numbers=left,numberstyle={\small},breaklines=true]
pLtGt = Integrate[
	Boole[(y1 x2 < y2 x1) && (y1 (x2 + l) > y2 (x1 + l))],
    {x1, 0, 1}, {y1, x1, 1}, 
    {x2, 0, 1}, {y2, x2, 1},
	Assumptions -> l > 0
] / tot
pLtGt = FullSimplify[pLtGt, Assumptions -> l > 0]
tau = 1 - 4 * pLtGt
\end{lstlisting}

And we have
\begin{equation}
    \kendall(\scoreFBeta; \scoreRecall) = \frac{2}{3} \ell \left(-6 \ell^2+6 \left(\ell^2-1\right) \ell \log \left(\frac{1}{\ell}+1\right)+3 \ell+4\right)
    \comma
\end{equation}
as shown with the code (see \cref{fig:distri_4_tau_Fbeta_Re_Lt_Gt}):
\begin{lstlisting}[backgroundcolor = \color{lightgray},language=Mathematica,numbers=left,numberstyle={\small},breaklines=true]
pLtGt = Integrate[
	Boole[(y1 (x2 + l) < y2 (x1 + l)) && (y1 > y2)], 
    {x1, 0, 1}, {y1, x1, 1}, 
    {x2, 0, 1}, {y2, x2, 1},
	Assumptions -> l > 0
] / tot
tau = 1 - 4 * pLtGt
tau = FullSimplify[tau, Assumptions -> l > 0]
\end{lstlisting}

\input{figs/distri_IV_proof}

\input{figs/detailed_results_distri_IV}
The plots are provided in \cref{fig:detailed_results_distri_IV}.

%% file: figs/distri_IV_proof.tex
\begin{figure*}[p]
\begin{centering}
    \subfloat[To compute $\kendall(\scorePrecision; \scoreRecall)$ with \cref{{eq:tau-from-pLtGt}}, we determine the probability that $\scorePrecision(\aPerformance_1)<\scorePrecision(\aPerformance_2)$ (the performance $\aPerformance_1$ is under the blue line) and $\scoreRecall(\aPerformance_1)>\scoreRecall(\aPerformance_2)$ (the performance $\aPerformance_1$ is above the green line), which is the pink area averaged for all positions of $\aPerformance_2$.\label{fig:distri_4_tau_Pr_Re_Lt_Gt}]{
        \resizebox{0.8\linewidth}{!}{
            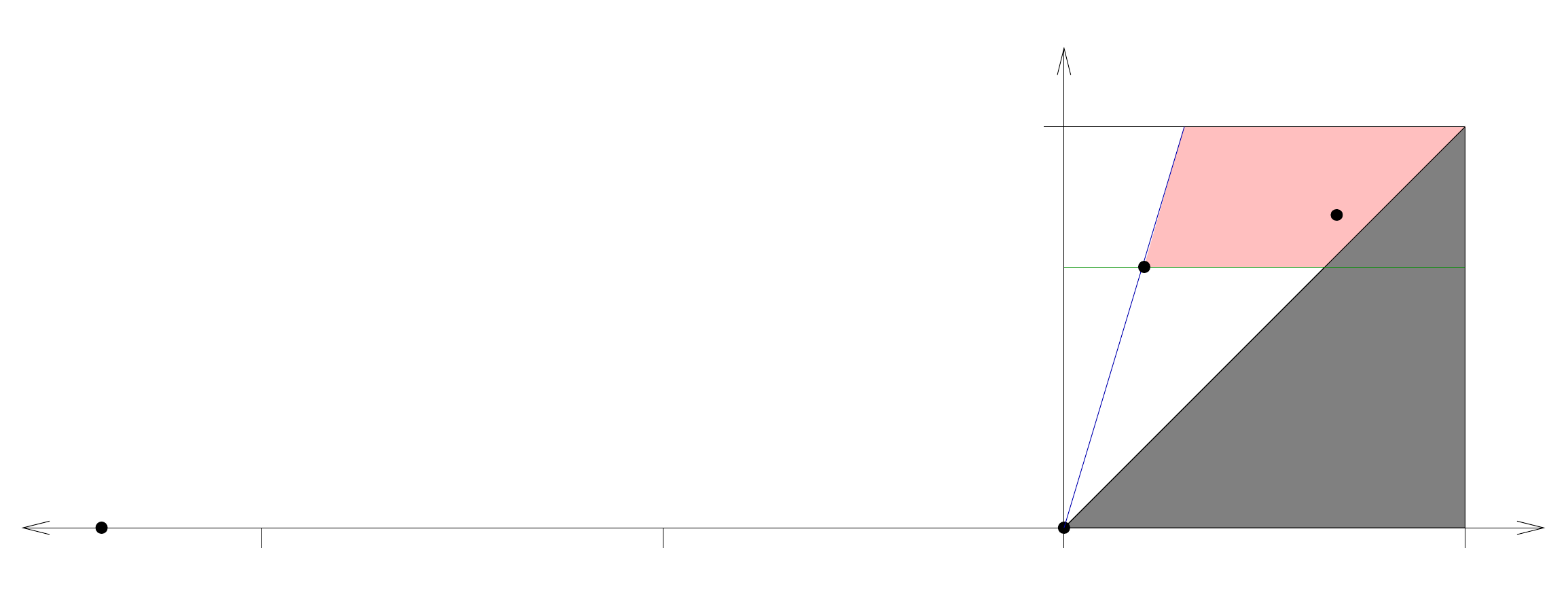
        }
    }
\par\end{centering}
\begin{centering}
    \subfloat[To compute $\kendall(\scorePrecision; \scoreFBeta)$ with \cref{{eq:tau-from-pLtGt}}, we determine the probability that $\scorePrecision(\aPerformance_1)<\scorePrecision(\aPerformance_2)$ (the performance $\aPerformance_1$ is under the blue line) and $\scoreFBeta(\aPerformance_1)>\scoreFBeta(\aPerformance_2)$ (the performance $\aPerformance_1$ is above the red line), which is the pink area averaged for all positions of $\aPerformance_2$.\label{fig:distri_4_tau_Pr_Fbeta_Lt_Gt}]{
        \resizebox{0.8\linewidth}{!}{
            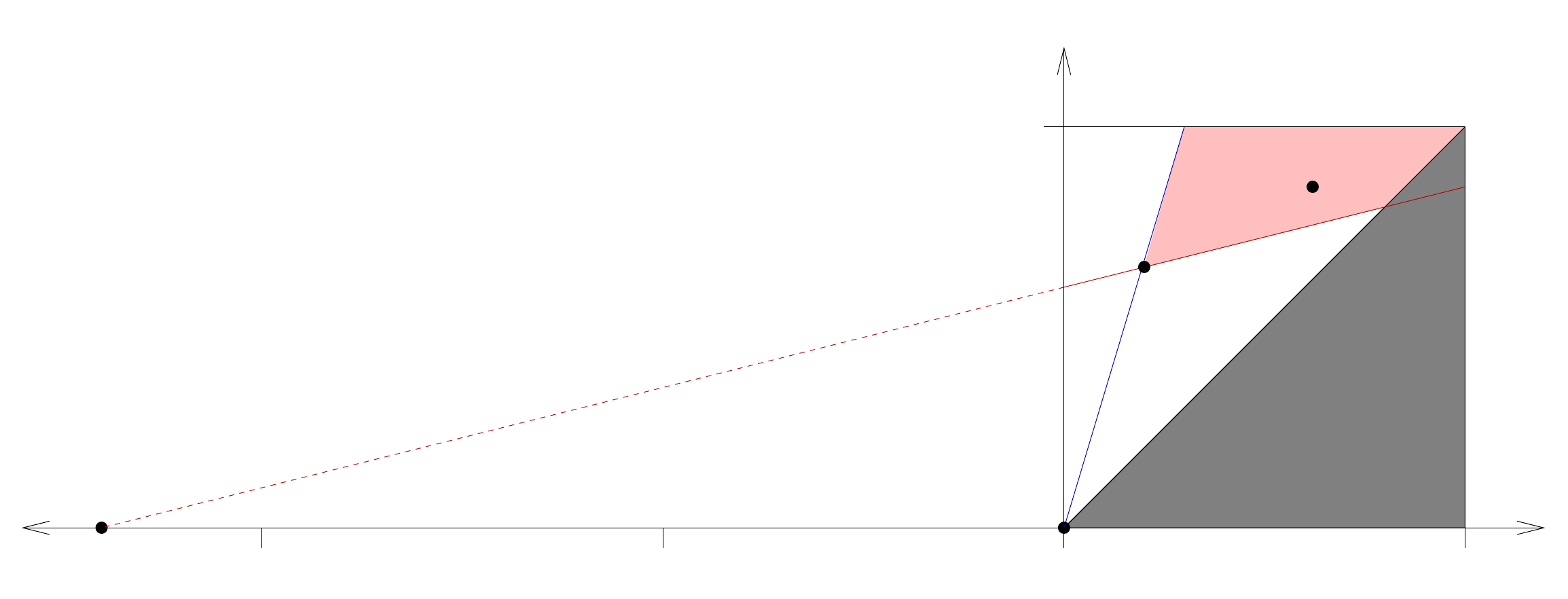
        }
    }
\par\end{centering}
\begin{centering}
    \subfloat[To compute $\kendall(\scoreFBeta; \scoreRecall)$ with \cref{{eq:tau-from-pLtGt}}, we determine the probability that $\scoreFBeta(\aPerformance_1)<\scoreFBeta(\aPerformance_2)$ (the performance $\aPerformance_1$ is under the red line) and $\scoreRecall(\aPerformance_1)>\scoreRecall(\aPerformance_2)$ (the performance $\aPerformance_1$ is above the green line), which is the pink area averaged for all positions of $\aPerformance_2$.\label{fig:distri_4_tau_Fbeta_Re_Lt_Gt}]{
        \resizebox{0.8\linewidth}{!}{
            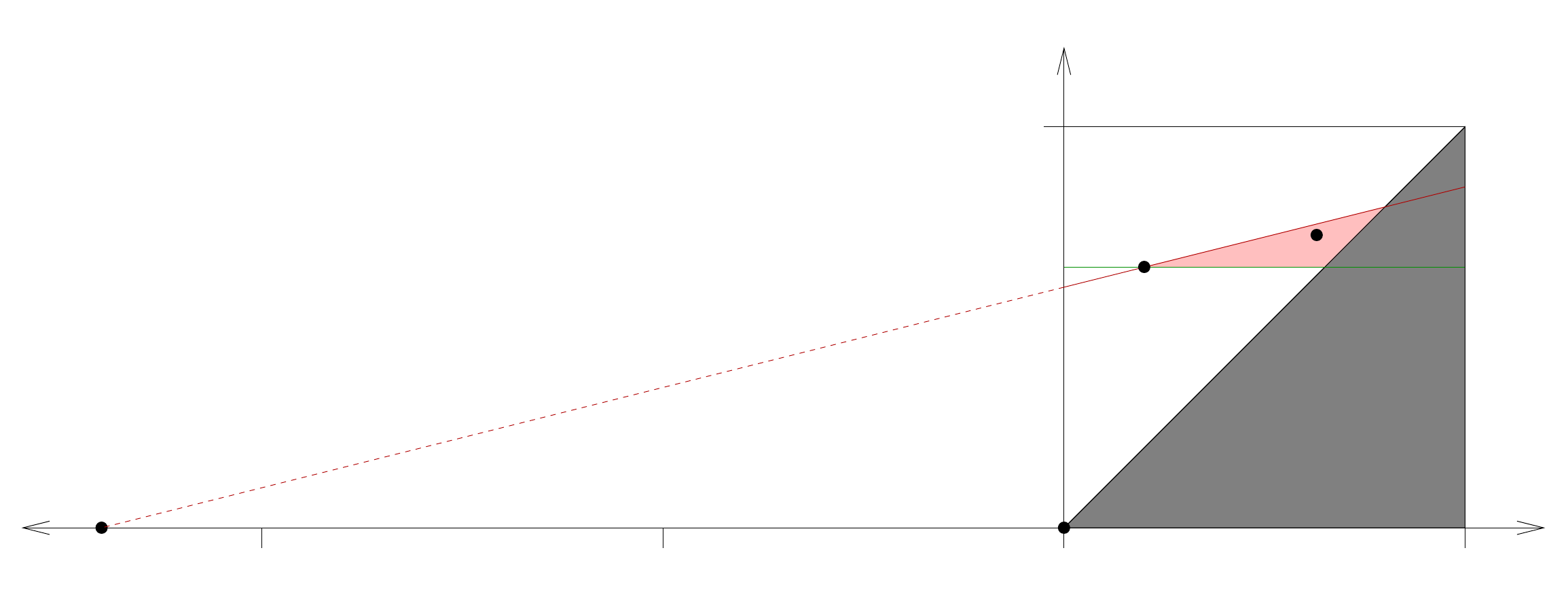
        }
    }
\par\end{centering}
\caption{Graphical representation, in and around the \roc space, of the principle that we use to derive the analytical expression for the optimal tradeoff between precision and recall for the purpose of ranking, in the case of uniform distributions with fixed class priors above no-skill. Note that the unshaded area of \roc space is a linear mapping of the triangles depicted in \cref{fig:setIV}.}

\end{figure*}

%% file: drawings/distri_4_tau_Pr_Re_Lt_Gt.pdf_t
\begin{picture}(0,0)%
\includegraphics{distri_4_tau_Pr_Re_Lt_Gt.pdf}%
\end{picture}%
\setlength{\unitlength}{4144sp}%
\begin{picture}(17580,6709)(-11939,-4454)
\put(2836, 74){\makebox(0,0)[rb]{\smash{\fontsize{24}{28.8}\usefont{T1}{ptm}{m}{n}{\color[rgb]{0,0,0}$P_1$}%
}}}
\put(5626,-3661){\makebox(0,0)[lb]{\smash{\fontsize{24}{28.8}\usefont{T1}{ptm}{m}{n}{\color[rgb]{0,0,0}$x$}%
}}}
\put(2251,1064){\makebox(0,0)[b]{\smash{\fontsize{24}{28.8}\usefont{T1}{ptm}{m}{n}{\color[rgb]{0,0,0}ROC space}%
}}}
\put(  1,1964){\makebox(0,0)[b]{\smash{\fontsize{24}{28.8}\usefont{T1}{ptm}{m}{n}{\color[rgb]{0,0,0}$y$}%
}}}
\put(-11924,-3661){\makebox(0,0)[rb]{\smash{\fontsize{24}{28.8}\usefont{T1}{ptm}{m}{n}{\color[rgb]{0,0,0}$\ell$}%
}}}
\put(  1,-4336){\makebox(0,0)[b]{\smash{\fontsize{24}{28.8}\usefont{T1}{ptm}{m}{n}{\color[rgb]{0,0,0}$(0,0,0)$}%
}}}
\put(-449,839){\makebox(0,0)[rb]{\smash{\fontsize{24}{28.8}\usefont{T1}{ptm}{m}{n}{\color[rgb]{0,0,0}$1$}%
}}}
\put(4501,-4336){\makebox(0,0)[b]{\smash{\fontsize{24}{28.8}\usefont{T1}{ptm}{m}{n}{\color[rgb]{0,0,0}$1$}%
}}}
\put(-4499,-4336){\makebox(0,0)[b]{\smash{\fontsize{24}{28.8}\usefont{T1}{ptm}{m}{n}{\color[rgb]{0,0,0}$1$}%
}}}
\put(-8999,-4336){\makebox(0,0)[b]{\smash{\fontsize{24}{28.8}\usefont{T1}{ptm}{m}{n}{\color[rgb]{0,0,0}$2$}%
}}}
\put(-10799,-4336){\makebox(0,0)[b]{\smash{\fontsize{24}{28.8}\usefont{T1}{ptm}{m}{n}{\color[rgb]{0,0,0}$\beta^2 \frac{\pi_-}{\pi_+}$}%
}}}
\put(676,-511){\makebox(0,0)[rb]{\smash{\fontsize{24}{28.8}\usefont{T1}{ptm}{m}{n}{\color[rgb]{0,0,0}$P_2$}%
}}}
\end{picture}%

%% file: drawings/distri_4_tau_Pr_Fbeta_Lt_Gt.pdf_t
\begin{picture}(0,0)%
\includegraphics{distri_4_tau_Pr_Fbeta_Lt_Gt.pdf}%
\end{picture}%
\setlength{\unitlength}{4144sp}%
\begin{picture}(17580,6709)(-11939,-4454)
\put(2566,389){\makebox(0,0)[rb]{\smash{\fontsize{24}{28.8}\usefont{T1}{ptm}{m}{n}{\color[rgb]{0,0,0}$P_1$}%
}}}
\put(5626,-3661){\makebox(0,0)[lb]{\smash{\fontsize{24}{28.8}\usefont{T1}{ptm}{m}{n}{\color[rgb]{0,0,0}$x$}%
}}}
\put(2251,1064){\makebox(0,0)[b]{\smash{\fontsize{24}{28.8}\usefont{T1}{ptm}{m}{n}{\color[rgb]{0,0,0}ROC space}%
}}}
\put(  1,1964){\makebox(0,0)[b]{\smash{\fontsize{24}{28.8}\usefont{T1}{ptm}{m}{n}{\color[rgb]{0,0,0}$y$}%
}}}
\put(-11924,-3661){\makebox(0,0)[rb]{\smash{\fontsize{24}{28.8}\usefont{T1}{ptm}{m}{n}{\color[rgb]{0,0,0}$\ell$}%
}}}
\put(  1,-4336){\makebox(0,0)[b]{\smash{\fontsize{24}{28.8}\usefont{T1}{ptm}{m}{n}{\color[rgb]{0,0,0}$(0,0,0)$}%
}}}
\put(-449,839){\makebox(0,0)[rb]{\smash{\fontsize{24}{28.8}\usefont{T1}{ptm}{m}{n}{\color[rgb]{0,0,0}$1$}%
}}}
\put(4501,-4336){\makebox(0,0)[b]{\smash{\fontsize{24}{28.8}\usefont{T1}{ptm}{m}{n}{\color[rgb]{0,0,0}$1$}%
}}}
\put(-4499,-4336){\makebox(0,0)[b]{\smash{\fontsize{24}{28.8}\usefont{T1}{ptm}{m}{n}{\color[rgb]{0,0,0}$1$}%
}}}
\put(-8999,-4336){\makebox(0,0)[b]{\smash{\fontsize{24}{28.8}\usefont{T1}{ptm}{m}{n}{\color[rgb]{0,0,0}$2$}%
}}}
\put(-10799,-4336){\makebox(0,0)[b]{\smash{\fontsize{24}{28.8}\usefont{T1}{ptm}{m}{n}{\color[rgb]{0,0,0}$\beta^2 \frac{\pi_-}{\pi_+}$}%
}}}
\put(676,-511){\makebox(0,0)[rb]{\smash{\fontsize{24}{28.8}\usefont{T1}{ptm}{m}{n}{\color[rgb]{0,0,0}$P_2$}%
}}}
\end{picture}%

%% file: drawings/distri_4_tau_Fbeta_Re_Lt_Gt.pdf_t
\begin{picture}(0,0)%
\includegraphics{distri_4_tau_Fbeta_Re_Lt_Gt.pdf}%
\end{picture}%
\setlength{\unitlength}{4144sp}%
\begin{picture}(17580,6709)(-11939,-4454)
\put(2611,-151){\makebox(0,0)[rb]{\smash{\fontsize{24}{28.8}\usefont{T1}{ptm}{m}{n}{\color[rgb]{0,0,0}$P_1$}%
}}}
\put(5626,-3661){\makebox(0,0)[lb]{\smash{\fontsize{24}{28.8}\usefont{T1}{ptm}{m}{n}{\color[rgb]{0,0,0}$x$}%
}}}
\put(2251,1064){\makebox(0,0)[b]{\smash{\fontsize{24}{28.8}\usefont{T1}{ptm}{m}{n}{\color[rgb]{0,0,0}ROC space}%
}}}
\put(  1,1964){\makebox(0,0)[b]{\smash{\fontsize{24}{28.8}\usefont{T1}{ptm}{m}{n}{\color[rgb]{0,0,0}$y$}%
}}}
\put(-11924,-3661){\makebox(0,0)[rb]{\smash{\fontsize{24}{28.8}\usefont{T1}{ptm}{m}{n}{\color[rgb]{0,0,0}$\ell$}%
}}}
\put(  1,-4336){\makebox(0,0)[b]{\smash{\fontsize{24}{28.8}\usefont{T1}{ptm}{m}{n}{\color[rgb]{0,0,0}$(0,0,0)$}%
}}}
\put(-449,839){\makebox(0,0)[rb]{\smash{\fontsize{24}{28.8}\usefont{T1}{ptm}{m}{n}{\color[rgb]{0,0,0}$1$}%
}}}
\put(4501,-4336){\makebox(0,0)[b]{\smash{\fontsize{24}{28.8}\usefont{T1}{ptm}{m}{n}{\color[rgb]{0,0,0}$1$}%
}}}
\put(-4499,-4336){\makebox(0,0)[b]{\smash{\fontsize{24}{28.8}\usefont{T1}{ptm}{m}{n}{\color[rgb]{0,0,0}$1$}%
}}}
\put(-8999,-4336){\makebox(0,0)[b]{\smash{\fontsize{24}{28.8}\usefont{T1}{ptm}{m}{n}{\color[rgb]{0,0,0}$2$}%
}}}
\put(-10799,-4336){\makebox(0,0)[b]{\smash{\fontsize{24}{28.8}\usefont{T1}{ptm}{m}{n}{\color[rgb]{0,0,0}$\beta^2 \frac{\pi_-}{\pi_+}$}%
}}}
\put(676,-511){\makebox(0,0)[rb]{\smash{\fontsize{24}{28.8}\usefont{T1}{ptm}{m}{n}{\color[rgb]{0,0,0}$P_2$}%
}}}
\end{picture}%

%% file: figs/detailed_results_distri_IV.tex
\begin{figure*}
    \begin{centering}
        \hfill
        \subfloat[$\scoreFOne$, the traditional (balanced) F-score, is not the optimal tradeoff: $\kendall(\scorePrecision; \scoreFOne) \ne \kendall(\scoreFOne; \scoreRecall)$.]{
            \includegraphics[scale=0.5]{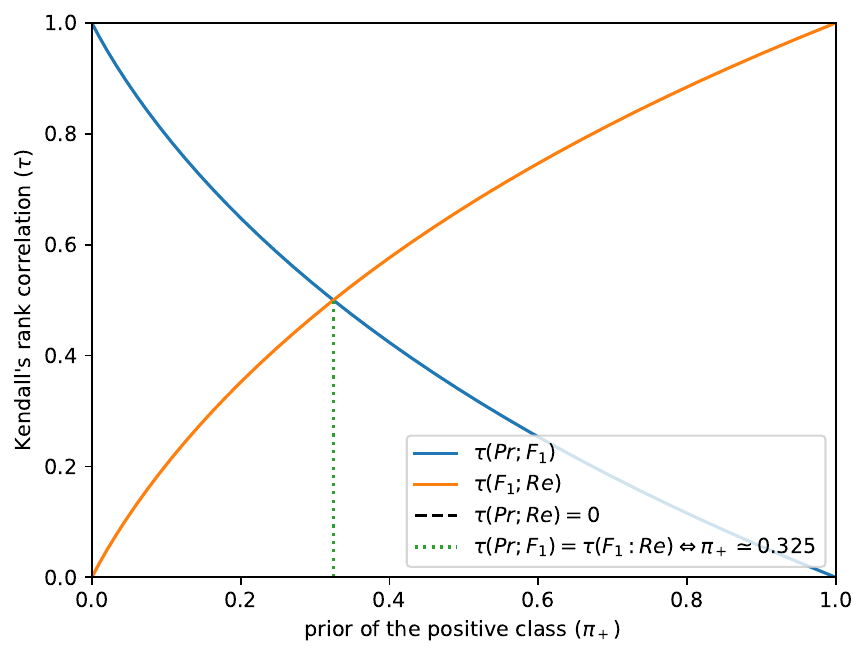}
        }
        \hfill
        \hfill
        \subfloat[$\scoreSkewInsensitiveVersionFOne$, the skew-insensitive version of $\scoreFOne$~\cite{Flach2003TheGeometry}, is not the optimal tradeoff: $\kendall(\scorePrecision; \scoreSkewInsensitiveVersionFOne) \ne \kendall(\scoreSkewInsensitiveVersionFOne; \scoreRecall)$.]{
            \includegraphics[scale=0.5]{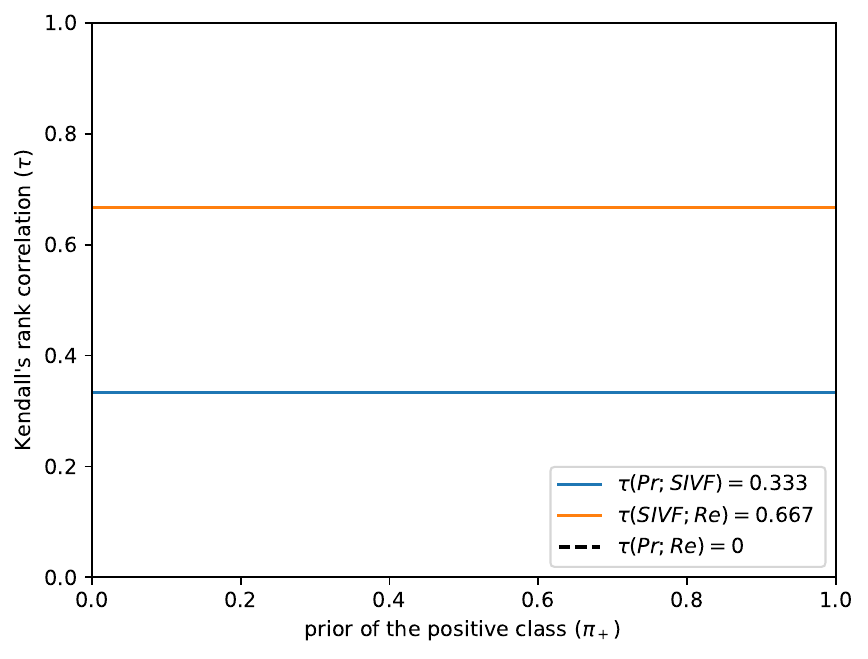}
        }
        \hfill
    \par\end{centering}
    
    \begin{centering}
        \hfill
        \subfloat[Fréchet variance for various priors. It is defined at \cref{eq:frechet-variance} and should be minimized to obtain the optimal tradeoff.]{
            \includegraphics[scale=0.5]{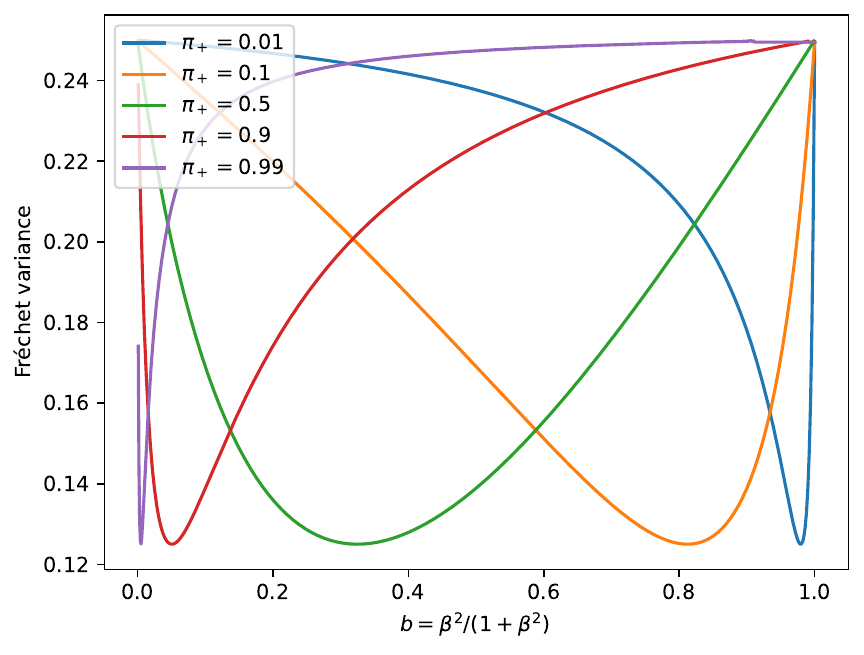}
        }
        \hfill
        \hfill
        \subfloat[$\scoreOptimalTradeoff$ if the optimal tradeoff: $\kendall(\scorePrecision; \scoreOptimalTradeoff) = \kendall(\scoreOptimalTradeoff; \scoreRecall)$.]{
            \includegraphics[scale=0.5]{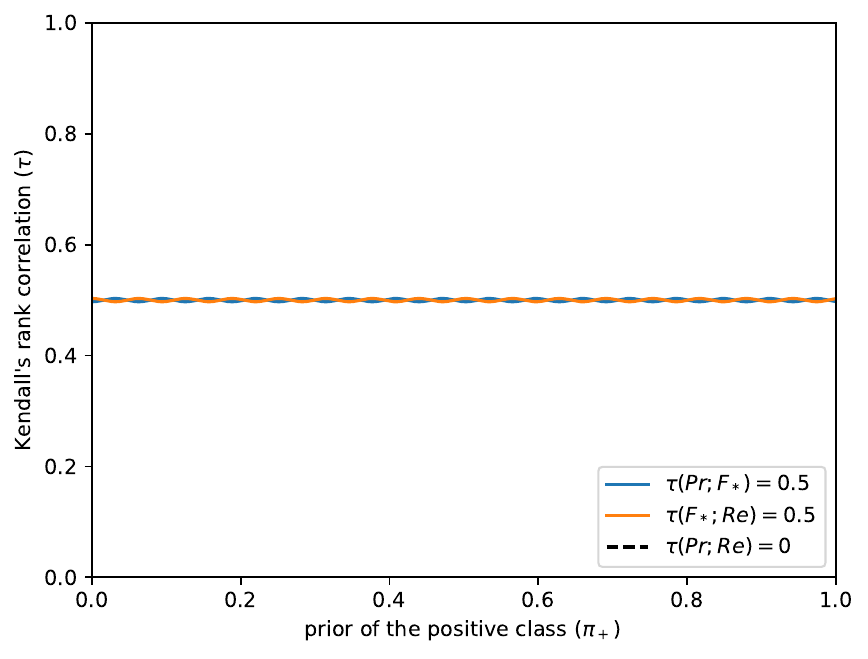}
        }
        \hfill
    \par\end{centering}
    
    \begin{centering}
        \hfill
        \subfloat[Adaptation of $\beta$ \wrt class priors.]{
            \includegraphics[scale=0.5]{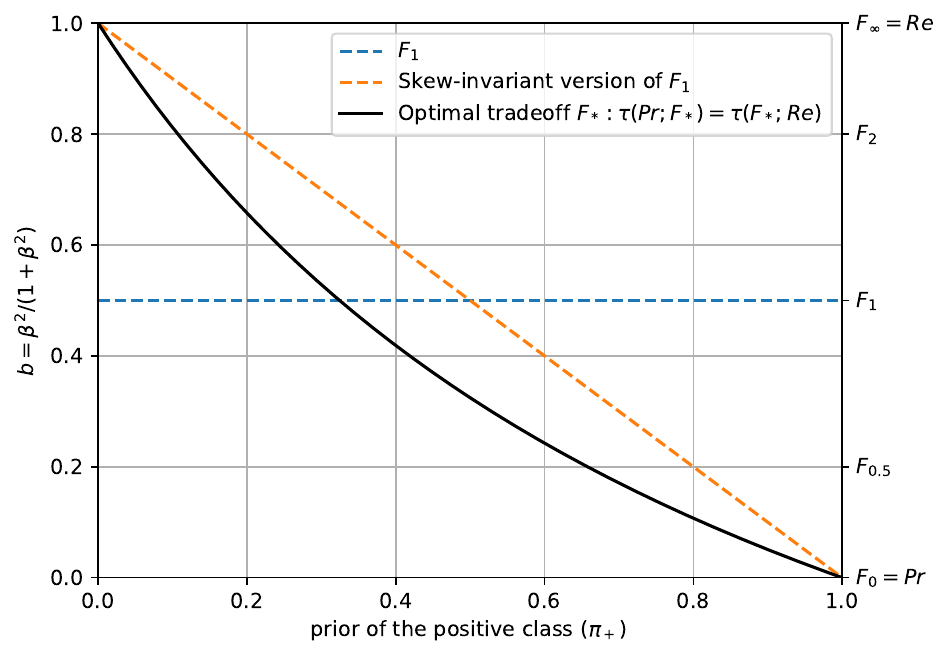}
        }
        \hfill
        \hfill
        \subfloat[\pca of the manifold (for $\priorpos=0.1$).]{
            \includegraphics[scale=0.5]{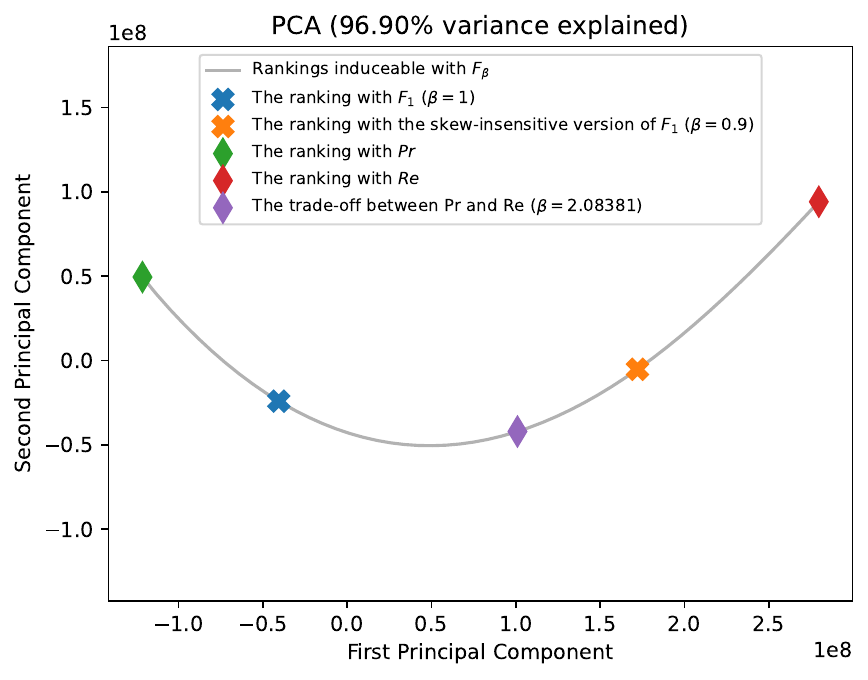}
        }
        \hfill
    \par\end{centering}
        
    \caption{
        Results for uniform distributions over the performances with fixed class priors, \ie $\setIV$.
        \label{fig:detailed_results_distri_IV}
    }
\end{figure*}

%% file: sections/A_3_CS5_detailed_distribution_V.tex
\subsubsection{Detailed Results for the Uniform Distributions With Fixed Class Priors, Close to Oracle}

To find the probabilities necessary to compute $\kendall$, we integrate over a rectangle in the \roc space. Let us denote by $(\scoreFPR,\scoreTPR)=(x,y)$ the coordinates of a performance in this space. We have
\begin{equation}
    \scorePrecision(\aPerformance_1) < \scorePrecision(\aPerformance_2) \Leftrightarrow \frac{y_1}{x_1} < \frac{y_2}{x_2}
\end{equation}
\begin{equation}
    \scoreRecall(\aPerformance_1) < \scoreRecall(\aPerformance_2) \Leftrightarrow y_1 < y_2
\end{equation}
\begin{equation}
    \scoreFBeta(\aPerformance_1) < \scoreFBeta(\aPerformance_2) \Leftrightarrow \frac{y_1}{x_1+\ell} < \frac{y_2}{x_2+\ell}
    \qquad
    \textrm{with}
    \qquad
    \ell = \beta^2 \frac{\priorpos}{\priorneg} = \frac{b}{1-b} \frac{\priorpos}{1-\priorpos}
\end{equation}

Unlike the previous cases, here we can no longer group $\priorpos$ and $\beta$ together to form $\ell$, as $\priorpos$ appears alone in the integration bounds of $x$ and $y$. More precisely, we cannot anymore minimize Fréchet variance as a function of just $\ell$. The optimal $\ell$ is not a constant anymore: it is now a function of $\priorpos$. This explains why the look of the curve for the adaptation differs from what we had before.

We start with a few initializations.

\begin{lstlisting}[backgroundcolor = \color{lightgray},language=Mathematica,numbers=left,numberstyle={\small},breaklines=true]
tot = Integrate[
    1, 
    {x1, 0, p}, {y1, p, 1}, 
    {x2, 0, p}, {y2, p, 1},
	Assumptions -> 0 < p < 1
]
\end{lstlisting}

We have
\begin{equation}
    \kendall(\scorePrecision; \scoreRecall) = 1-\frac{-\priorpos^4+2 \priorpos^4 \log (\priorpos)+\priorpos^2}{2 (1-\priorpos)^2 \priorpos^2}
    \comma
\end{equation}
as shown with the code (see \cref{fig:distri_5_tau_Pr_Re_Lt_Gt}):
\begin{lstlisting}[backgroundcolor = \color{lightgray},language=Mathematica,numbers=left,numberstyle={\small},breaklines=true]
pLtGt = Integrate[
	Boole[(y1 x2 < y2 x1) && (y1 > y2)], 
    {x1, 0, p}, {y1, p, 1}, 
    {x2, 0, p}, {y2, p, 1},
	Assumptions -> 0 < p < 1
] / tot
tau = 1 - 4 * pLtGt
\end{lstlisting}

With Wolfram, we were unable to obtain the analytical expression of $\kendall(\scorePrecision; \scoreFBeta)$ and $\kendall(\scoreFBeta; \scoreRecall)$ for this family of distributions. 
Wolfram was unable to provide you with the analytical expression in a reasonable time. 
We therefore report the results obtained using the Monte-Carlo technique.

\input{figs/distri_V_proof}

\input{figs/detailed_results_distri_V}
The plots are provided in \cref{fig:detailed_results_distri_V}.

%% file: figs/distri_V_proof.tex
\begin{figure*}[p]
\begin{centering}
    \subfloat[To compute $\kendall(\scorePrecision; \scoreRecall)$ with \cref{{eq:tau-from-pLtGt}}, we determine the probability that $\scorePrecision(\aPerformance_1)<\scorePrecision(\aPerformance_2)$ (the performance $\aPerformance_1$ is under the blue line) and $\scoreRecall(\aPerformance_1)>\scoreRecall(\aPerformance_2)$ (the performance $\aPerformance_1$ is above the green line), which is the pink area averaged for all positions of $\aPerformance_2$.\label{fig:distri_5_tau_Pr_Re_Lt_Gt}]{
        \resizebox{0.8\linewidth}{!}{
                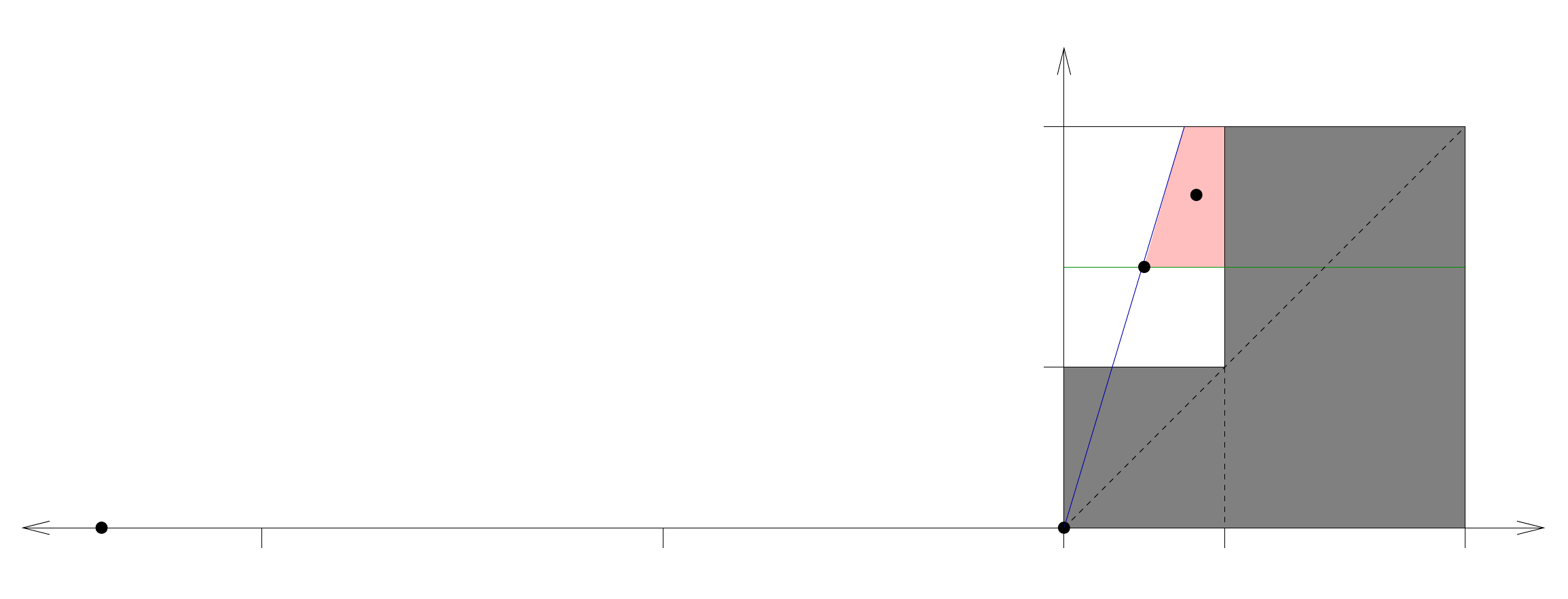
        }
    }
\par\end{centering}
\begin{centering}
    \subfloat[To compute $\kendall(\scorePrecision; \scoreFBeta)$ with \cref{{eq:tau-from-pLtGt}}, we determine the probability that $\scorePrecision(\aPerformance_1)<\scorePrecision(\aPerformance_2)$ (the performance $\aPerformance_1$ is under the blue line) and $\scoreFBeta(\aPerformance_1)>\scoreFBeta(\aPerformance_2)$ (the performance $\aPerformance_1$ is above the red line), which is the pink area averaged for all positions of $\aPerformance_2$.\label{fig:distri_5_tau_Pr_Fbeta_Lt_Gt}]{
        \resizebox{0.8\linewidth}{!}{
            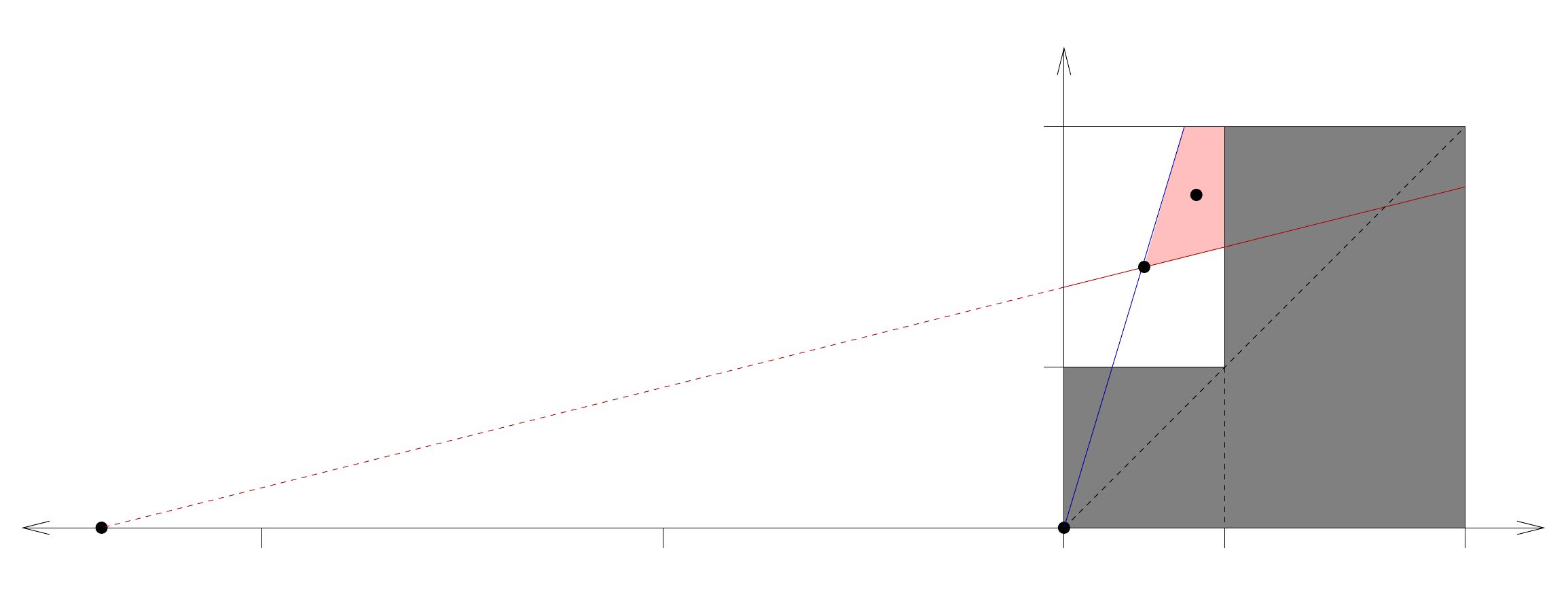
        }
    }
\par\end{centering}
\begin{centering}
    \subfloat[To compute $\kendall(\scoreFBeta; \scoreRecall)$ with \cref{{eq:tau-from-pLtGt}}, we determine the probability that $\scoreFBeta(\aPerformance_1)<\scoreFBeta(\aPerformance_2)$ (the performance $\aPerformance_1$ is under the red line) and $\scoreRecall(\aPerformance_1)>\scoreRecall(\aPerformance_2)$ (the performance $\aPerformance_1$ is above the green line), which is the pink area averaged for all positions of $\aPerformance_2$.\label{fig:distri_5_tau_Fbeta_Re_Lt_Gt}]{
        \resizebox{0.8\linewidth}{!}{
            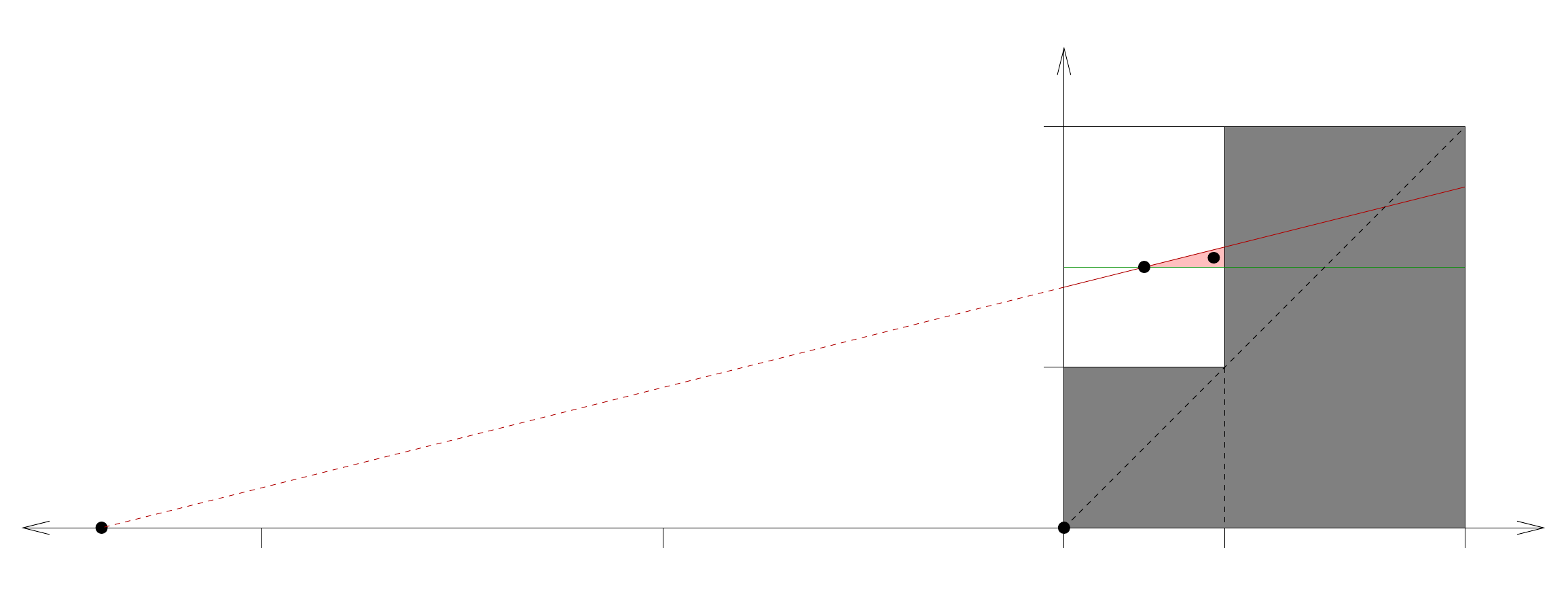
        }
    }
\par\end{centering}
\caption{Graphical representation, in and around the \roc space, of the principle that we use to derive the analytical expression for the optimal tradeoff between precision and recall for the purpose of ranking, in the case of uniform distributions with fixed class priors close to oracle. Note that the unshaded area of \roc space is a linear mapping of the squares depicted in \cref{fig:setV}.}

\end{figure*}

%% file: drawings/distri_5_tau_Pr_Re_Lt_Gt.pdf_t
\begin{picture}(0,0)%
\includegraphics{distri_5_tau_Pr_Re_Lt_Gt.pdf}%
\end{picture}%
\setlength{\unitlength}{4144sp}%
\begin{picture}(17580,6709)(-11939,-4454)
\put(1261,299){\makebox(0,0)[rb]{\smash{\fontsize{24}{28.8}\usefont{T1}{ptm}{m}{n}{\color[rgb]{0,0,0}$P_1$}%
}}}
\put(5626,-3661){\makebox(0,0)[lb]{\smash{\fontsize{24}{28.8}\usefont{T1}{ptm}{m}{n}{\color[rgb]{0,0,0}$x$}%
}}}
\put(2251,1064){\makebox(0,0)[b]{\smash{\fontsize{24}{28.8}\usefont{T1}{ptm}{m}{n}{\color[rgb]{0,0,0}ROC space}%
}}}
\put(  1,1964){\makebox(0,0)[b]{\smash{\fontsize{24}{28.8}\usefont{T1}{ptm}{m}{n}{\color[rgb]{0,0,0}$y$}%
}}}
\put(-11924,-3661){\makebox(0,0)[rb]{\smash{\fontsize{24}{28.8}\usefont{T1}{ptm}{m}{n}{\color[rgb]{0,0,0}$\ell$}%
}}}
\put(  1,-4336){\makebox(0,0)[b]{\smash{\fontsize{24}{28.8}\usefont{T1}{ptm}{m}{n}{\color[rgb]{0,0,0}$(0,0,0)$}%
}}}
\put(-449,839){\makebox(0,0)[rb]{\smash{\fontsize{24}{28.8}\usefont{T1}{ptm}{m}{n}{\color[rgb]{0,0,0}$1$}%
}}}
\put(4501,-4336){\makebox(0,0)[b]{\smash{\fontsize{24}{28.8}\usefont{T1}{ptm}{m}{n}{\color[rgb]{0,0,0}$1$}%
}}}
\put(-4499,-4336){\makebox(0,0)[b]{\smash{\fontsize{24}{28.8}\usefont{T1}{ptm}{m}{n}{\color[rgb]{0,0,0}$1$}%
}}}
\put(-8999,-4336){\makebox(0,0)[b]{\smash{\fontsize{24}{28.8}\usefont{T1}{ptm}{m}{n}{\color[rgb]{0,0,0}$2$}%
}}}
\put(1801,-4336){\makebox(0,0)[b]{\smash{\fontsize{24}{28.8}\usefont{T1}{ptm}{m}{n}{\color[rgb]{0,0,0}$\pi_+$}%
}}}
\put(-449,-1861){\makebox(0,0)[rb]{\smash{\fontsize{24}{28.8}\usefont{T1}{ptm}{m}{n}{\color[rgb]{0,0,0}$\pi_+$}%
}}}
\put(-10799,-4336){\makebox(0,0)[b]{\smash{\fontsize{24}{28.8}\usefont{T1}{ptm}{m}{n}{\color[rgb]{0,0,0}$\beta^2 \frac{\pi_-}{\pi_+}$}%
}}}
\put(676,-511){\makebox(0,0)[rb]{\smash{\fontsize{24}{28.8}\usefont{T1}{ptm}{m}{n}{\color[rgb]{0,0,0}$P_2$}%
}}}
\end{picture}%

%% file: drawings/distri_5_tau_Pr_Fbeta_Lt_Gt.pdf_t
\begin{picture}(0,0)%
\includegraphics{distri_5_tau_Pr_Fbeta_Lt_Gt.pdf}%
\end{picture}%
\setlength{\unitlength}{4144sp}%
\begin{picture}(17580,6709)(-11939,-4454)
\put(1261,299){\makebox(0,0)[rb]{\smash{\fontsize{24}{28.8}\usefont{T1}{ptm}{m}{n}{\color[rgb]{0,0,0}$P_1$}%
}}}
\put(5626,-3661){\makebox(0,0)[lb]{\smash{\fontsize{24}{28.8}\usefont{T1}{ptm}{m}{n}{\color[rgb]{0,0,0}$x$}%
}}}
\put(2251,1064){\makebox(0,0)[b]{\smash{\fontsize{24}{28.8}\usefont{T1}{ptm}{m}{n}{\color[rgb]{0,0,0}ROC space}%
}}}
\put(  1,1964){\makebox(0,0)[b]{\smash{\fontsize{24}{28.8}\usefont{T1}{ptm}{m}{n}{\color[rgb]{0,0,0}$y$}%
}}}
\put(-11924,-3661){\makebox(0,0)[rb]{\smash{\fontsize{24}{28.8}\usefont{T1}{ptm}{m}{n}{\color[rgb]{0,0,0}$\ell$}%
}}}
\put(  1,-4336){\makebox(0,0)[b]{\smash{\fontsize{24}{28.8}\usefont{T1}{ptm}{m}{n}{\color[rgb]{0,0,0}$(0,0,0)$}%
}}}
\put(-449,839){\makebox(0,0)[rb]{\smash{\fontsize{24}{28.8}\usefont{T1}{ptm}{m}{n}{\color[rgb]{0,0,0}$1$}%
}}}
\put(4501,-4336){\makebox(0,0)[b]{\smash{\fontsize{24}{28.8}\usefont{T1}{ptm}{m}{n}{\color[rgb]{0,0,0}$1$}%
}}}
\put(-4499,-4336){\makebox(0,0)[b]{\smash{\fontsize{24}{28.8}\usefont{T1}{ptm}{m}{n}{\color[rgb]{0,0,0}$1$}%
}}}
\put(-8999,-4336){\makebox(0,0)[b]{\smash{\fontsize{24}{28.8}\usefont{T1}{ptm}{m}{n}{\color[rgb]{0,0,0}$2$}%
}}}
\put(1801,-4336){\makebox(0,0)[b]{\smash{\fontsize{24}{28.8}\usefont{T1}{ptm}{m}{n}{\color[rgb]{0,0,0}$\pi_+$}%
}}}
\put(-449,-1861){\makebox(0,0)[rb]{\smash{\fontsize{24}{28.8}\usefont{T1}{ptm}{m}{n}{\color[rgb]{0,0,0}$\pi_+$}%
}}}
\put(-10799,-4336){\makebox(0,0)[b]{\smash{\fontsize{24}{28.8}\usefont{T1}{ptm}{m}{n}{\color[rgb]{0,0,0}$\beta^2 \frac{\pi_-}{\pi_+}$}%
}}}
\put(676,-511){\makebox(0,0)[rb]{\smash{\fontsize{24}{28.8}\usefont{T1}{ptm}{m}{n}{\color[rgb]{0,0,0}$P_2$}%
}}}
\end{picture}%

%% file: drawings/distri_5_tau_Fbeta_Re_Lt_Gt.pdf_t
\begin{picture}(0,0)%
\includegraphics{distri_5_tau_Fbeta_Re_Lt_Gt.pdf}%
\end{picture}%
\setlength{\unitlength}{4144sp}%
\begin{picture}(17580,6709)(-11939,-4454)
\put(1456,-406){\makebox(0,0)[rb]{\smash{\fontsize{24}{28.8}\usefont{T1}{ptm}{m}{n}{\color[rgb]{0,0,0}$P_1$}%
}}}
\put(5626,-3661){\makebox(0,0)[lb]{\smash{\fontsize{24}{28.8}\usefont{T1}{ptm}{m}{n}{\color[rgb]{0,0,0}$x$}%
}}}
\put(2251,1064){\makebox(0,0)[b]{\smash{\fontsize{24}{28.8}\usefont{T1}{ptm}{m}{n}{\color[rgb]{0,0,0}ROC space}%
}}}
\put(  1,1964){\makebox(0,0)[b]{\smash{\fontsize{24}{28.8}\usefont{T1}{ptm}{m}{n}{\color[rgb]{0,0,0}$y$}%
}}}
\put(-11924,-3661){\makebox(0,0)[rb]{\smash{\fontsize{24}{28.8}\usefont{T1}{ptm}{m}{n}{\color[rgb]{0,0,0}$\ell$}%
}}}
\put(  1,-4336){\makebox(0,0)[b]{\smash{\fontsize{24}{28.8}\usefont{T1}{ptm}{m}{n}{\color[rgb]{0,0,0}$(0,0,0)$}%
}}}
\put(-449,839){\makebox(0,0)[rb]{\smash{\fontsize{24}{28.8}\usefont{T1}{ptm}{m}{n}{\color[rgb]{0,0,0}$1$}%
}}}
\put(4501,-4336){\makebox(0,0)[b]{\smash{\fontsize{24}{28.8}\usefont{T1}{ptm}{m}{n}{\color[rgb]{0,0,0}$1$}%
}}}
\put(-4499,-4336){\makebox(0,0)[b]{\smash{\fontsize{24}{28.8}\usefont{T1}{ptm}{m}{n}{\color[rgb]{0,0,0}$1$}%
}}}
\put(-8999,-4336){\makebox(0,0)[b]{\smash{\fontsize{24}{28.8}\usefont{T1}{ptm}{m}{n}{\color[rgb]{0,0,0}$2$}%
}}}
\put(1801,-4336){\makebox(0,0)[b]{\smash{\fontsize{24}{28.8}\usefont{T1}{ptm}{m}{n}{\color[rgb]{0,0,0}$\pi_+$}%
}}}
\put(-449,-1861){\makebox(0,0)[rb]{\smash{\fontsize{24}{28.8}\usefont{T1}{ptm}{m}{n}{\color[rgb]{0,0,0}$\pi_+$}%
}}}
\put(-10799,-4336){\makebox(0,0)[b]{\smash{\fontsize{24}{28.8}\usefont{T1}{ptm}{m}{n}{\color[rgb]{0,0,0}$\beta^2 \frac{\pi_-}{\pi_+}$}%
}}}
\put(676,-511){\makebox(0,0)[rb]{\smash{\fontsize{24}{28.8}\usefont{T1}{ptm}{m}{n}{\color[rgb]{0,0,0}$P_2$}%
}}}
\end{picture}%

%% file: figs/detailed_results_distri_V.tex
\begin{figure*}
    \begin{centering}
        \hfill
        \subfloat[$\scoreFOne$, the traditional (balanced) F-score, is not the optimal tradeoff: $\kendall(\scorePrecision; \scoreFOne) \ne \kendall(\scoreFOne; \scoreRecall)$.]{
            \includegraphics[scale=0.5]{images/distri_5_f1_is_not_between_ppv_tpr.pdf}
        }
        \hfill
        \hfill
        \subfloat[$\scoreSkewInsensitiveVersionFOne$, the skew-insensitive version of $\scoreFOne$~\cite{Flach2003TheGeometry}, is not the optimal tradeoff: $\kendall(\scorePrecision; \scoreSkewInsensitiveVersionFOne) \ne \kendall(\scoreSkewInsensitiveVersionFOne; \scoreRecall)$.]{
            \includegraphics[scale=0.5]{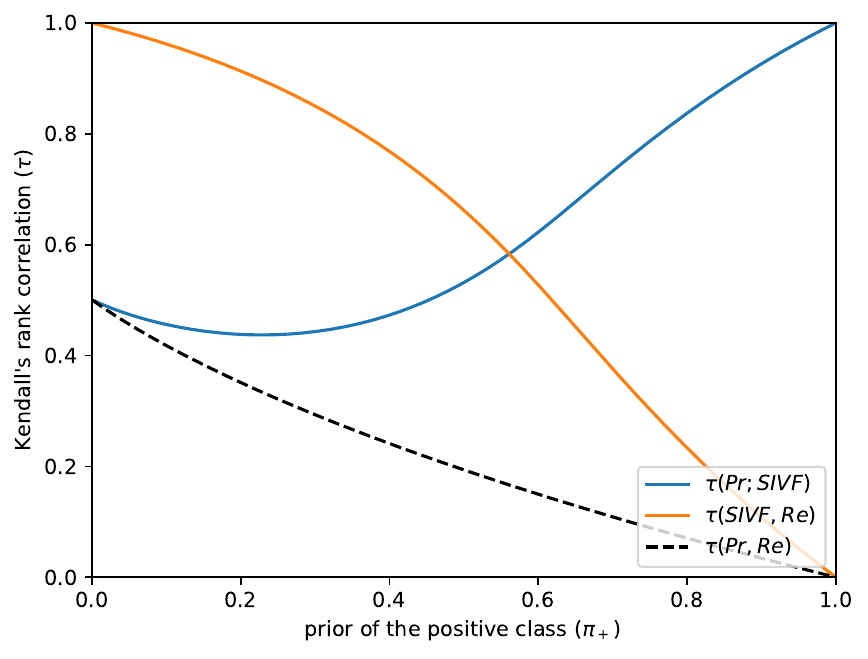}
        }
        \hfill
    \par\end{centering}
    
    \begin{centering}
        \hfill
        \subfloat[Fréchet variance for various priors. It is defined at \cref{eq:frechet-variance} and should be minimized to obtain the optimal tradeoff.]{
            \includegraphics[scale=0.5]{images/distri_5_frechet_variance.pdf}
        }
        \hfill
        \hfill
        \subfloat[$\scoreOptimalTradeoff$ if the optimal tradeoff: $\kendall(\scorePrecision; \scoreOptimalTradeoff) = \kendall(\scoreOptimalTradeoff; \scoreRecall)$.]{
            \includegraphics[scale=0.5]{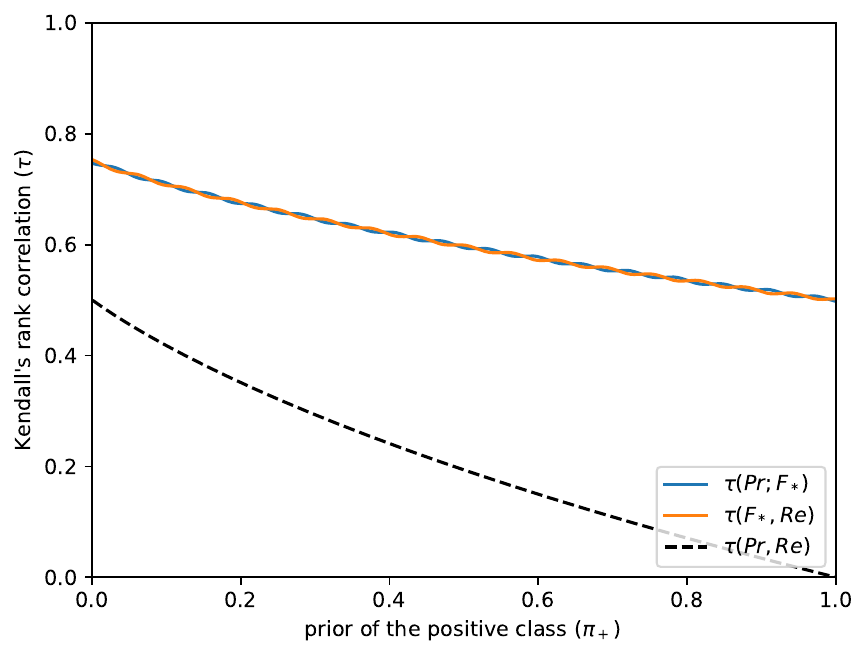}
        }
        \hfill
    \par\end{centering}
    
    \begin{centering}
        \hfill
        \subfloat[Adaptation of $\beta$ \wrt class priors.]{
            \includegraphics[scale=0.5]{images/distri_5_adaptation.pdf}
        }
        \hfill
        \hfill
        \subfloat[\pca of the manifold (for $\priorpos=0.1$).]{
            \includegraphics[scale=0.5]{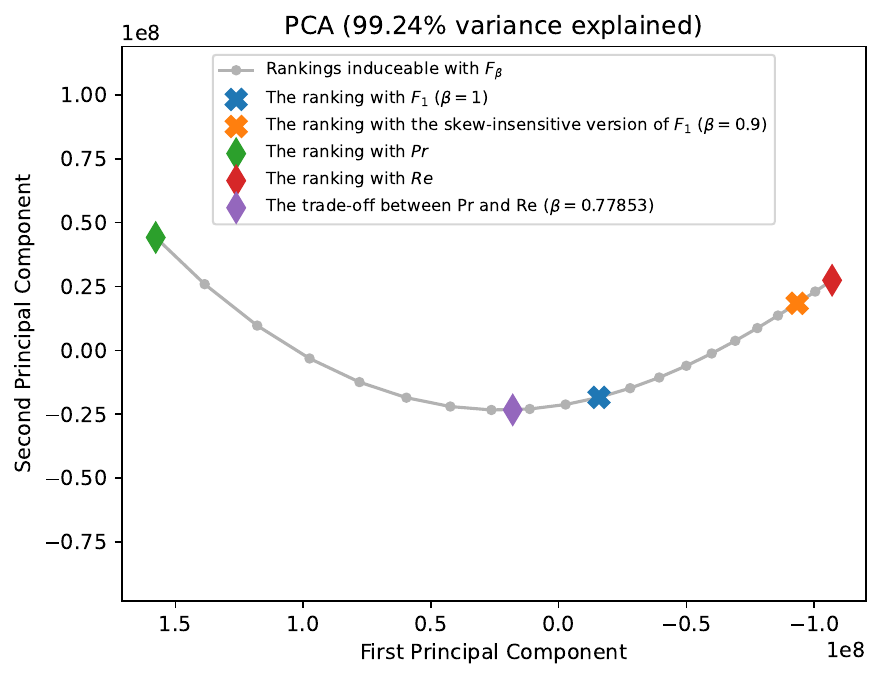}
        }
        \hfill
    \par\end{centering}
        
    \caption{
        Results for uniform distributions over the performances with fixed class priors, \ie $\setV$.
        \label{fig:detailed_results_distri_V}
    }
\end{figure*}

%% file: sections/A_3_CS6_all_results_CDnet.tex
\subsubsection{All Results for Some Real Sets of Performances}

\newcommand{\itemResultCDnet}[4]{
    \item The results for video "#2" (ranking of #4 methods; $\priorpos=#3$) can be found at \cref{fig:cdnet_2014_#1}.
}

We provide here the detailed results for each video of \CDnetMMXIV.

\paragraph{Category "Baseline"}
\begin{itemize}
    \itemResultCDnet{00}{pedestrians}{0.0098}{57}
    \itemResultCDnet{01}{PETS2006}{0.0130}{57}
    \itemResultCDnet{02}{office}{0.0690}{57}
    \itemResultCDnet{03}{highway}{0.0593}{57}
\end{itemize}

\paragraph{Category "Dynamic Background"}
\begin{itemize}
    \itemResultCDnet{04}{overpass}{0.0134}{57}
    \itemResultCDnet{05}{canoe}{0.0354}{57}
    \itemResultCDnet{06}{fountain01}{0.0008}{57}
    \itemResultCDnet{07}{fountain02}{0.0022}{57}
    \itemResultCDnet{08}{fall}{0.0177}{57}
    \itemResultCDnet{09}{boats}{0.0063}{57}
\end{itemize}

\paragraph{Category "Camera Jitter"}
\begin{itemize}
    \itemResultCDnet{10}{boulevard}{0.0469}{57}
    \itemResultCDnet{11}{sidewalk}{0.0261}{57}
    \itemResultCDnet{12}{badminton}{0.0343}{57}
    \itemResultCDnet{13}{traffic}{0.0623}{57}
\end{itemize}

\paragraph{Category "Intermittent Object Motion"}
\begin{itemize}
    \itemResultCDnet{14}{abandonedBox}{0.0481}{57}
    \itemResultCDnet{15}{winterDriveway}{0.0075}{57}
    \itemResultCDnet{16}{sofa}{0.0437}{57}
    \itemResultCDnet{17}{tramstop}{0.1795}{57}
    \itemResultCDnet{18}{parking}{0.0773}{57}
    \itemResultCDnet{19}{streetLight}{0.0485}{57}
\end{itemize}

\paragraph{Category "Shadow"}
\begin{itemize}
    \itemResultCDnet{20}{copyMachine}{0.0693}{57}
    \itemResultCDnet{21}{bungalows}{0.0600}{57}
    \itemResultCDnet{22}{busStation}{0.0369}{57}
    \itemResultCDnet{23}{peopleInShade}{0.0564}{57}
    \itemResultCDnet{24}{backdoor}{0.0199}{57}
    \itemResultCDnet{25}{cubicle}{0.0196}{57}
\end{itemize}

\paragraph{Category "Thermal"}
\begin{itemize}
    \itemResultCDnet{26}{lakeSide}{0.0192}{57}
    \itemResultCDnet{27}{diningRoom}{0.0859}{57}
    \itemResultCDnet{28}{park}{0.0203}{57}
    \itemResultCDnet{29}{corridor}{0.0331}{57}
    \itemResultCDnet{30}{library}{0.1928}{57}
\end{itemize}

\paragraph{Category "Bad Weather"}
\begin{itemize}
    \itemResultCDnet{31}{skating}{0.0397}{57}
    \itemResultCDnet{32}{wetSnow}{0.0150}{57}
    \itemResultCDnet{33}{snowFall}{0.0105}{57}
    \itemResultCDnet{34}{blizzard}{0.0115}{57}
\end{itemize}

\paragraph{Category "Low Framerate"}
\begin{itemize}
    \itemResultCDnet{35}{tunnelExit\_0\_35fps}{0.0195}{57}
    \itemResultCDnet{36}{port\_0\_17fps}{0.0002}{57}
    \itemResultCDnet{37}{tramCrossroad\_1fps}{0.0288}{57}
    \itemResultCDnet{38}{turnpike\_0\_5fps}{0.0581}{57}
\end{itemize}

\paragraph{Category "Night Videos"}
\begin{itemize}
    \itemResultCDnet{39}{tramStation}{0.0339}{58}
    \itemResultCDnet{40}{busyBoulvard}{0.0847}{58}
    \itemResultCDnet{41}{streetCornerAtNight}{0.0062}{58}
    \itemResultCDnet{42}{fluidHighway}{0.0175}{58}
    \itemResultCDnet{43}{bridgeEntry}{0.0200}{58}
    \itemResultCDnet{44}{winterStreet}{0.0689}{58}
\end{itemize}

\paragraph{Category "PTZ"}
\begin{itemize}
    \itemResultCDnet{45}{twoPositionPTZCam}{0.0117}{56}
    \itemResultCDnet{46}{zoomInZoomOut}{0.0019}{56}
    \itemResultCDnet{47}{continuousPan}{0.0056}{56}
    \itemResultCDnet{48}{intermittentPan}{0.0094}{56}
\end{itemize}

\paragraph{Category "Turbulence"}
\begin{itemize}
    \itemResultCDnet{49}{turbulence2}{0.0008}{57}
    \itemResultCDnet{50}{turbulence3}{0.0121}{57}
    \itemResultCDnet{51}{turbulence0}{0.0015}{57}
    \itemResultCDnet{52}{turbulence1}{0.0026}{57}
\end{itemize}

\newcommand{\displayResultCDnet}[4]{
    \begin{figure}[p!]
        \begin{centering}
            \hfill
            \subfloat[The performances of #4 classifiers (\bgs methods) depicted as points in the \roc space, with the isometrics of the optimal tradeoff score, from the ranking point of view, between precision and recall. See \cref{eq:closed-form-expression}.]{
                \begin{minipage}[t]{0.45\linewidth}%
                    \begin{center}
                        \includegraphics[scale=0.5]{images/cdnet_2014_#1_#2_roc.pdf}
                    \par\end{center}%
                \end{minipage}
            }
            \hfill
            \hfill
            \subfloat[The rank correlations $\kendall(\scorePrecision;\scoreFBeta)$ and $\kendall(\scoreFBeta;\scoreRecall)$ \wrt $\beta$. The optimal value (or range of optimal values) for $\beta$ is where the two curves intersect.]{
                \begin{minipage}[t]{0.45\linewidth}%
                    \begin{center}
                        \includegraphics[scale=0.5]{images/cdnet_2014_#1_#2_correlations.pdf}
                    \par\end{center}%
                \end{minipage}
            }
            \hfill
        \par\end{centering}
        
        \begin{centering}
            \hfill
            \subfloat[The ranks of each classifier \wrt $\beta$. The optimal value (or range of optimal values) for $\beta$, shown here by the vertical line, is such that the number of swaps on its left is equal to the number of swaps on its right.]{
                \begin{minipage}[t]{0.45\linewidth}%
                    \begin{center}
                        \includegraphics[scale=0.5]{images/cdnet_2014_#1_#2_rankings.pdf}
                    \par\end{center}%
                \end{minipage}
            }
            \hfill
            \hfill
            \subfloat[The Fréchet variance $\sigma^2(\beta) = \distKendall^2(\scorePrecision;\scoreFBeta) + \distKendall^2(\scoreFBeta ; \scoreRecall)$ \wrt $\beta$. The optimal value (or range of optimal values) for $\beta$ is where the curve has its minimum.]{
                \begin{minipage}[t]{0.45\linewidth}%
                    \begin{center}
                        \includegraphics[scale=0.5]{images/cdnet_2014_#1_#2_frechet_variance.pdf}
                    \par\end{center}%
                \end{minipage}
            }
            \hfill
        \par\end{centering}
        
        \begin{centering}
            \hfill
            \subfloat[Linear projection (\pca) of the manifold of the rankings induced by the $\scoreFBeta$ scores. The color points indicate the precision, the recall, $\scoreFOne$, $\scoreSkewInsensitiveVersionFOne$, as well as the optimal tradeoff. The optimal tradeoff is at the same distance of the two extremities when the distance is measured along the manifold, with Kendall's distance $\distKendall$.]{
                \begin{minipage}[t]{0.45\linewidth}%
                    \begin{center}
                        \includegraphics[scale=0.5]{images/cdnet_2014_#1_#2_pca.pdf}
                    \par\end{center}%
                \end{minipage}
            }
            \hfill
            \hfill
            \subfloat[The degree of optimality $\optimality(\beta)$ \wrt $\beta$. It is the probability to optimally ordering a pair of classifiers (\bgs methods) given that it is not trivial (\ie, that $\scorePrecision$ and $\scoreRecall$ are in contradiction). The optimal value (or range of optimal values) for $\beta$ is where the curve reaches $100\%$.]{
                \begin{minipage}[t]{0.45\linewidth}%
                    \begin{center}
                        \includegraphics[scale=0.5]{images/cdnet_2014_#1_#2_optimality}
                    \par\end{center}%
                \end{minipage}
            }
            \hfill
        \par\end{centering}
        
        \caption{
            Ranking of #4 \bgs methods evaluated on the video "#2" ($\priorpos=#3$).
            \label{fig:cdnet_2014_#1}
        }
    \end{figure}
}

\clearpage
\displayResultCDnet{00}{pedestrians}{0.0098}{57}
\displayResultCDnet{01}{PETS2006}{0.0130}{57}
\displayResultCDnet{02}{office}{0.0690}{57}
\displayResultCDnet{03}{highway}{0.0593}{57}

\clearpage
\displayResultCDnet{04}{overpass}{0.0134}{57}
\displayResultCDnet{05}{canoe}{0.0354}{57}
\displayResultCDnet{06}{fountain01}{0.0008}{57}
\displayResultCDnet{07}{fountain02}{0.0022}{57}
\displayResultCDnet{08}{fall}{0.0177}{57}
\displayResultCDnet{09}{boats}{0.0063}{57}

\clearpage
\displayResultCDnet{10}{boulevard}{0.0469}{57}
\displayResultCDnet{11}{sidewalk}{0.0261}{57}
\displayResultCDnet{12}{badminton}{0.0343}{57}
\displayResultCDnet{13}{traffic}{0.0623}{57}

\clearpage
\displayResultCDnet{14}{abandonedBox}{0.0481}{57}
\displayResultCDnet{15}{winterDriveway}{0.0075}{57}
\displayResultCDnet{16}{sofa}{0.0437}{57}
\displayResultCDnet{17}{tramstop}{0.1795}{57}
\displayResultCDnet{18}{parking}{0.0773}{57}
\displayResultCDnet{19}{streetLight}{0.0485}{57}

\clearpage
\displayResultCDnet{20}{copyMachine}{0.0693}{57}
\displayResultCDnet{21}{bungalows}{0.0600}{57}
\displayResultCDnet{22}{busStation}{0.0369}{57}
\displayResultCDnet{23}{peopleInShade}{0.0564}{57}
\displayResultCDnet{24}{backdoor}{0.0199}{57}
\displayResultCDnet{25}{cubicle}{0.0196}{57}

\clearpage
\displayResultCDnet{26}{lakeSide}{0.0192}{57}
\displayResultCDnet{27}{diningRoom}{0.0859}{57}
\displayResultCDnet{28}{park}{0.0203}{57}
\displayResultCDnet{29}{corridor}{0.0331}{57}
\displayResultCDnet{30}{library}{0.1928}{57}

\clearpage
\displayResultCDnet{31}{skating}{0.0397}{57}
\displayResultCDnet{32}{wetSnow}{0.0150}{57}
\displayResultCDnet{33}{snowFall}{0.0105}{57}
\displayResultCDnet{34}{blizzard}{0.0115}{57}

\clearpage
\displayResultCDnet{35}{tunnelExit\_0\_35fps}{0.0195}{57}
\displayResultCDnet{36}{port\_0\_17fps}{0.0002}{57}
\displayResultCDnet{37}{tramCrossroad\_1fps}{0.0288}{57}
\displayResultCDnet{38}{turnpike\_0\_5fps}{0.0581}{57}

\clearpage
\displayResultCDnet{39}{tramStation}{0.0339}{58}
\displayResultCDnet{40}{busyBoulvard}{0.0847}{58}
\displayResultCDnet{41}{streetCornerAtNight}{0.0062}{58}
\displayResultCDnet{42}{fluidHighway}{0.0175}{58}
\displayResultCDnet{43}{bridgeEntry}{0.0200}{58}
\displayResultCDnet{44}{winterStreet}{0.0689}{58}

\clearpage
\displayResultCDnet{45}{twoPositionPTZCam}{0.0117}{56}
\displayResultCDnet{46}{zoomInZoomOut}{0.0019}{56}
\displayResultCDnet{47}{continuousPan}{0.0056}{56}
\displayResultCDnet{48}{intermittentPan}{0.0094}{56}

\clearpage
\displayResultCDnet{49}{turbulence2}{0.0008}{57}
\displayResultCDnet{50}{turbulence3}{0.0121}{57}
\displayResultCDnet{51}{turbulence0}{0.0015}{57}
\displayResultCDnet{52}{turbulence1}{0.0026}{57}

%% file: sections/A_3_H_our_heuristic.tex
\subsubsection{More information on our simple heuristic for ROC users}

\newcommand{\summarizedPerformance}{\overline{\aPerformance}}

We provide here additional information about the simple heuristic introduced in \cref{sec:cdnet}. 
\begin{equation}
    \beta^2 = \frac{
        \expectedValueSymbol[\scorePFP]
    }{
        \expectedValueSymbol[\scorePFN]
    }
    \Leftrightarrow
    b = \frac{
        \expectedValueSymbol[\scorePFP]
    }{
        \expectedValueSymbol[\scorePFN] + \expectedValueSymbol[\scorePFP]
    }
    \label{eq:heuristic-with-expected-values}
\end{equation}
This heuristic was designed for the typical performances that researchers in the computer vision community (and more specifically those working on ``change detection'' and ``background subtraction'') have found useful to report in public benchmarks. For this community, we are fortunate to have a large number of domains (videos) in which a wide range of methods have been evaluated and their performances reported in public rankings. There is however no guarantee that this heuristic performs well in all cases. Nevertheless, as shown in \cref{tbl:summary}, our heuristic performs reasonably well for all the parametric distributions studied in this paper. It is therefore worth describing it in depth.

\input{figs/heuristic_in_ROC}

As shown in \cref{fig:heuristic-in-ROC}, it is possible to depict the heuristic in the Receiver Operating Characteristic (ROC) space, which is by definition $ROC=(\scoreFPR, \scoreTPR)$. We start by discussing the particular case in which the class priors are fixed, and then discuss the general case.
\begin{itemize}
    
    \item When the class priors $(\priorneg,\priorpos)$ are fixed, as $\scorePFP = \priorneg \, \scoreFPR$ and $\scorePFN = \priorpos \, \scoreFNR$, 
    our heuristic can be rewritten as
    \begin{equation}
        \beta^2 = \frac{
            \priorneg \, \expectedValueSymbol[\scoreFPR]
        }{
            \priorpos \, \expectedValueSymbol[\scoreFNR]
        }
        \Leftrightarrow
        b = \frac{
            \priorneg \, \expectedValueSymbol[\scoreFPR]
        }{
            \priorpos \, \expectedValueSymbol[\scoreFNR] + \priorneg \, \expectedValueSymbol[\scoreFPR]
        }
        \point
    \end{equation}
    Moreover, as $\scoreFNR = 1 - \scoreTPR$, we have
    \begin{equation}
        \beta^2 = \frac{
            \priorneg \, \expectedValueSymbol[\scoreFPR]
        }{
            \priorpos ( 1 - \expectedValueSymbol[\scoreTPR] )
        }
        \Leftrightarrow
        b = \frac{
            \priorneg \, \expectedValueSymbol[\scoreFPR]
        }{
            \priorpos ( 1 - \expectedValueSymbol[\scoreTPR] ) + \priorneg \, \expectedValueSymbol[\scoreFPR]
        }
        \point
        \label{eq:heuristic-in-roc-fixed-priors}
    \end{equation}
    This last form allows one to see the recommended $b$ (or $\beta$) as a function of $\expectedValueSymbol[\scoreFPR]$ and $\expectedValueSymbol[\scoreTPR]$, or in other words as a function of the position of the centroid in ROC.

    \item In the general case, it is also possible to use the same visualization of the heuristic in ROC. However, instead of looking at the centroid of the performance points, one should look at where the \emph{summarized performance} $\summarizedPerformance$ is projected. Let us remind that, when working with a finite set $\aSetOfPerformances \subset \allPerformances$ of performances, the works~\cite{Pierard2020Summarizing,Pierard2025Multidomain-arxiv} proposed a consistent way of averaging the score values for all scores. The summarized value for a score $\aScore$ is $\aScore(\summarizedPerformance)$, where $\summarizedPerformance\in\allPerformances$ is the performance given by $\frac{1}{|\aSetOfPerformances|} \sum_{\aPerformance \in \aSetOfPerformances} \aPerformance$. In other words, instead of arithmetically averaging score values, it has been proposed to average, or summarize, performances.
    \begin{itemize}
        
        \item For all unconditional probabilistic scores $\aScore$ (such as $\scorePFP$ and $\scorePFN$), we have $\expectedValueSymbol[\aScore] = \frac{1}{|\aSetOfPerformances|} \sum_{\aPerformance \in \aSetOfPerformances} \aScore(\aPerformance) = \aScore(\summarizedPerformance)$. The recommended $\beta$ depends only on the average, summarized, performance $\summarizedPerformance$.
        \begin{equation}
            \beta^2 = \frac{
                \scorePFP(\summarizedPerformance)
            }{
                \scorePFN(\summarizedPerformance)
            }
            \Leftrightarrow
            b = \frac{
                \scorePFP(\summarizedPerformance)
            }{
                \scorePFN(\summarizedPerformance) + \scorePFP(\summarizedPerformance)
            }
            \label{eq:heuristic-with-summarized-performance}
        \end{equation}

        \item In general, for the conditional probabilistic scores $\aScore$, one does not have $\aScore(\summarizedPerformance) = \frac{1}{|\aSetOfPerformances|} \sum_{\aPerformance \in \aSetOfPerformances} \aScore(\aPerformance)$. So, $\aScore(\summarizedPerformance)$ is not necessarily equal to $\expectedValueSymbol[\aScore]$. This is the case for $\scoreTNR$, $\scoreFPR$, $\scoreFNR$, and $\scoreTPR$. But as $\scorePFP(\summarizedPerformance) = \priorneg(\summarizedPerformance) \, \scoreFPR(\summarizedPerformance)$ and $\scorePFN(\summarizedPerformance) = \priorpos(\summarizedPerformance) \, \scoreFNR(\summarizedPerformance)$,
        \begin{equation}
            \beta^2 = \frac{
                \priorneg(\summarizedPerformance) \, \scoreFPR(\summarizedPerformance)
            }{
                \priorpos(\summarizedPerformance) \, \scoreFNR(\summarizedPerformance)
            }
            \Leftrightarrow
            b = \frac{
                \priorneg(\summarizedPerformance) \, \scoreFPR(\summarizedPerformance)
            }{
                \priorpos(\summarizedPerformance) \, \scoreFNR(\summarizedPerformance) + \priorneg(\summarizedPerformance) \, \scoreFPR(\summarizedPerformance)
            }
            \point
        \end{equation}
        Moreover, as $\scoreFNR(\summarizedPerformance) = 1 - \scoreTPR(\summarizedPerformance)$, we have
        \begin{equation}
            \beta^2 = \frac{
                \priorneg(\summarizedPerformance) \, \scoreFPR(\summarizedPerformance)
            }{
                \priorpos(\summarizedPerformance) ( 1 - \scoreTPR(\summarizedPerformance) )
            }
            \Leftrightarrow
            b = \frac{
                \priorneg(\summarizedPerformance) \, \scoreFPR(\summarizedPerformance)
            }{
                \priorpos(\summarizedPerformance) ( 1 - \scoreTPR(\summarizedPerformance) ) + \priorneg(\summarizedPerformance) \, \scoreFPR(\summarizedPerformance)
            }
            \point
            \label{eq:heuristic-in-roc-general-case}
        \end{equation}
        It should be stressed that, when the priors differ from one performance to another, the point $(\scoreFPR(\summarizedPerformance), \scoreTPR(\summarizedPerformance))$ does not coincide with $(\expectedValueSymbol[\scoreFPR], \expectedValueSymbol[\scoreTPR])$.
        By comparing \cref{eq:heuristic-in-roc-fixed-priors} with \cref{eq:heuristic-in-roc-general-case}, we see that the same visualization can be used. However, instead of looking at $(\expectedValueSymbol[\scoreFPR], \expectedValueSymbol[\scoreTPR])$, one has to look at $(\scoreFPR(\summarizedPerformance), \scoreTPR(\summarizedPerformance))$.
        
    \end{itemize}
    
\end{itemize}

\paragraph{Our heuristic recommends to rank \wrt $\scoreFOne$ when the average performance is unbiased (\ie, $\scorePrecision=\scoreRecall$).}
In the sense of Byrt \etal \cite{Byrt1993Bias}, a classifier is unbiased when the classifier predicts as much positive samples as there are in reality. The probability for the predicted class to be positive $\aPerformance(\{\sampleFP,\sampleTP\})$ equals the probability for the ground-truth class to be positive $\aPerformance(\{\sampleFN,\sampleTP\})$. Thus, $\scorePFP=\scorePFN$ and $\scorePrecision=\scoreRecall$. In ROC, this corresponds to a line passing through $(\scoreFPR,\scoreTPR)=(0,1)$ and $(\scoreFPR,\scoreTPR)=(\priorpos,\priorpos)$. When the performance is unbiased in average, the heuristic recommends $\beta^2 = 1$. This can be seen by taking $\expectedValueSymbol[\scorePFP]=\expectedValueSymbol[\scorePFN]$ in \cref{eq:heuristic-with-expected-values} or $\scorePFP(\summarizedPerformance)=\scorePFN(\summarizedPerformance)$ in \cref{eq:heuristic-with-summarized-performance}.

\paragraph{Our heuristic recommends to rank \wrt $\scoreSkewInsensitiveVersionFOne$ when the priors are fixed, and \wrt $\scoreFBeta$ with $\beta^2=\nicefrac{\priorneg}{\priorpos}$ in the general case, when the average performance is on the descending diagonal of ROC.} If the average performance is on the descending diagonal of ROC, we have $\expectedValueSymbol[\scoreFPR]=1-\expectedValueSymbol[\scoreTPR]$ when the priors are fixed, or $\scoreFPR(\summarizedPerformance)=1-\scoreTPR(\summarizedPerformance)$ in the general case. Thus, by looking at \cref{eq:heuristic-in-roc-fixed-priors,eq:heuristic-in-roc-general-case}, we see that the recommended value for $\beta^2$ is $\nicefrac{\priorneg}{\priorpos}$. When the priors are fixed, the F-score corresponding to this value induces the same performance ordering as the skew-insensitive version of $\scoreFOne$, $\scoreSkewInsensitiveVersionFOne$, defined in \cite{Flach2003TheGeometry} as $\scoreSkewInsensitiveVersionFOne = \frac{
    2 \scoreTPR
}{
    \scoreTPR + \scoreFPR + 1
}$.

A detailed use of our heuristic on the CADA-RRE example is shown, step by step, in \cref{fig:cada-rre-heuristic}.
\input{figs/cada_rre_heuristic}

%% file: figs/heuristic_in_ROC.tex
\begin{figure}
    \centering
    \includegraphics[width=0.32\linewidth]{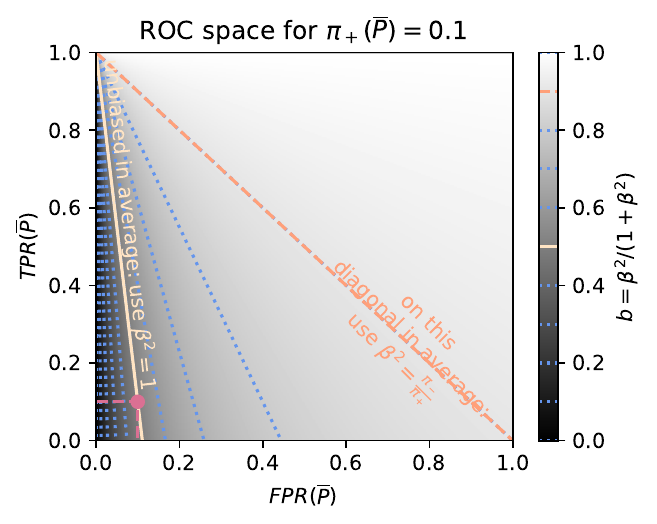}
    \includegraphics[width=0.32\linewidth]{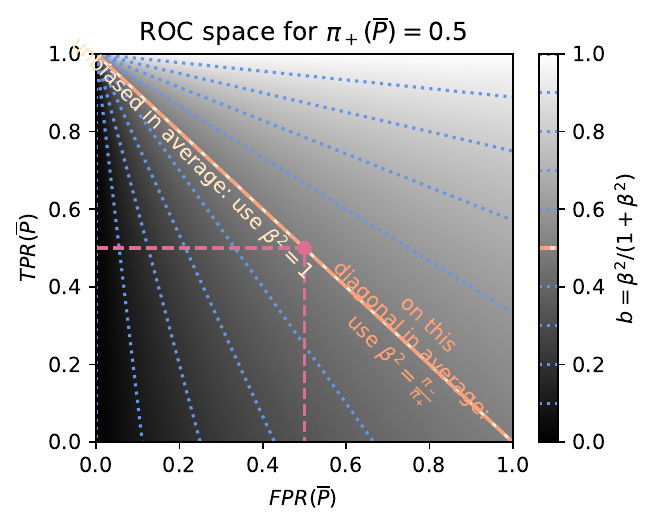}
    \includegraphics[width=0.32\linewidth]{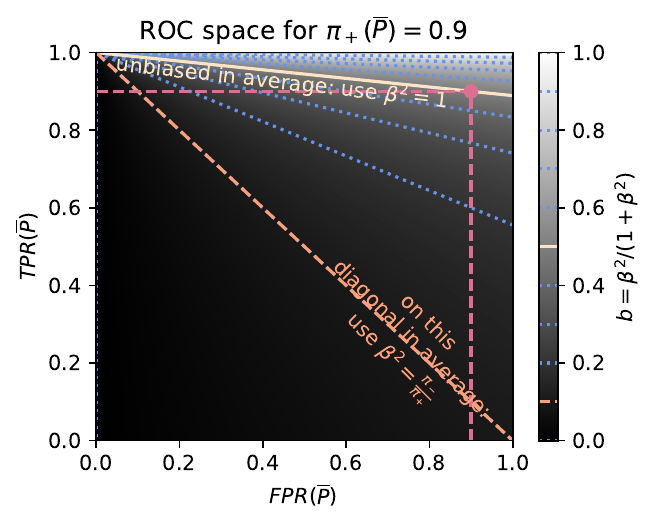}
    \caption{Visualization of our heuristic in ROC space. The recommended value for $b=\frac{\beta^2}{1+\beta^2}$ is the value at the point $(\expectedValueSymbol[\scoreFPR], \expectedValueSymbol[\scoreTPR])$ when the priors are fixed, and the value at the point $(\scoreFPR(\overline{\aPerformance}), \scoreTPR(\overline{\aPerformance}))$ in the general case. This visualization is specific to certain priors: the common priors when the priors are fixed, and $(\priorneg(\overline{\aPerformance}), \priorpos(\overline{\aPerformance}))$ in the general case. The more the prior $\priorpos$ of the positive class increases, the more the recommended value for $\beta$ decreases (compare with \cref{fig:roc_isometrics}).}
    \label{fig:heuristic-in-ROC}
\end{figure}

%% file: figs/cada_rre_heuristic.tex
\begin{figure}
    \centering
    \,
    \hfill
    \subfloat[Step 1. Point cloud in the ROC space showing the performances of the methods to rank. Note that all these performances correspond to fixed class priors.]{
        \includegraphics[width=0.45\linewidth]{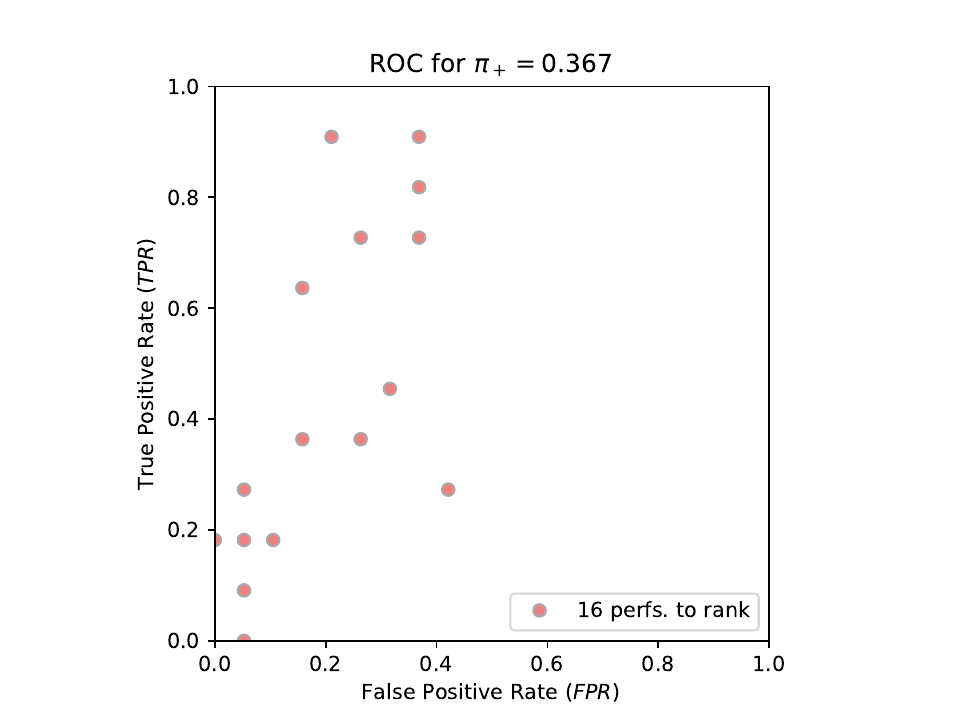}
    }
    \hfill
    \hfill
    \subfloat[Step 2. Computation of the summarized performance $\summarizedPerformance$ \cite{Pierard2020Summarizing}. As all performances correspond to fixed class priors, it is located at the centroid of the point cloud (bisque star).]{
        \includegraphics[width=0.45\linewidth]{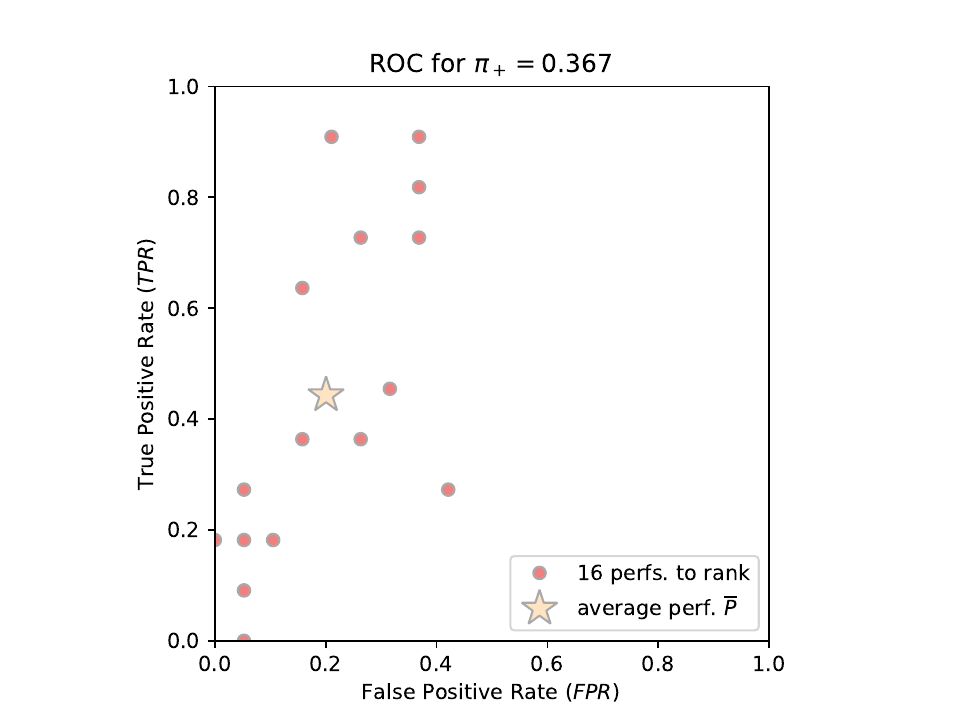}
    }
    \hfill
    \,
    \\
    \,
    \hfill
    \subfloat[Step 3. Use of our heuristic. One reads the recommended value for $b=\frac{\beta^2}{1+\beta^2}$ at $(\scoreFPR(\summarizedPerformance),\scoreTPR(\summarizedPerformance))$.]{
        \includegraphics[width=0.45\linewidth]{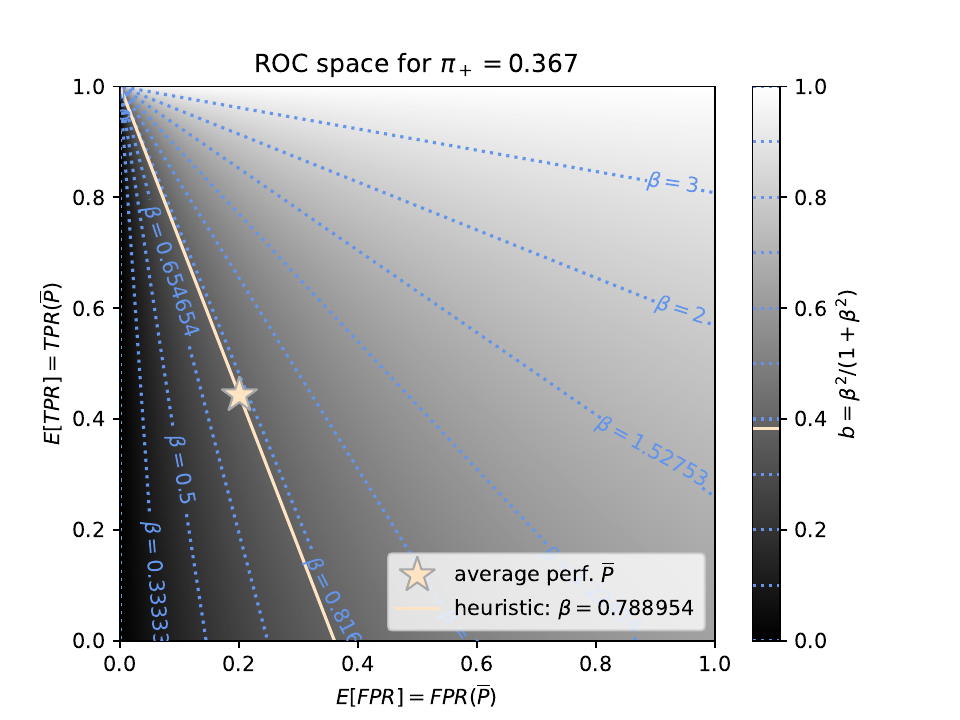}
    }
    \hfill
    \hfill
    \subfloat[Step 4. Drawing the score $\scoreFBeta$ for the recommended $\beta$ in the background of ROC, with superimposed isometrics depicting the performance ordering induced by it.]{
        \includegraphics[width=0.45\linewidth]{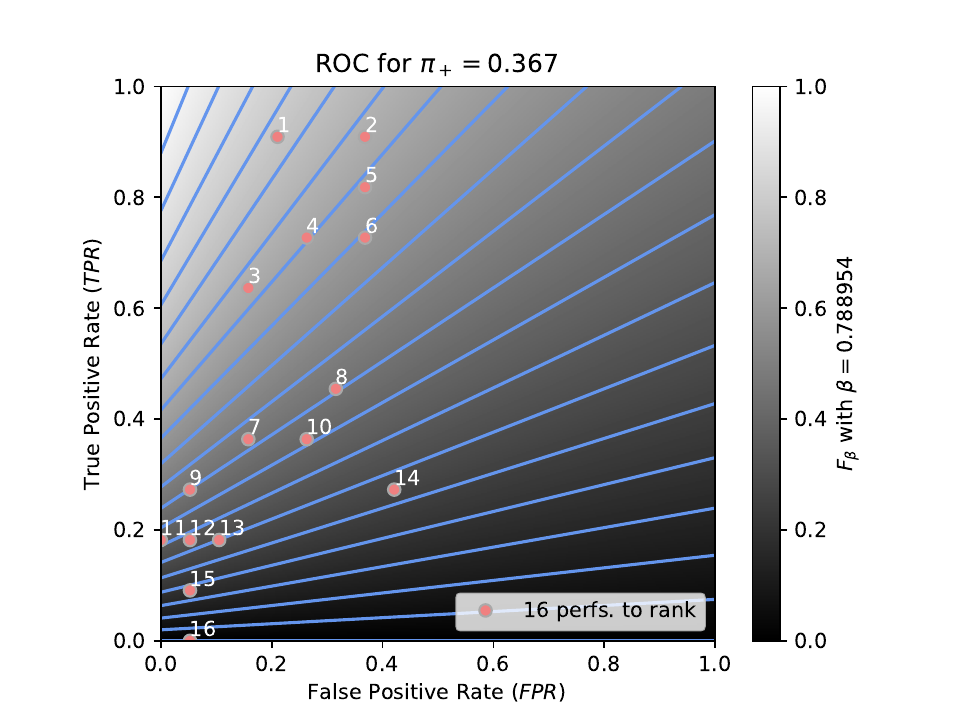}
    }
    \hfill
    \,
    \caption{Example of use of our heuristic our \linkCADARRE example (see \cref{sec:CADA-RRE}).}
    \label{fig:cada-rre-heuristic}
\end{figure}